\documentclass[aps,prc,onecolumn]{revtex4}
\usepackage{epsf,epsfig}

\newcommand{\be}{\begin{equation}}
\newcommand{\ee}{\end{equation}}
\newcommand{\bea}{\begin{eqnarray}}
\newcommand{\eea}{\end{eqnarray}}
\newcommand{\mybibitem}{\bibitem}

\newcommand{\ch}{{\rm ch}}
\newcommand{\sh}{{\rm sh}}
\newcommand{\acos}{{\rm acos}}
\newcommand{\Foo}{{_1\! F_1}}
\renewcommand{\vec}[1]{{\bf #1}}

\newcommand{\gton}{\mathrel{\lower.9ex \hbox{$\stackrel{\displaystyle 
>}{\sim}$}}} 
\newcommand{\lton}{\mathrel{\lower.9ex \hbox{$\stackrel{\displaystyle 
<}{\sim}$}}}

\newcommand{\ve}{{\bf e}}

\newcommand{\vx}{{\bf x}}

\newcommand{\vp}{{\bf p}}

\newcommand{\vo}{{\bf 0}}

\newcommand{\ttau}{{\tilde \tau}}
\newcommand{\tpi}{{\tilde\pi}}
\newcommand{\bs}{{\bar s}}
\newcommand{\cN}{{\cal N}}
\newcommand{\cO}{{\cal O}}

\newcommand{\etaOs} {\eta_s/s_{eq}}

\begin{document}

\title{The applicability of causal dissipative hydrodynamics to relativistic 
heavy ion collisions}

\author{Pasi Huovinen}
\affiliation{Department of Physics, University of Virginia, 
Charlottesville, VA 22904}
\affiliation{Physics Department, Purdue University, West Lafayette, IN 47907}

\author{Denes Molnar}
\affiliation{Physics Department, Purdue University, West Lafayette, IN 47907}
\affiliation{RIKEN BNL Research Center, Brookhaven National Laboratory, 
Upton, NY 11973}

\date{\today}

\begin{abstract}
We utilize nonequilibrium covariant transport theory to
determine the region of validity of causal Israel-Stewart dissipative
hydrodynamics (IS) and Navier-Stokes theory (NS) 
for relativistic heavy ion physics 
applications.
A massless ideal gas with $2\to 2$ interactions is considered in a 0+1D 
Bjorken scenario, appropriate
for the early longitudinal expansion stage of the collision.
In the scale invariant case of a constant 
shear viscosity to entropy
density ratio $\eta/s\approx const$, we find that Israel-Stewart theory is 
10\% accurate in calculating dissipative effects if {\em initially} the 
expansion timescale exceeds half the transport
mean free path $\tau_0/\lambda_{tr,0} \gton 2$. The same accuracy with
Navier-Stokes requires three times larger $\tau_0/\lambda_{tr,0} \gton 6$.
For dynamics driven by a constant cross section, on the other hand, 
about 50\% larger $\tau_0/\lambda_{tr,0} \gton 3$ (IS) and $9$ (NS) are needed.
For typical applications at RHIC energies 
$\sqrt{s_{NN}} \sim 100-200$~GeV, these limits imply that even the 
Israel-Stewart approach becomes marginal when
$\eta/s \gton 0.15$.
In addition, we find that the 'naive' approximation to Israel-Stewart theory,
which neglects products of gradients and dissipative quantities,
has an even smaller range of applicability than Navier-Stokes.
We also obtain analytic Israel-Stewart and Navier-Stokes solutions in 0+1D,
and present further tests for numerical dissipative hydrodynamics codes in
1+1, 2+1, and 3+1D based on generalized conservation laws.
\end{abstract}

\maketitle

\section{Introduction}

The realization that shear viscosity
is likely nonzero in 
general\cite{GyulassyDanielewicz85,etalimit,etalimitrevised},
and therefore the perfect (Euler) fluid 
paradigm\cite{uliQGP3,GyulassyMcLerranWP,ShuryakWP,StockerWP} 
of nuclear collisions at the Relativistic
Heavy Ion Collider (RHIC) 
could have significant viscous corrections\cite{dissipv2},
has fuelled great interest in studying dissipative
hydrodynamics\cite{Muronga,TeaneyQM2004,BRW,Heinzviscous,Romatschke,Kodama,%
Romatschkev2pt,Heinz2D,Teaney2D,QM2008}. 
Causality and stability problems\cite{unstableNS} exhibited by 
standard first-order 
relativistic Navier-Stokes hydrodynamics\cite{LandauHydro,DeGroot} 
steered most effort toward
application of the second-order Israel-Stewart (IS) approach\cite{IS1,IS2}.

However, unlike Navier-Stokes that comes from a rigorous 
expansion\cite{ChapmanEnskog} in
small gradients near equilibrium, 
the IS formulation is {\em not} a controlled expansion in some small parameter
(see Section~\ref{Sec:hytr}).
Moreover, though causality is restored in a region of hydrodynamic 
parameters, the stability of IS solutions 
is not necessarily guaranteed\cite{unstableIS}.
Therefore it is imperative to test the applicability of the IS approach
against a stable, nonequilibrium theory.

In this work we perform such a test utilizing the fully stable and causal 
covariant transport approach\cite{ZPC,Bin:Et,nonequil,v2}. 
We focus on the special case
of $2\to 2$ transport and a
longitudinally boost invariant system\cite{Bjorkenhydro} 
with transverse translational
symmetry, i.e, 0+1 dimensions. Follow-up studies in higher dimensions,
such as our earlier comparison between transport and ideal hydrodynamics
in 2+1D\cite{dissipv2}, will be pursued in the future.

A similar study by Zhang and 
Gyulassy\cite{Bin:Et} compared kinetic theory and
Navier-Stokes results. Here we compare to the causal IS solutions.
In addition, we provide a series of tests and semi-analytic 
approximations that demonstrate the general behavior of IS solutions,
which can be utilized to verify the accuracy of numerical IS 
solutions.

The paper is structured as follows. We start with reviewing the relationship
between hydrodynamics
and covariant transport (Sec.~\ref{Sec:hytr}), then proceed to discuss the 
Israel-Stewart equations (Sec.~\ref{Sec:IS}). The basic observables studied 
here
are introduced in Sec.~\ref{Sec:obs}, while the main results from the 
hydro-transport comparison
are presented in Sec.~\ref{Sec:IS_validity}, together with implications
for heavy-ion collisions. Many details are deferred to Appendices 
\ref{App:ai}-\ref{App:transp}. We highlight here 
the generalized conservation
laws derived in App.~\ref{App:ConsLaws}, 
and the detailed study of Israel-Stewart
and Navier-Stokes solutions in App.~\ref{App:IS} utilizing numerical and
analytic methods.

\section{Hydrodynamics and covariant transport}
\label{Sec:hytr}

Hydrodynamics describes a system in terms of a few local, 
macroscopic variables\cite{LandauHydro}, such as energy
density $\varepsilon(x)$, pressure $p(x)$, 
charge density $n(x)$ and flow velocity 
$u^\mu(x)$. The equations of motion are energy-momentum and charge
conservation
\be
\partial_\mu T^{\mu\nu}(x) = 0 \quad , \qquad \partial_\mu N^\mu(x) = 0 \ ,
\label{hydroeq}
\ee
and the equation of state $p(e,n)$.
{\em Ideal} (Euler) hydrodynamics assumes local equilibrium
in which case
\be
T^{\mu\nu}_{LR,id} = diag(\varepsilon,p,p,p) \quad , \qquad  N^\mu_{LR,id} 
= (n, \vec 0) 
\qquad\qquad [u^\mu_{LR} = (1,\vec 0)]
\label{idealTandN}
\ee 
in the fluid rest frame LR. 
Extension of the theory with additive corrections {\em linear}
in flow and temperature 
gradients\cite{LandauHydro}
\bea
\delta T^{\mu\nu}_{NS} &=&
\eta_s (\nabla^\mu u^\nu + \nabla^\nu u^\mu 
        - \frac{2}{3}\Delta ^{\mu\nu} \partial^\alpha u_\alpha) 
+ \zeta \Delta^{\mu\nu} \partial^\alpha u_\alpha \quad , \quad
\label{NS_T}
\\
\delta N^\mu_{NS} &=& 
\kappa_q \left(\frac{n T}{\varepsilon+p}\right)^2 
\nabla^\mu \left(\frac{\mu}{T}\right)
\qquad\qquad\qquad\qquad\qquad 
(\Delta^{\mu\nu} \equiv g^{\mu\nu} - u^\mu u^\nu \quad , \quad 
\Delta^\mu \equiv \Delta^{\mu\nu} \partial_\nu)
\label{NS_N}
\eea
leads via (\ref{hydroeq}) to the relativistic Navier-Stokes (NS) equations.
(We use the Landau frame convention $u_\mu \delta T^{\mu\nu} \equiv 0$ 
throughout this paper,
i.e., the flow velocity is chosen such that momentum flow vanishes in the LR 
frame.)
Here $\eta_s(e,n)$ and $\zeta(e,n)$ 
are the shear and bulk viscosities,
while $\kappa_q(e,n)$ is the heat conductivity of the matter.
The most notable feature of NS theory relative to the ideal case
is dissipation, i.e., 
entropy production.
For consistency, the dissipative corrections (\ref{NS_T})-(\ref{NS_N})
must be small, otherwise nonlinear terms
and higher gradients should also be considered.

It is crucial that the above hydrodynamic equations 
can indeed be obtained from a general nonequilibrium theory,
namely on-shell covariant transport\cite{DeGroot,Bin:Et,nonequil,v2}.
For a one-component system
the covariant transport
equation reads
\be
p^\mu \partial_\mu f(x,\vp) = S(x, \vp) + C[f,f](x,\vp)
\label{BTE}
\ee
where the source term 
$S$ specifies the initial
conditions and $C$ is the collision term.
Throughout this paper we consider the Boltzmann limit%
\footnote{Bose (+) or Fermi (-) statistics can be included in a straightforward
manner via substituting
$f_1 g_2 \to f_1 g_2 (1\pm f_3) (1\pm g_4)$ and 
$f_3 g_4 \to f_3 g_4 (1\pm f_1) (1\pm g_2)$ in the collision term
(\ref{BTE_C}). The various hydrodynamic limits can
then be derived analogously to the Boltzmann case,
if one makes the convenient replacement
$\phi\to (1\pm f_{eq})\phi$ in (\ref{phiDef}).
}
with binary $2\to 2$ rates
\be
C[f,g](x,\vp_1) \equiv \int\limits_2\!\!\!\!
\int\limits_3\!\!\!\!
\int\limits_4\!\!
\left(f_3 g_4 - f_1 g_2\right)
W_{12\to 34} \  \delta^4(p_1{+}p_2{-}p_3{-}p_4)
\label{BTE_C}
\ee
where $f_i \equiv f(x,\vp_i)$ and 
$\int_i \equiv \int d^3 p_i / (2E_i)$.
For dilute systems, $f$ is the phasespace distribution of 
quasi-particles, while the transition probability 
$W = (1/\pi) s (s-4m^2) d\sigma/dt$
is given by the scattering matrix 
element\cite{DeGroot}.
Our interest here, on the other hand, is the theory
{\em near its hydrodynamic limit}, $W \to \infty$.
In this case, ``particles'' and ``interactions'' do not necessarily 
have to be physical but could simply be
mathematical constructs adjusted to reproduce the transport properties
of the system near equilibrium\cite{minvisc}.
The main advantage of transport theory is its 
ability to dynamically interpolate between the dilute and opaque limits.

The Euler and Navier-Stokes hydrodynamic equations 
follow from a rigorous expansion of (\ref{BTE}) 
in {\em small gradients} near local equilibrium
\be
f(x,\vp) = f_{eq}(x,\vp) [ 1 + \phi(x,\vp)]
\qquad\qquad\qquad |\phi| \ll 1 \ , 
\quad |p^\mu \partial_\mu \phi| \ll |p^\mu \partial_\mu f_{eq}| / f_{eq} \ ,
\label{phiDef}
\ee
and substitution of {\em moments} of the solutions
\be
N^\mu(x) \equiv \int \frac{d^3p}{p_0} p^\mu \; f(x,\vp) \quad , \qquad 
T^{\mu\nu}(x) \equiv \int \frac{d^3p}{p_0} p^\mu p^\nu \; f(x,\vp) \ ,
\ee
into (\ref{hydroeq}).
The $0$-th order $\phi = 0$ reproduces ideal
hydrodynamics. The first order result is the solution to the linear 
integral equation
\be
p^\mu 
\partial_\mu f_{eq}(x,\vp) 
= 2 C[f_{eq},f_{eq}\phi_{NS}](x,\vp) 
\label{firstCE}
\ee
and leads to the Navier-Stokes equations.

Unfortunately the relativistic Navier-Stokes equations 
are parabolic and therefore 
acausal. A solution proposed by Mueller\cite{Mueller} and later
extended by Israel and Stewart \cite{IS1,IS2} converts the NS equations
into relaxation equations for the shear stress
$\pi^{\mu\nu}$, bulk pressure $\Pi$, and heat flow $q^\mu$.
The dissipative corrections
\be
\delta T^{\mu\nu} \equiv \pi^{\mu\nu} - \Pi \Delta^{\mu\nu} 
\quad , \qquad
\delta N^{\mu} \equiv - \frac{n}{\varepsilon+p} q^\mu
\qquad (u_\mu q^\mu = 0, \ u_\mu \pi^{\mu\nu} = u_\mu \pi^{\nu\mu} = 0)
\ee
dynamically relax on microscopic time scales
$\tau_\pi(e,n)$, $\tau_\Pi(e,n)$, $\tau_q(e,n)$ towards
values dictated by gradients in flow 
and temperature.
Causality is satisfied in a region of parameter space, however,
stability is not guaranteed\cite{unstableIS}.

More importantly, unlike the Euler and NS equations, the 
Israel-Stewart approach is not a controlled approximation 
to the transport theory (\ref{BTE}). Instead of an expansion in some
small parameter, it corresponds to a
quadratic ansatz \cite{Grad,IS2} for the deviation from local equilibrium
\be
\phi_{G}(x,\vp) = D^\mu(x) \frac{p_\mu}{T} 
+ C^{\mu\nu}(x) \frac{p_\mu p_\nu}{T^2} \ ,  \qquad 
(u_\mu D^\mu = 0 = u_\mu C^{\mu\nu} u_\nu)
\label{phiG}
\ee
where $D^\mu$ and $C^{\mu\nu}$ are determined by the local dissipative
corrections $\pi^{\mu\nu}$, $\Pi$, and $q^\mu$%
\footnote{The alternative formulation based on transient 
thermodynamics\cite{IS1,IS2} also lacks a small expansion parameter.}.
In contrast, the Chapman-Enskog solution (\ref{firstCE}) 
contains all orders in momentum.
An evident pathology of the quadratic form (\ref{phiG}) 
is that, in general, $\phi_G$
is not bounded from below and thus the phase space density becomes
negative at large momenta (cf. (\ref{phiDef}) and (\ref{phiGpi})).
Furthermore, the two approaches give different results not
only for the relaxation times\cite{IS2,DeGroot}, e.g., 
\be
\tau_\pi^{NS} = 0 \qquad , \qquad \qquad \tau_\pi^{IS} = \frac{3\eta_s}{2p} \ ,
\label{taupi}
\ee 
but also for the transport coefficients themselves.
For an energy-independent isotropic cross section
and ultrarelativistic particles ($T\gg m$) the difference is 
small\cite{DeGroot},
e.g.,
\be
\eta_{s}^{NS} \approx 0.8436 \frac{T}{\sigma_{tr}}  \qquad , \qquad\qquad
\eta_{s}^{IS} = \frac{4T}{5\sigma_{tr}} \ ,
\label{etashear}
\ee
but can be large for more complicated interactions. 
Here $\sigma_{tr}\equiv \int 
d\Omega_{cm} \sin^2 \theta_{cm} d\sigma/d\Omega_{cm}$
 is the transport cross section 
(for isotropic, $\sigma_{tr} = 2\sigma_{TOT}/3$).

In the following Sections we analyze IS hydrodynamic solutions
analytically and numerically, and test the accuracy of the IS approximation
via comparison to solutions from full $2\to 2$ transport theory.

\section{Israel-Stewart hydrodynamics and boost invariance}
\label{Sec:IS}

\subsection{Israel-Stewart equations}

There seems to be some confusion regarding Israel-Stewart theory\cite{IS1,IS2} 
in the recent 
literature, therefore we start with reviewing the key ingredients.
The starting point of Israel and Stewart (IS)
is an entropy current that includes terms up to quadratic order in
dissipative quantities%
\footnote{Unlike we here, Israel and Stewart choose
$g^{\mu\nu} = diag(-1,1,1,1)$, $\Delta^{\mu\nu} = g^{\mu\nu} + u^\mu u^\nu$.}
\be
 S^\mu = u^\mu\left[s_{eq} - \frac{1}{2T}
                             \left(\beta_0\Pi^2 - \beta_1 q_\nu q^\nu
                                 + \beta_2 \pi^{\lambda\nu}\pi_{\lambda\nu}
                             \right)\right]
        + \frac{q^\mu}{T}\left(\frac{\mu n}{\varepsilon+p}
                     + \alpha_0 \Pi \right)
        - \frac{\alpha_1 q_\nu\pi^{\nu\mu}}{T} 
\label{SmuIS}
\ee
(we follow the Landau frame convention).
Here $\mu$ is the chemical potential, and $s_{eq}$ is the entropy
density in local equilibrium.
The coefficients $\{\alpha_i(e,n)\}$ and $\{\beta_i(e,n)\}$ encode additional
matter properties that complement the equation of state and the transport 
coefficients. Most importantly, $\{\beta_i\}$ control 
the relaxation times for dissipative quantities:
\be
\tau_\Pi = \zeta \beta_0 \ , \qquad  \tau_q = \kappa_q T \beta_1 \ , \qquad
\tau_\pi = 2\eta_s \beta_2 \ .
\label{reltimes_IS}
\ee
 The entropy current 
and relaxation times in Navier-Stokes theory are recovered
when all the coefficients are set to zero
$\beta_0 = \beta_1 = \beta_2 = 0 = \alpha_0 = \alpha_1$
(but as discussed previously, the IS and NS transport coefficients differ
in general).

The requirement of entropy
non-decrease ($\partial_\mu S^\mu \ge 0$), which IS satisfy via a 
positive semi-definite%
\footnote{Positive semi-definiteness follows from the general properties
$q^\mu q_\mu \le 0$ and $\pi^{\mu\nu} \pi_{\mu\nu} \ge 0$.}
quadratic ansatz
\be
 T \partial_\mu S^\mu = \frac{\Pi^2}{\zeta} - \frac{q_\mu q^\mu}{\kappa_q T}
                    + \frac{\pi_{\mu\nu}\pi^{\mu\nu}}{2\eta_s}  \ge 0 \ ,
\label{SdivIS}
\ee
leads to the identification
of the dissipative currents:
\bea
 \Pi & = & \zeta\left[-\nabla_\mu u^\mu
           - \frac{1}{2}\Pi T\partial_\mu\left(\frac{\beta_0 u^\mu}{T}\right)
           - \beta_0 D\Pi
           + \alpha_0 \partial_\mu q^\mu
           - a_0^\prime q^\mu D u_\mu \right]    \label{bulkdef}       \\
 q^\mu &=& -\kappa_q T \Delta^{\mu\nu}\left[
           \frac{Tn}{\varepsilon + p} \nabla_\nu \left(\frac{\mu}{T}\right)
           +\frac{1}{2} q_\nu T
             \partial_\lambda\left(\frac{\beta_1 u^\lambda}{T}\right)
           +\beta_1 D q_\nu
           + \alpha_0 \nabla_\nu \Pi
\label{heatdef}
           - \alpha_1 \partial^\lambda \pi_{\lambda\nu}
           - a_0 \Pi D u_\nu + a_1 \pi_{\lambda\nu} D u^\lambda \right] \\
 \pi^{\mu\nu} & = & 2\eta_s\left[\nabla^{\langle\mu} u^{\nu\rangle}
                   -\frac{1}{2}\pi^{\mu\nu}T
                    \partial_\lambda\left(\frac{\beta_2 u^\lambda}{T}\right)
                   -\beta_2 \langle D\pi^{\mu\nu} \rangle
                   - \alpha_1 \nabla^{\langle\mu} q^{\nu\rangle}
                   +a_1^\prime q^{\langle\mu} D u^{\nu\rangle}\right] 
                          \label{sheardef}
\\ 
a'_i &\equiv& 
\left.\frac{\partial (\alpha_i/T)}{\partial (1/T)}\right|_{\mu/T = const} 
  - a_i \ .
\label{ai_prime}
\eea
Here $D \equiv u^\mu \partial_\mu$ 
and the $\langle\rangle$ brackets denote
traceless symmetrization and projection orthogonal to the flow
\be
A^{\langle\mu\nu\rangle}
\equiv 
\frac{1}{2} \Delta^{\mu\alpha}\Delta^{\nu\beta} 
    (A_{\alpha\beta} + A_{\beta\alpha}) 
- \frac{1}{3} \Delta^{\mu\nu} \Delta_{\alpha\beta} A^{\alpha\beta} 
\ .
\ee
The new matter coefficients $\{a_i(e,n)\}$
are needed to describe how contributions from the $q^\mu \Pi$ and $q_\nu \pi^{\mu\nu}$ terms in (\ref{SmuIS}) are split 
between the bulk pressure and heat flow, and the heat flow 
and shear stress evolution equations, 
respectively
(in other words, 
a whole class of equations of motion generates the same amount of entropy 
- see Appendix \ref{App:ai}).

Notice that the time-derivatives of heat flow,
$q^\mu$, and shear stress tensor, $\pi^{\mu\nu}$ are not expressed 
{\em explicitly} in
(\ref{heatdef})-(\ref{sheardef}) - instead, orthogonal projections
to the flow velocity vector appear (cf. Eqs.~(8a)-(8c) in \cite{IS1}). 
Reordering the 
equations 
explicitly for the time derivatives 
gives rise to terms $-u^\mu q_\nu D u^\nu$ and
$-(\pi^{\lambda\mu} u^\nu + \pi^{\lambda\nu} u^\mu) D u_\lambda$.
There is therefore no need for a kinetic theory
treatment~\cite{Romatschkebig} to derive these terms. They were missed
in Ref.~\cite{Heinzviscous}, but
they are already present in standard IS theory as
a trivial consequence of the product rule of differentiation and
the orthogonality of the flow velocity and shear stress/heat flow.

As we saw above, the Israel-Stewart procedure 
only determines the equations of motion up to 
nonequilibrium terms that do not contribute to entropy production.
In kinetic theory, 
further such terms arise~\cite{IS2} when the vorticity
\be
\omega^{\mu\nu} \equiv \frac{1}{2} \Delta^{\mu\alpha}\Delta^{\nu\beta}
(\partial_\beta u_\alpha - \partial_\alpha u_\beta) 
\ee
is nonzero.
Including the vorticity terms, the complete set of
evolution equations for the dissipative currents are:
\bea
 D\Pi & = & -\frac{1}{\tau_\Pi}\left(\Pi +\zeta\nabla_\mu u^\mu\right) 
\label{IS_EOMPi}
\\
      &   & -\frac{1}{2}\Pi\left(\nabla_\mu u^\mu
                                 + D \ln \frac{\beta_0}{T}\right) \nonumber
\\
      &   & +\frac{\alpha_0}{\beta_0}\partial_\mu q^\mu
            -\frac{a_0^\prime}{\beta_0} q^\mu D u_\mu \nonumber \\
 Dq^\mu & = & -\frac{1}{\tau_q}\left[q^\mu
              +\kappa_q\frac{T^2 n}{\varepsilon + p}\nabla^\mu 
                   \left(\frac{\mu}{T}\right)
                               \right]
               -u^\mu q_\nu D u^\nu     
\label{IS_EOMq} 
\\
        &   &  -\frac{1}{2}q^\mu\left(\nabla_\lambda u^\lambda      \nonumber
                                 + D \ln \frac{\beta_1}{T}\right)
               -\omega^{\mu\lambda} q_\lambda  \\
        &   &  -\frac{\alpha_0}{\beta_1} \nabla^\mu \Pi
               +\frac{\alpha_1}{\beta_1}(\partial_\lambda \pi^{\lambda\mu} 
                     + u^\mu   \pi^{\lambda\nu} \partial_\lambda u_\nu)
               +\frac{a_0}{\beta_1} \Pi D u^\mu
\nonumber
               -\frac{a_1}{\beta_1} \pi^{\lambda\mu} D u_\lambda \\
 D\pi^{\mu\nu} & = & -\frac{1}{\tau_\pi}
                     \left(\pi^{\mu\nu}-2\eta\nabla^{\langle\mu}u^{\nu\rangle}
                           \right)
                   -(\pi^{\lambda\mu}u^\nu+\pi^{\lambda\nu}u^\mu)D u_\lambda
\label{IS_EOMpi}
\\
              &   & -\frac{1}{2}\pi^{\mu\nu}\left(\nabla_\lambda u^\lambda
\nonumber
                                            + D \ln \frac{\beta_2}{T}\right)
                    - 2 \pi_\lambda^{\ \langle\mu}\omega^{\nu\rangle\lambda}
\\
              &   &-\frac{\alpha_1}{\beta_2}
                          \nabla^{\langle\mu} q^{\nu\rangle}
                   +\frac{a_1^\prime}{\beta_2}
                          q^{\langle\mu} D u^{\nu\rangle} \ .       \nonumber
\eea
We will refer to these equations as ``complete IS''. 
If we ignore their tensorial structure, the equations have the
general form
\be
\dot X = - \frac{1}{\tau_X} (X - X_{NS}) + X\,Y_X + Z_X
\label{IS_struct}
\ee
for each dissipative quantity $X$, where $X_{NS}$ is the value of $X$ in 
Navier-Stokes theory and $Y_X$, $Z_X$ are 
given by the ideal hydrodynamic fields and
 dissipative quantities other than $X$. Therefore, Israel-Steward theory 
describes relaxation towards Navier-Stokes on a characteristic time $\tau_{X}$,
 {\em provided} $|Y_X|\tau_X \ll 1$ and $|Z_X| \tau_X \ll |X_{NS}|$.

In the last step of their derivation, Israel and Stewart neglect
the first terms of the second lines, the ones
with a factor 1/2 
(this gives the Landau frame equivalent to (7.1a)-(7.1c) in \cite{IS2}), 
because they expect to study astrophysical systems 
with small gradients $|\partial^\mu u^\nu + \partial^\nu u^\mu| / T \ll 1$,
$|\partial^\mu e| / (Te) \ll 1$, $|\partial^\mu n| / (Tn) \ll 1$,
near a global (possibly rotating) equilibrium state. The neglected terms are
then products of small gradients and the dissipative quantities.
We will refer to this approximation as ``naive IS''%
\footnote{Note that in Ref.~\cite{Song}, the equivalent set of equations are
  called "full IS" and "simplified IS".}.
{\em In heavy ion physics applications,
on the other hand, gradients $\partial^\mu u^\nu / T$,
$|\partial^\mu e| / (Te)$,
$|\partial^\mu n| / (Tn)$ 
at early times $\tau \sim 1$~fm are large $\sim \cO(1)$,
and therefore cannot be ignored. 
Nevertheless, hydrodynamics may still be applicable, provided
the viscosities are unusually small $\etaOs \sim 0.1$, 
$\zeta/s_{eq} \sim 0.1$, where $s_{eq}$ is the entropy density in 
local equilibrium.} In this case, dissipative effects
 are still moderate, for example, pressure corrections from
Navier-Stokes theory (\ref{NS_T})
\be
\frac{\delta T^{\mu\nu}_{NS}}{p} \approx 
\left( 2 \frac{\eta_s}{s_{eq}} \frac{\nabla^{\langle \mu} u^{\nu\rangle}}{T}
+ \frac{\zeta}{s_{eq}} \frac{\nabla_\alpha u^\alpha}{T} \right) 
\frac{\varepsilon + p}{p} 
\sim \cO\left(\frac{8\eta_s}{s_{eq}},\frac{4\zeta}{s_{eq}}\right) \ .
\ee
Heat flow effects can also be estimated based on 
(\ref{NS_N})
\be
\frac{\delta N^{\mu}_{NS}}{n} \approx
\frac{\kappa_q T}{s_{eq}} \frac{n}{s_{eq}} \frac{\nabla^\mu (\mu/T)}{T} \ .
\ee
For RHIC energies and above, at midrapidity, the correction is rather small 
even for large $\kappa_q$
because the baryon density and therefore $\mu_B/T$ is very low. 
For example, in a recent ideal fluid calculation
at RHIC energy~\cite{Huovinen:2007xh}, these ratios were
$n_B/s \approx 2.2\cdot 10^{-3}$ and $\mu_B/T \approx 0.2$ 
in order to reproduce
the observed net baryon spectra.
These choices are also supported by thermal model analyses of particle
ratios which lead to $\mu_B/T \approx 0.17$~\cite{Manninen:2008mg}.

\subsection{Viscous equations of motion for  
longitudinally boost-invariant 0+1D dynamics}

At this point we specialize the equations of motion 
to a viscous, longitudinally 
boost-invariant%
\footnote{By a boost-invariant system we mean a
system which obeys the scaling flow, $\vec v=(0, 0, z/t)$, where all scalar
quantities are independent of coordinate rapidity 
$\eta\equiv (1/2) \ln [(t+z)/(t-z)]$, and where all
vector and tensor quantities can be obtained from their values at
$\eta=0$ by an appropriate Lorentz boost.}
system with transverse translation invariance and
vanishing bulk viscosity:
\bea
\dot n + \frac{n}{\tau} &=& 0 \quad \Leftrightarrow 
\quad n(\tau) = \frac{\tau_0\, n(\tau_0)}{\tau}
\label{n_bj1D}
\\
\dot e + \frac{e + p}{\tau} &=& 
 - \frac{\pi_L}{\tau} 
\label{e_bj1D}
\\
\tau_\pi \dot \pi_{L} 
+ \pi_{L}
  \left[1 + \frac{\tau_\pi}{2\tau} 
        + \frac{\eta_s T}{2} \dot{\left(\frac{\tau_\pi}{\eta_s T}\right)}
  \right]
 &=&  -\frac{4\eta_s}{3\tau} 
\label{piL_bj1D}
\\
\pi_T &=& -\frac{\pi_L}{2} \ .
\label{piT_bj1D}
\eea 
This special case is well known in the
literature~\cite{Muronga,Romatschkebig,Dumitru:2007qr}
as a useful approximation to the early longitudinal expansion stage of 
a heavy ion collision for observables near midrapidity $\eta \approx 0$.
Here $\tau \equiv \sqrt{t^2-z^2}$ is the Bjorken proper time, and
the 'dot' denotes $d/d\tau$. 
$\pi_L$ and $\pi_T$ are the viscous corrections to the
longitudinal and transverse pressure, i.e. the $\pi_{zz}$ and
$\pi_{xx}=\pi_{yy}$ components of the shear stress tensor evaluated at
local rest frame%
\footnote{I.e., in the often 
employed curvilinear $\tau-\eta-x-y$ coordinates we have
$\pi_{\eta\eta} = \tau^2  \pi_L$.}%
, respectively. All the other components of the stress
tensor are zero due to symmetry. There is no equation for heat
flow because the symmetries of the system--- longitudinal 
boost-invariance,
axial symmetry in the transverse plane and $\eta \to -\eta$ reflection
symmetry---force the heat flow to be zero everywhere. We have
chosen to ignore bulk viscosity since shear viscosity is expected to
dominate at RHIC. In the following we also concentrate on a system
of massless particles, where bulk viscosity is zero in
general. It is worth noticing that these equations are identical in
both Eckart and Landau frames, but in less restricted systems where
heat flow is nonzero, Eckart and Landau frames differ.

To simplify the discussion and to facilitate comparison with transport
results, from here on we concentrate on a system of massless particles
with only $2\to 2$ interactions. Particle number is then conserved and
the equation of state is
\be
e = 3p \ , \qquad T = \frac{p}{n}  \ .
\label{EOS}
\ee
Now the density equation decouples entirely and we end up with two
coupled equations for the equilibrium pressure and the viscous
correction $\pi_L$. The shear stress relaxation time (\ref{taupi})
and the shear
viscosity (\ref{etashear}) can be recast with the {\em transport} 
mean free path  $\lambda_{tr} \equiv 1/(n \sigma_{tr})$
as
\be
\eta_s = C n T \lambda_{tr} \ , \qquad 
\tau_\pi = \frac{3C}{2} \lambda_{tr} \ ,\qquad C\approx \frac{4}{5} \ ,
\ee
and (\ref{e_bj1D})-(\ref{piL_bj1D}) can then be
written as
\bea
\dot p + \frac{4p}{3\tau} &=& - \frac{\pi_L}{3\tau}
\label{EOMp}
\\
\dot \pi_{L} 
+ \frac{\pi_{L}}{\tau}
  \left(\frac{2\kappa(\tau)}{3} + \frac{4}{3} + \frac{\pi_L}{3p} \right) 
 &=&  -\frac{8p}{9 \tau}  \ ,
\label{EOMpiL}
\eea
where
\be
\kappa(\tau) \equiv \frac{K(\tau)}{C} = \frac{nT\tau}{\eta_s}\ , \qquad 
K(\tau) \equiv \frac{\tau}{\lambda_{tr}(\tau)}
\label{def_K} \ .
\ee
The ratio of expansion and scattering
timescales $K$ controls how well ideal and/or dissipative
hydrodynamics applies. This is essentially the {\em inverse} 
of the ratio of shear stress relaxation time to 
hydrodynamic timescales $\tau_\pi / \tau = 3/(2\kappa)$.
$K$ is as well a generalization of the Knudsen number 
$L/\lambda$,
since the shortest spatial length scale is given by gradients 
in longitudinal direction $L \sim 1/ (\partial_z u_z) \sim \tau$.
It is also a measure of the shear viscosity to 
entropy density ratio because for a system in chemical equilibrium
$s_{eq} = 4 n$ and thus
\be
\frac{\eta_s}{s_{eq}} =  \frac{T\tau}{4\kappa}
\ee
(see Sec.~\ref{Sec:HI} for the general case).

Similar treatment to relativistic Navier-Stokes theory leads to
\be
  \pi_L = -\frac{4\eta_s}{3\tau} = - \frac{4p}{3\kappa}
\label{piL_NS}
\ee
and the equation of motion
\be
\dot p + \frac{4p}{3\tau} = \frac{4}{9\kappa(\tau)}\frac{p}{\tau} \ .
\label{EOM_NS}
\ee
As discussed in the previous Section, the viscosities in NS and IS theories
differ and, therefore, $\kappa$ in (\ref{EOM_NS}) is not identical
to the one in (\ref{EOMpiL}). We will ignore the difference because
in our case it is only $\approx 5$\%.

Finally, we note that in the ``naive'' Israel-Stewart approximation 
(\ref{EOMpiL}) changes to
\be
\dot \pi_{L} 
+ \frac{2\kappa(\tau) \pi_{L}}{3 \tau}
 =  -\frac{8p}{9 \tau} \ .
\label{EOMpiL_naive}
\ee

\section{Basic observables}
\label{Sec:obs}

Here we introduce the basic observables investigated in this study, and discuss
their evolution based on the analytic Israel-Stewart and Navier-Stokes
solutions of Appendix~\ref{App:IS}.
It is important to emphasize that our observations will hold only during 
the longitudinal expansion stage of heavy ion collisions.
After some time $\tau \sim R/c_s$, transverse expansion sets in 
and hydrodynamics, whether Israel-Stewart or Navier-Stokes,
eventually breaks down because for expansion in three dimensions
$\lambda_{tr} \sim \tau^3 / \sigma$,
i.e., $\kappa \sim \sigma / \tau^2 \to 0$
in the hadronic world where cross sections are bounded.
It is interesting to note that $\etaOs \approx const$ 
would not decouple even for a three-dimensional expansion 
(because in that case $T \propto 1/\tau$ and thus 
$\lambda_{tr} \propto \eta/p \propto \tau$, while 
$\tau_{exp} \equiv 1/(\partial_\mu u^\mu) \propto \tau$, i.e., 
$\kappa \sim const$).

Throughout this section and the rest of the paper, 
the subscript '0' refers to the value of
quantities at the initial time $\tau_0$ (e.g., $A_0 \equiv A(\tau_0)$).
The most important parameters in the problem
are 
the initial Knudsen number $K_0$, or the corresponding $\kappa_0$,
and the initial shear stress to pressure ratio $\xi_0 \equiv \pi_{L,0}/p_0$.

\subsection{Pressure anisotropy}
\label{Sec:Rp}

The magnitude of dissipative corrections can be quantified
through the ratio of viscous longitudinal shear and equilibrium pressure
\be
 \xi \equiv \frac{\pi_L}{p} \ .
\ee
A suitable equivalent measure 
is the pressure anisotropy coefficient
\be
R_p \equiv \frac{p_L}{p_T}  = \frac{1+\xi}{1-\xi/2} \ , 
\ee
which is the ratio of the transverse and longitudinal pressures
$p_T \equiv p - \pi_L/2$, $p_L \equiv p + \pi_L$.
In the ideal hydro limit
the anisotropy is unity $R_p \to 1$.

The time-evolution of the anisotropy coefficient is given by
the equations of motion (\ref{EOMp}) and
(\ref{EOMpiL}):
\be
\dot R_p = -\frac{4}{3\tau} \, \frac{4+3\,\kappa\, \xi}{(2-\xi)^2} 
 \ .
\label{Rdot_IS}
\ee
Thus, in IS theory the pressure
anisotropy is a constant of motion
when the viscous stress is equal to its Navier-Stokes value (\ref{piL_NS}),
or at asymptotically late times $\tau \to \infty$.
In contrast, from NS
\be
R_p^{NS} = \frac{3 \kappa - 4}{3\kappa + 2} \ ,
\label{R_NS}
\ee
which is only constant for $\kappa(\tau) = const$ (constant cross section), 
or in the ideal hydro limit $\kappa \to \infty$
(in which case $R_p\to 1$).
From the above it also follows that in the special case of 
our boost invariant scenario, if the cross section is constant 
and the shear stress starts from its Navier-Stokes value, 
Navier-Stokes and Israel-Stewart theory {\em coincide}.

\subsection{Longitudinal work}
\label{Sec:work}

Dissipation also affects the evolution of the equilibrium (or average) 
pressure. From (\ref{EOMp}), for ideal hydro evolution the pressure drops as 
$p(\tau) \propto \tau^{-4/3}$ due to longitudinal work. 
In the viscous case, the work done by the system is smaller because
the viscous correction to the longitudinal pressure is usually 
negative $\pi_L < 0$.
Therefore, pressure decreases slower than in ideal hydro, and deviations from 
the ideal evolution, such as the ratio 
\be
\frac{p(\tau)}{p_{ideal}(\tau)} \equiv \frac{T(\tau)}{T_{ideal}}
\qquad\qquad ({\rm for\ conserved\ particle\ number})
\ee
can be used to quantify dissipative effects.

Studies in the past\cite{Bin:Et,nonequil} 
have analyzed a closely related quantity, 
the transverse energy per unit rapidity, $dE_T/d\eta$. 
This is simply a combination
of
the pressure anisotropy and deviation from ideal pressure
\be
\frac{dE_T}{d\eta} = \frac{3\pi T}{4} \frac{dN}{d\eta} 
\left(1 - \frac{5}{16}\xi\right) 
= \frac{3\pi T_0}{4} 
\frac{dN}{d\eta} \left(\frac{\tau_0}{\tau}\right)^{-1/3} 
\frac{p(\tau)}{p_{ideal}(\tau)} \frac{3[7 + R_p(\tau)]}{8[2 + R_p(\tau)]}
\ee
provided the quadratic ansatz (\ref{phiG}) is applicable 
(see Appendix \ref{App:Et}).

We can make a few generic observations based on 
the analytic Israel-Stewart and Navier-Stokes 
results (\ref{piL_constsi})-(\ref{p_constsi}), (\ref{p_constsi_NS}), 
(\ref{p_tau23})-(\ref{piL_tau23}), and (\ref{p_tau23_NS})
from Appendix~\ref{App:IS}. 
For a constant cross section,
$p/p_{ideal}$ grows without bound - dissipative
corrections keep accumulating forever. The influence of the
initial shear stress,
or equivalently shear stress to pressure ratio $\xi_0 \equiv \xi(\tau_0)$, 
is of $\cO(\xi_0/\kappa_0$) and thus vanishes in the large $\kappa_0$ limit.
At late times $\tau \gg \tau_0$, for $K_0 \gton 2$ and
not too large initial shear stress to pressure ratio $|\xi_0| \ll 2\kappa_0$
\be
\left(\frac{p}{p_{ideal}}\right)_{\sigma=const} 
\approx N \left(\frac{\tau}{\tau_0}\right)^\beta
\ , \qquad \beta \approx \frac{4}{9\kappa_0} 
\left(1 - \frac{2}{3\kappa_0^2}\right) \ , \quad
N \approx 1 - \frac{2}{3\kappa_0^2} + \frac{4}{3\kappa_0^4} 
            - \frac{\xi_0}{2\kappa_0} \ ,
\label{p_approx_constsi}
\ee  
i.e., for $\tau \approx 10 \tau_0$ and $K_0 = 2$ the 
accumulated pressure increase is
$p/p_{ideal} \approx 1.3$, while $p/p_{ideal} \approx 1.15$ if $K_0 = 5$.
For a scale invariant system with $\etaOs \approx const$, 
on the other hand, dissipative effects are more moderate for 
the same $K_0$ and at late times
approach a finite upper bound 
\be
\left(\frac{p}{p_{ideal}}\right)_{\eta/s \approx const} 
\approx \left[1-\frac{2}{3\kappa_0} 
                \left(\frac{\tau_0}{\tau}\right)^{2/3}
        \right]
        \left(1+\frac{2}{3\kappa_0} - \frac{\xi_0}{2\kappa_0}\right)
\ \to \ 
1 + \frac{2}{3\kappa_0} - \frac{\xi_0}{2\kappa_0} \ .
\label{p_approx_etas}
\ee 
This is because scale invariant systems 
turn more and more ideal hydrodynamic as they evolve
(as long as their expansion is only longitudinal).
For the same $K_0 = 2$ and 5 with $\xi_0 \approx 0$, the bounds are modest, 
$p/p_{ideal} \lton 1.25$ and $\lton 1.1$,
respectively.

\subsection{Entropy}
\label{Sec:S}

Another quantitative measure of the importance of 
dissipative effects is entropy production.
Here we consider an ultra-relativistic system 
(thus $\Pi = 0$ and $\beta_2 = 3/(4p)$) with
$2\to 2$ interactions, 1D Bjorken boost invariance, 
and transverse translational,
axial, and $\eta \to -\eta$ reflectional symmetries (imply $q^\mu =0$).
Therefore, the entropy current (\ref{SmuIS}) simplifies to
\be
S^\mu 
= \bs u^\mu  \ , \qquad \bs = s_{eq} - \frac{9\pi_L^2}{16 p T}
\label{Smu1D}
\ee
where
\be
s_{eq} = n (4 - \chi) \ , \qquad
\chi \equiv \ln \frac{n}{n_{eq}(T)} = \frac{\mu}{T}
\label{muT}
\ee
and
\be
n_{eq}(T) = \frac{g}{\pi^2} T^3
\label{neq}
\ee
is the particle density in chemical equilibrium at temperature $T$
for massless particles of degeneracy $g$ in the Boltzmann limit.
Dissipative contributions in the entropy density $\bs$ 
are {\em negative}, in accordance with the fundamental 
principle of maximal entropy in equilibrium.

The equations of motion (\ref{EOMp})-(\ref{EOMpiL}) 
imply an entropy production rate
\be
\partial_\mu S^\mu = \frac{1}{\tau}\partial_\tau(\tau \bar s) = 
\frac{3\kappa n}{4\tau}\xi^2 \ge 0 \ .
\label{entropy_production}
\ee
Equivalently, the entropy per unit rapidity
\be
\frac{dS}{d\eta} \equiv \tau A_T \, \bs
\label{dSdeta}
\ee
never decreases
\be
\partial_\tau\left(\frac{dS}{d\eta}\right) 
= \frac{3\kappa}{4\tau} 
\frac{dN}{d\eta} \xi^2  \ge 0 \ .
\ee
Here $A_T$
is the transverse area of the system, and in 
the last step we substituted the rapidity density
$dN/d\eta = \tau A_T n$.
Equation (\ref{dSdeta}) is a special case of
a {\em generalized conservation law} (\ref{conslaw-rap})
applied to the entropy current $S^\mu$
\be
\tau \int dx_T^2\,  \partial_\mu S^\mu
= \partial_ \tau
\left(\tau \, \int dx_T^2\,  S_0^{LR} \right) 
- \partial_\eta 
\int dx_T^2 \, S_3^{LR} \ .
\ee 
Analogous relations can be obtained for the energy, momentum,
and charge density. In 0+1D these are quite trivial 
- they respectively  
reproduce (\ref{EOMp}),
give identically zero, and $dN/d\eta = const$.
In higher dimensions, however, the generalized conservation laws 
present important constraints that {\em any} solution must satisfy at 
{\em all} times and, 
therefore,
they are ideal for testing the accuracy of 
numerical solutions at each time step (see Appendix~\ref{App:ConsLaws}).

{\em Only the complete set of Israel-Stewart equations gives the correct rate 
of entropy production}.
The 'naive' approximation 
does not guarantee a monotonically increasing entropy
\be 
(\partial_\mu S^\mu)^{naive\ IS} 
= \frac{3\kappa n}{4\tau} \xi^2 \left(1 - \frac{\xi+4}{2\kappa}\right) \ ,
\ee
unless $\kappa$ is sufficiently large and, away from equilibrium, 
it underpredicts for a given $\xi$ the entropy production rate%
\footnote{
This however 
does {\em not} imply that the 'naive' IS equations always underpredict
the {\em total} integrated entropy change over a finite time interval.
The time evolution of $\xi(\tau)$ in the 'naive' approach differs in general 
from that in the complete theory.
}
(since $\xi < -1$ is unphysical).
In contrast, the second law of thermodynamics does hold for Navier-Stokes
\be 
(\partial_\mu S^\mu)^{NS} =   \frac{3\kappa n}{4\tau} \xi_{NS}^2 \ge 0 \ .
\ee
The NS result is the same as (\ref{entropy_production}) but with the 
shear stress restricted to its
Navier-Stokes value.
We note that in Israel-Stewart theory 
the naive entropy per unit rapidity,
defined using the {\em equilibrium} 
entropy density
\be
\frac{dS'}{d\eta} = s_{eq} \, \tau A_T 
\ee
does {\em not} increase monotonically.
Rather, it increases (decreases) for negative (positive) $\pi_L$.

Based on 
the analytic Israel-Stewart and Navier-Stokes 
results in Appendix~\ref{App:IS}, we can outline 
general expectations for the entropy evolution.
For a constant cross section by late times $\tau \gg \tau_0$ the entropy
increase relative to the initial entropy
is logarithmic with time
\be
\left[\frac{(dS/d\eta)}{(dS/d\eta)_0}\right]_{\sigma = const} - 1
\approx 
\frac{1}{4-\chi_0}\left(3\ln\frac{p}{p_{ideal}} - \frac{9\xi^2}{16}\right)
\approx  \frac{1}{4-\chi_0} 
             \left(3\beta \ln \frac{\tau}{\tau_0} 
                   - \frac{3}{\kappa_0^2} + \frac{16}{3\kappa_0^4} 
                   - \frac{3\xi_0}{2\kappa_0}\right)
\ee
where we considered
 initial conditions not too far from local equilibrium.
E.g., by $\tau \approx 10 \tau_0$
with $K_0 = 2$ and 
chemical equilibrium initial conditions 
$\approx 20$\% entropy is produced, while $\approx 10$\% with $K_0 = 5$.
For a scale invariant system with $\etaOs = const$, 
on the other hand, entropy production is slower
for the same $K_0$ and saturates at late times
\be
\left[\frac{(dS/d\eta)}{(dS/d\eta)_0}\right]_{\eta/s \approx const} 
- 1 \approx  \frac{1}{4-\chi_0}  
\frac{2}{\kappa_0} 
\left[1 - \left(\frac{\tau_0}{\tau}\right)^{2/3} - \frac{3\xi_0}{4}\right]
\ \to \  
\frac{1}{4-\chi_0}  
\frac{2}{\kappa_0} 
\left(1 - \frac{3\xi_0}{4}\right)
= \frac{2}{T_0\tau_0} \frac{\eta_s}{s_{eq}} \left(1 - \frac{3\xi_0}{4}\right)
\ee 
For the same $K_0 = 2$ and 5 (and $\xi_0 \approx 0$), 
the entropy increase by $\tau = 10 \tau_0$ is smaller,
$\approx 15$ and $\approx 6$\%,
respectively.
Based on this simple analytic formula for entropy production,
we also confirm the results of Ref.~\cite{Dumitru:2007qr},
which considered IS hydrodynamics with a unique initial condition 
$\xi_0 \approx - 16/(9T_0\tau_0) \times \etaOs$ where 
$T_0 \approx 0.39\ {\rm GeV} \times (0.14\ {\rm fm}/\tau_0)^{1/3}$ and $\tau_0$
was varied between 0.5 and 1.5~fm.

\section{Region of validity for dissipative hydrodynamics}
\label{Sec:IS_validity}

Here we determine the region of validity of dissipative hydrodynamics
via comparison to full nonequilibrium two-body transport 
theory\cite{ZPC,Bin:Et,nonequil,v2}. We consider two scenarios:
Scenario I with a constant cross section, 
which is least favorable for hydrodynamics; 
and Scenario II with a 
growing $\sigma \propto \tau^{2/3}$, which is
the most optimistic for applicability of hydrodynamics and is 
very close to $\etaOs = const$ as we show in Appendix \ref{App:IS}.
In the same Appendix we also study a scenario with 
$\sigma \propto 1/T^2$ that turns out to be close to Scenario II 
but with stronger dissipative effects,
and discuss analytic Navier-Stokes and (approximate) Israel-Stewart solutions.

{\em Due to scalings of the equations of motion, 
the results presented here are rather general.}
Equations (\ref{EOMp})-(\ref{EOMpiL}) 
are invariant under rescaling of time,
and/or joint rescaling of the pressures $p$ and $\pi_L$, provided the
dimensionless $\kappa$ depends only on $p$, $\pi_L$, $\tau/\tau_0$ and
no additional scales (all solutions studied here satisfy this
condition). The same scalings are exhibited by the transport\cite{nonequil}.
For a physically reasonable $p_0 > 0$, it is
therefore convenient to 
consider dimensionless pressure variables 
$\tilde p(\tau) \equiv p(\tau)/p_0$ and
$\pi_L(\tau)/p_0$, for which the solutions only depend on $\ttau
\equiv \tau/\tau_0$, $\kappa_0\equiv K_0/C$ and the initial condition
$\xi_0 \equiv  \pi_{L,0}/p_0$.

Unless stated otherwise, we initialize the transport based on
the quadratic form (\ref{phiG}). In our case of an
ultrarelativistic system ($e=3p$) in the Boltzmann limit
with vanishing bulk pressure
and heat flow
\be
D^\mu = 0 \ , \quad 
C^{\mu\nu} = \frac{\pi^{\mu\nu}}{8p} \qquad \qquad \Rightarrow \qquad \qquad
\phi_G(\eta=0,\vp) = \frac{\xi}{16} \frac{2 p_z^2 - p_\perp^2}{T^2} \ ,
\label{phiGpi}
\ee
where $p_\perp \equiv \sqrt{p_x^2 + p_y^2}$ is the transverse momentum.
We ensure nonnegativity of the phase space distribution via a 
$\Theta$-function
\be
f(\eta=0,\vp,\tau=\tau_0) 
= \frac{F(\xi)}{A_T\tau_0} \frac{dN}{d\eta} \frac{e^{-p/T}}{8\pi T^3} 
\, [1+\phi_G(\eta,\vp)] \, 
\Theta(1+\phi_G(\eta,\vp))
\ee
where $A_T$ is the transverse area of the system
(with the elimination of negative phase space contributions,
a normalization factor $F(\xi)\le 1$ 
 is needed to set a given $dN/d\eta$).
The cutoff does not affect the general scalings of transport solutions 
but does influence the initial pressure 
anisotropy (for example, values
$R_p = 0.3$ and 1.75 set based on (\ref{phiGpi})
change to $R_p \approx 0.476$ and 1.693 when the cutoff is applied).
Therefore, we 
initialize hydrodynamics with a shear stress $\pi_L$ 
that gives the same initial pressure anisotropy as the transport.

The transport solutions
were obtained using the MPC algorithm\cite{MPC},
which employs the particle subdivision 
technique to maintain covariance\cite{ZPC,nonequil}. 
Transverse translational invariance was maintained in the calculation through
periodic boundary conditions in the two transverse directions. A longitudinal
boost invariant system was initialized in a coordinate rapidity
interval $-5 < \eta < 5$, and observables were calculated via averaging
over $-2 < \eta < 2$ with proper Lorentz boosts of local 
quantities to $\eta = 0$.

\subsection{Pressure anisotropy}
\label{Sec:R_comp}

\begin{figure}
\epsfysize=10cm
\epsfbox{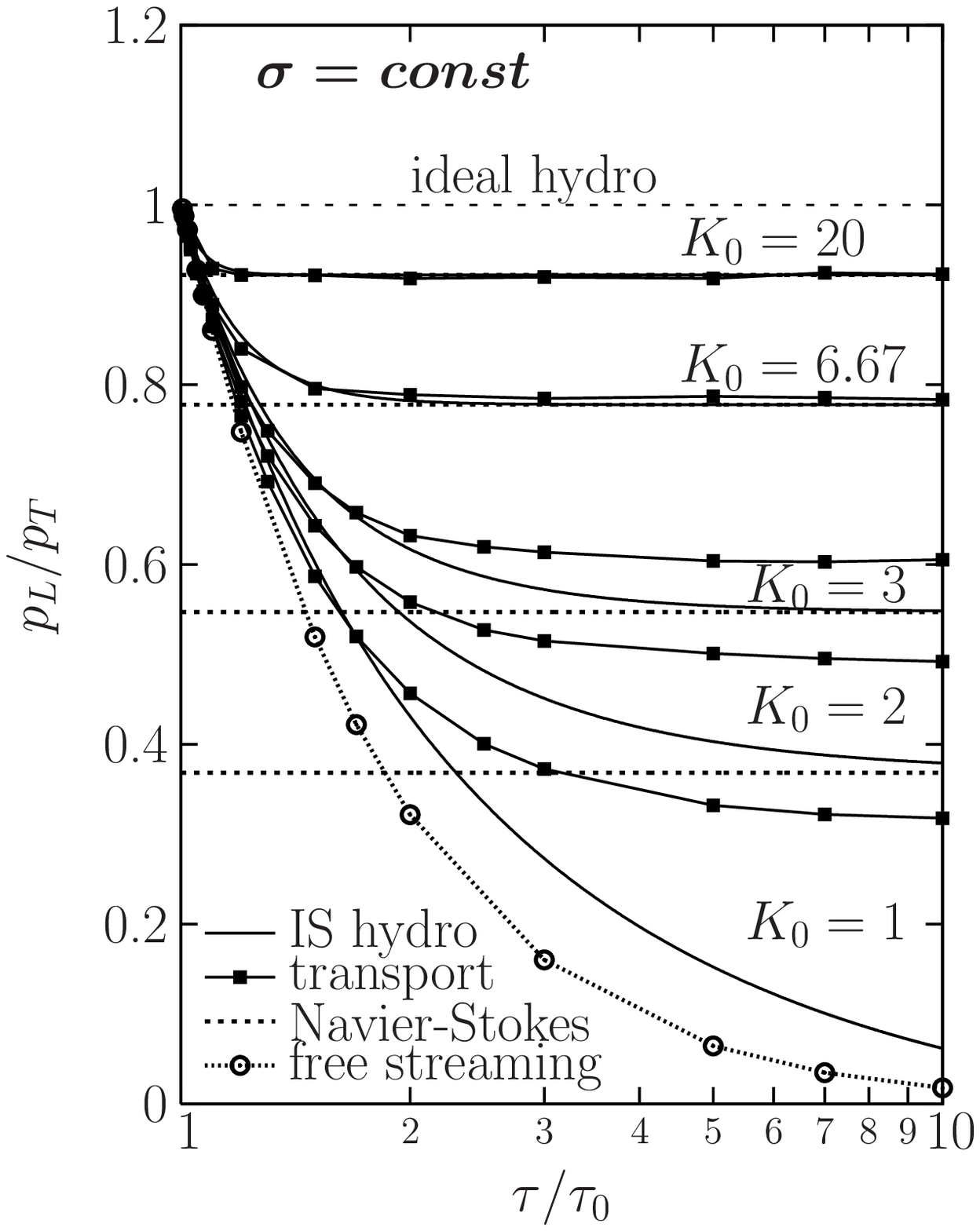}
\epsfysize=10cm
\epsfbox{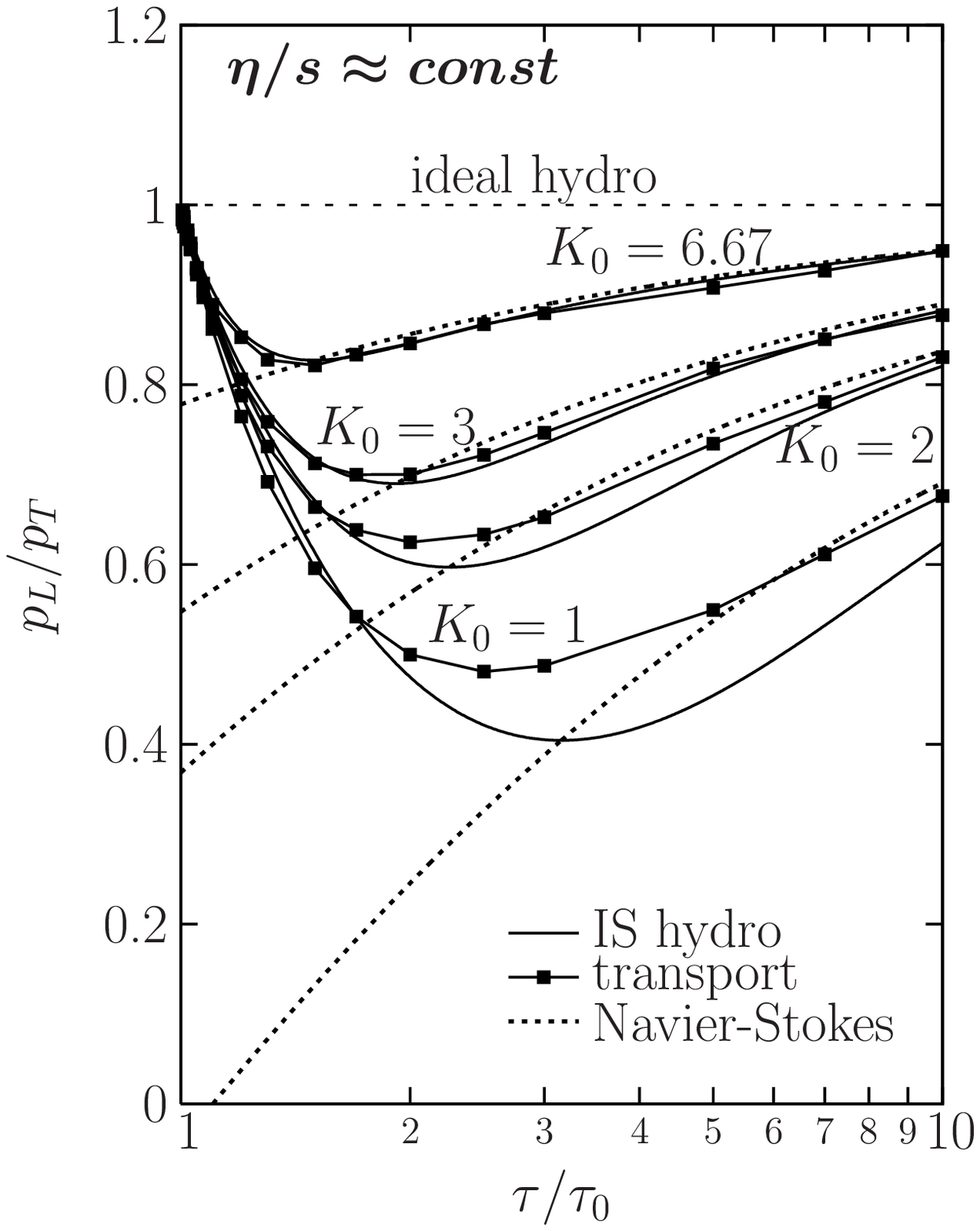}
\vskip 0.3cm
\caption{Time evolution of pressure anisotropy $R_p \equiv p_L/p_T$ 
from covariant
transport (solid lines with symbols) 
and Israel-Stewart dissipative hydrodynamics 
(solid lines) 
as a function of $K \equiv \tau / \lambda_{tr}(\tau)$,
from local equilibrium initial conditions $\pi_L(\tau_0) = 0$.
Results for Navier-Stokes (dotted lines) and free streaming 
(dotted line with circles) are also shown. 
{\em Left: } $\sigma = const$ scenario, for which
the curves are labeled by 
$K (\tau) = const = K_0 = 1$, 2, 3, 6.67, and 20. 
For $K = 1$, the Navier-Stokes result is negative and therefore not visible.
{\em Right:} $\sigma \propto \tau^{2/3}$ scenario, 
for which $\etaOs \approx const$ and the curves
are labeled by the initial $K_0 = K(\tau_0) = 1$, 2, 3, and 6.67. 
}
\label{Fig:R_comp}
\end{figure}

Figure~\ref{Fig:R_comp} shows the pressure anisotropy $p_L/p_T$ evolution
as a function of the rescaled proper time $\ttau = \tau/\tau_0$
from the transport (solid lines with symbols) and Israel-Stewart 
hydrodynamics (solid lines without symbols) with local equilibrium initial
condition. 
The left panel shows calculations for the $\sigma=const$ scenario. 
For $K_0 =1$, the anisotropy from IS hydro starts to
fall rapidly below the transport above $\tau \gton 2\tau_0$ and it is a 
factor $\sim 5$ smaller by late $\tau \sim 10 \tau_0$.
Clearly, the system cannot stay near equilibrium when the rate of 
scatterings equals the expansion rate.
With increasing $K_0$, the undershoot becomes smaller and gradually 
vanishes as $K_0 \to \infty$. The difference is only 
$\sim 10$\% already at $K_0 = 3$, and is rather small by $K_0 \approx 7$.

The right panel shows the same but for the growing cross section 
scenario with $\etaOs \approx const$. The situation of course improves
because in this case $K$ increases with time. For $K_0 = 1$, IS 
hydro undershoots the pressure anisotropy from the transport 
only by $\sim 20$\% and 
the differences vanish at late times (since in this case 
both theories converge to 
$R_p = 1$ as $\tau \to \infty$).
$\sim 10$\% accuracy is achieved already for $K_0 = 2$, while for $K_0 = 3$,
IS hydro is accurate to a few percent.

\begin{figure}
\epsfysize=10cm
\epsfbox{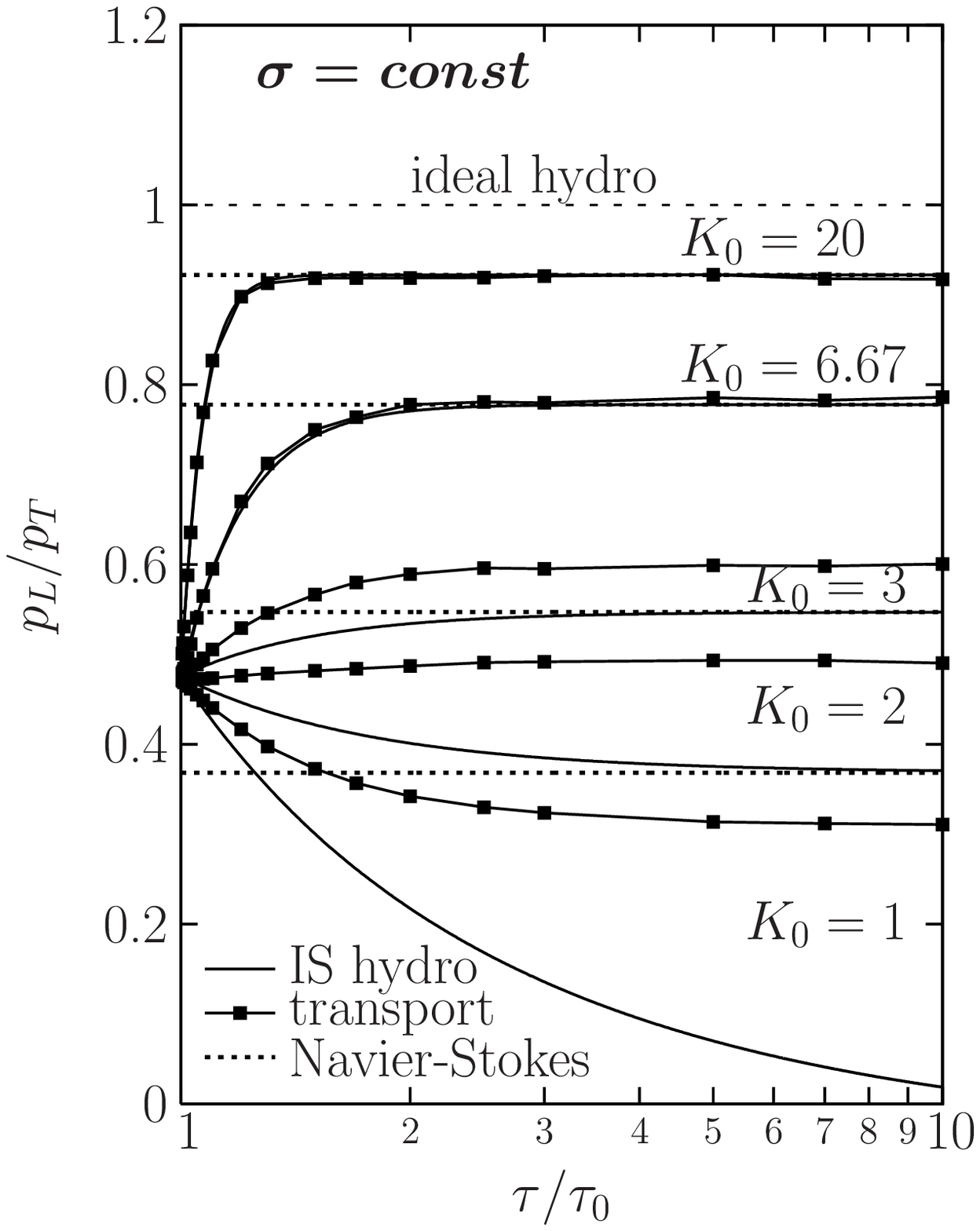}
\epsfysize=10cm
\epsfbox{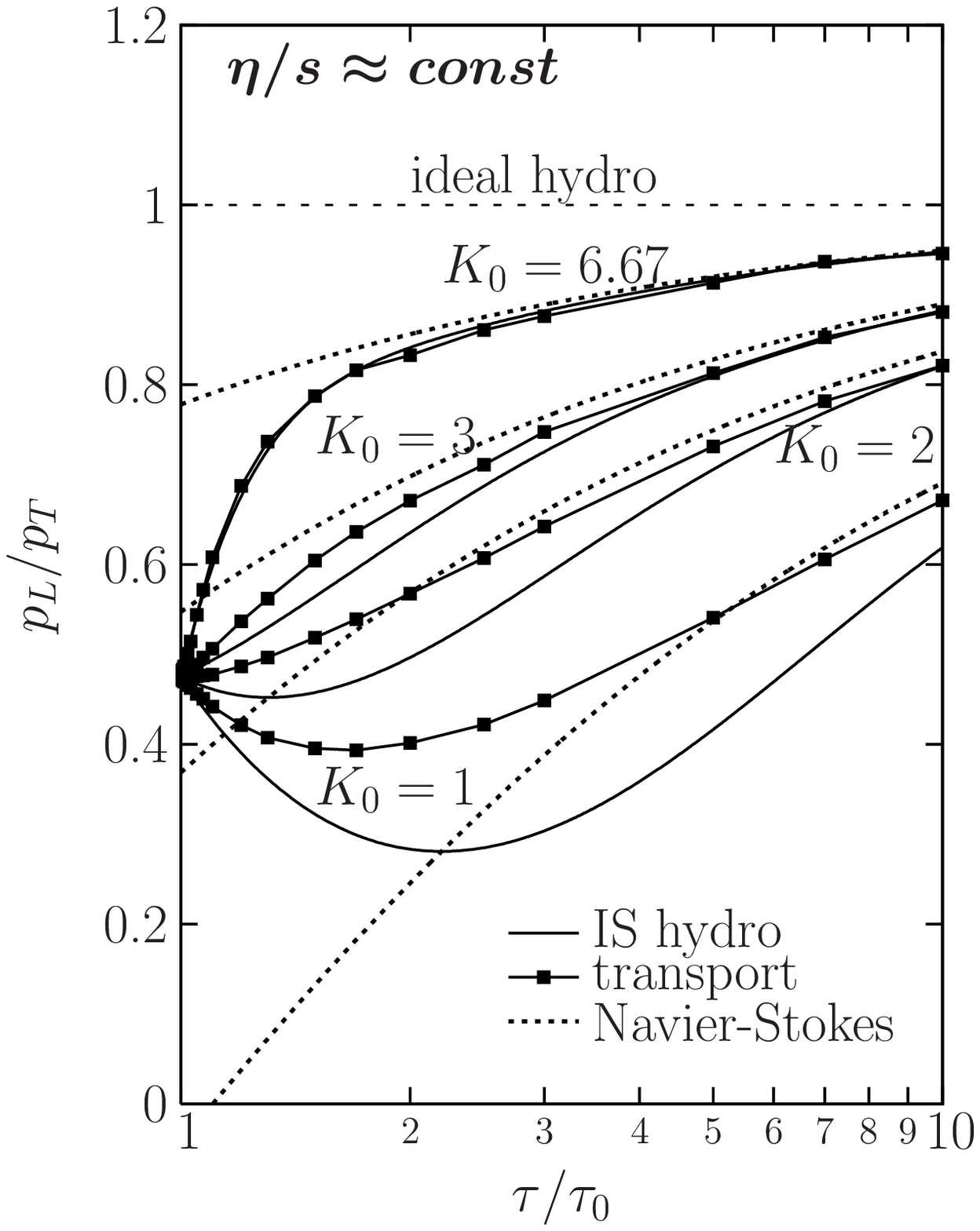}
\vskip 0.3cm
\caption{Same as Fig.~\ref{Fig:R_comp} but for an initial pressure anisotropy
 $R_p(\tau_0) = 0.476$ ($\xi_0 = -0.423$). In the left plot, the Navier-Stokes
curve for $K_0 = 1$ is negative and therefore not visible.}
\label{Fig:R03_comp}
\end{figure}

\begin{figure}
\epsfysize=10cm
\epsfbox{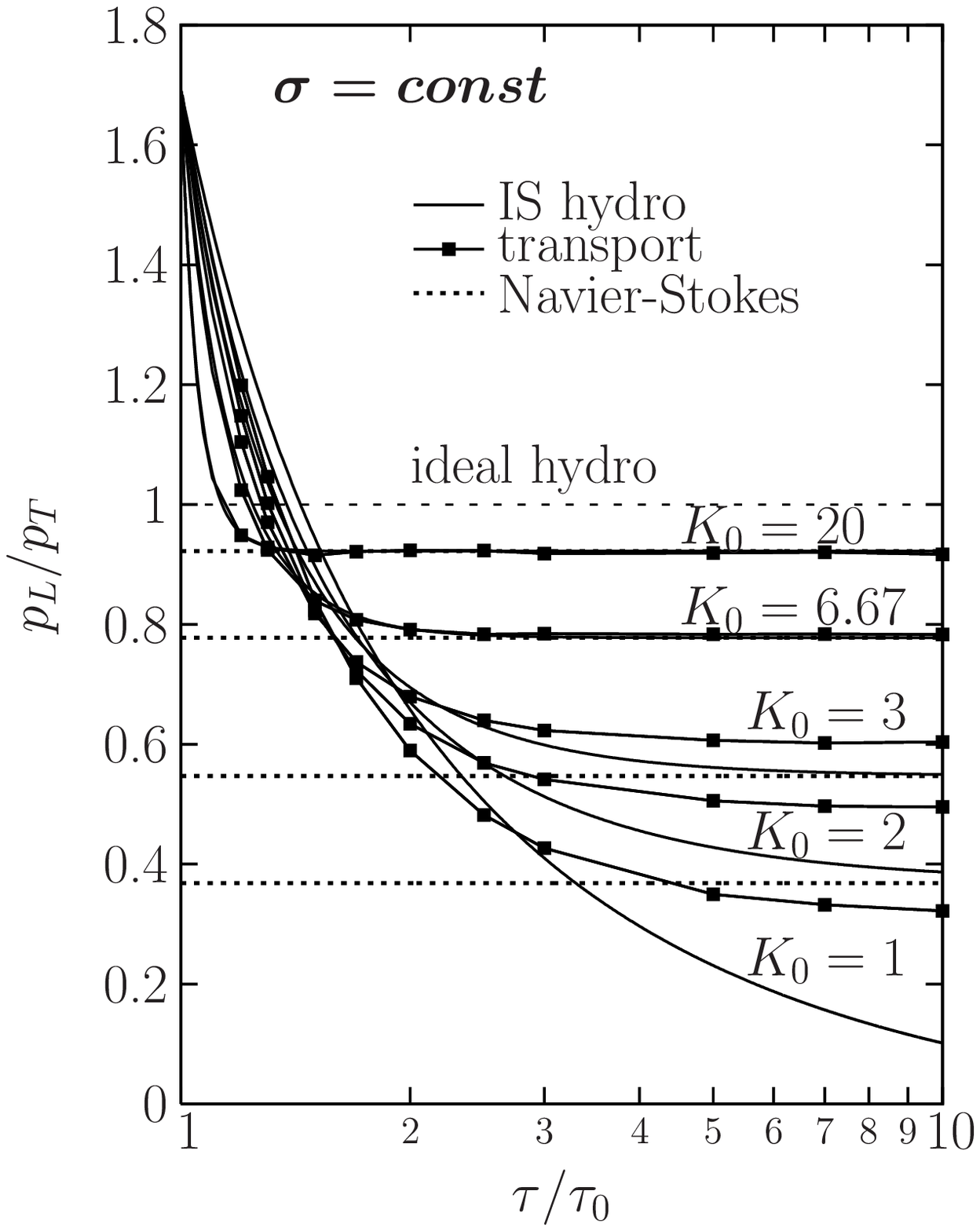}
\epsfysize=10cm
\epsfbox{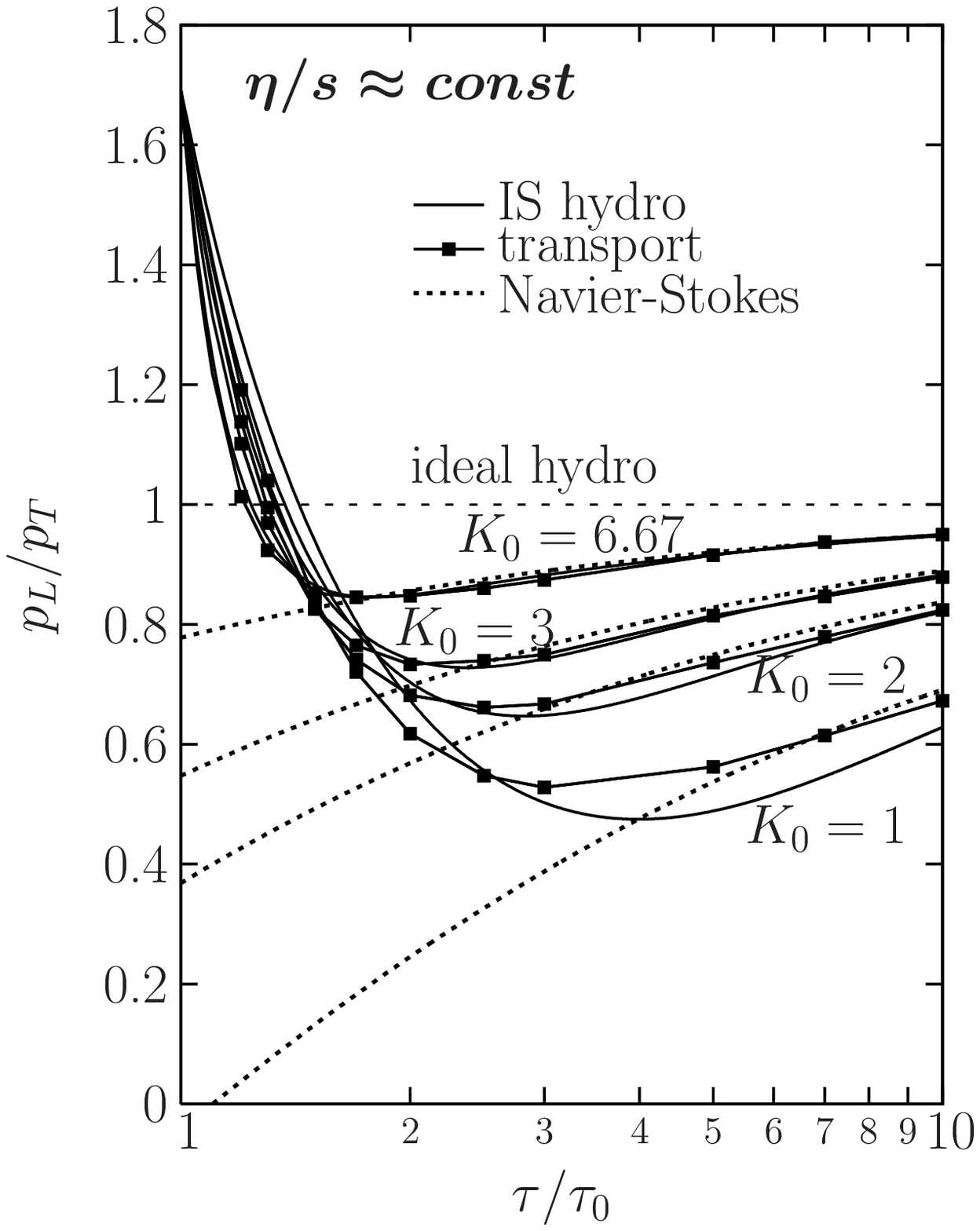}
\caption{Same as Fig.~\ref{Fig:R_comp} but for an initial pressure anisotropy 
$R_p(\tau_0) = 1.693$ ($\xi_0 = 0.375$).
In the left plot, the Navier-Stokes
curve for $K_0 = 1$ is negative and therefore not visible.}
\label{Fig:R175_comp}
\end{figure}

Moreover, the above findings hold for a wide range of the initial
conditions, including large initial pressure anisotropies,
as shown in 
Figures~\ref{Fig:R03_comp} and \ref{Fig:R175_comp}. These figures 
are for the same calculation but with $R_p(\tau_0) = 0.476$ and $1.693$, 
respectively (which correspond to $\xi_0 = -0.423$ and $0.375$).
We emphasize that the results hold only if nonequilibrium 
corrections are close to the form (\ref{phiG}) suggested by Grad. 
{\em For such class of initial conditions, however, we find that 
Israel-Stewart hydrodynamics can well approximate the 
transport ($\sim 10$\% accuracy) 
provided $K_0 \gton 3$, even for the most pessimistic constant cross 
section scenario. If $\etaOs = const$, only $K_0 \gton 2$ is needed.
We stress that in either case, 
there is no need for the initial conditions to be near the
Navier-Stokes limit.}

This is quite remarkable because
from Figs.~\ref{Fig:R_comp}-\ref{Fig:R175_comp} 
it is clear that already the early evolution differs
between IS hydrodynamics and transport. E.g., 
for an equilibrium initial condition ($\xi(\tau_0) = 0$), 
IS hydrodynamics (\ref{Rdot_IS}) gives
\be
R_p^{IS}(\tau) = 1 - \frac{4(\tau-\tau_0)}{3\tau_0} + \cO((\tau-\tau_0)^2)
\label{R_early_IS}
\ee
{\em for any} initial value and evolution scenario for $\kappa$.
From covariant transport, on the other hand (see Appendix \ref{App:early})
\be
R_p^{transp}(\tau) = 1 - \frac{8(\tau-\tau_0)}{5\tau_0} 
+ \cO((\tau-\tau_0)^2) \ .
\label{R_early_tr}
\ee
I.e., pressure anisotropy develops, universally, 
$20$\% faster from the transport than from IS hydrodynamics
(if the evolution starts from equilibrium).

This illustrates a limitation of the hydrodynamic description of transport
solutions. Similar discrepancies were observed in~\cite{dissipv2} in 
the early evolution of differential elliptic flow $v_2(p_T)$.
Remarkably, in our case, though the transport develops deviations from 
equilibrium faster, its rate of departure 
slows down quicker, which at intermediate times results in {\em smaller}
accumulated dissipative corrections to the pressure anisotropy 
than from IS hydrodynamics. Eventually,
the hydrodynamic evolution ``catches up'' to the transport,
except for low $K \lton 3$ in the $\sigma = const$ scenario.

Figures~\ref{Fig:R_comp}-\ref{Fig:R175_comp} 
also show the Navier-Stokes approximation (dotted lines without symbols)
for each of the Israel-Stewart 
results. By late
times, the Navier-Stokes and Israel-Stewart solutions converge for both cross 
section scenarios, independently of the initial pressure anisotropy
(for $\sigma = const$ and $K_0=1$, the NS anisotropy is negative and therefore
not visible in the plots). 
However, the applicability of Navier-Stokes theory at early times
depends, besides the value of $K_0$, 
strongly on how far the initial shear stress is from its
Navier-Stokes value (\ref{piL_NS}). Navier-Stokes assumes
that shear stress, and therefore the pressure anisotropy, 
relaxes immediately, but relaxation happens over
a finite time. 
The approach toward the Navier-Stokes limit is governed by 
$\tau_\pi = 3\tau/(2\kappa)$, therefore 
{\em Navier-Stokes becomes applicable
only after some time $\Delta\tau \sim  |R_0 - R_{NS}| \, \tau_0/ \kappa$}.
Note that the initial {\em slope} of the $R(\tau)$ curves does not 
always reflect $\tau_\pi$
directly because it is given by the initial
derivative of $\xi$
\be
\dot R(\tau) \sim \frac{3}{2}\dot \xi(\tau) 
= -\frac{3}{2\tau_\pi} (\xi - \xi_{NS}) + 
\cO(1) \frac{\xi}{\tau} 
\ee
where we combined (\ref{IS_struct}), (\ref{EOMp}), the observations that 
$Y \sim \cO(1)/\tau$ and $Z = 0$, and assumed $\xi$ is small. For local
equilibrium initial conditions the slope of $R(\tau)$ is therefore 
$\sim \cO(1)\xi_{NS}/\tau_\pi \sim \cO(1)/\tau$, independently of $K_0$
(cf. Figure~\ref{Fig:R_comp} and also (\ref{R_early_IS})).
For initial shear stresses far away from the Navier-Stokes limit,
on the other hand, the slope $\sim \cO(1) \xi / \tau_\pi \propto \kappa$ 
steepens with increasing $K$
as seen in Figures~\ref{Fig:R03_comp} and \ref{Fig:R175_comp}.

The inaccurate description of early shear stress evolution in Navier-Stokes
has a cumulative effect on the evolution of thermodynamic quantities, 
such as the pressure and the entropy, as we show in
the next two Sections. Of course, the errors are proportional to 
ratio of the time the system spends away from the NS limit and the 
hydrodynamic timescale, i.e., 
$\Delta \tau / \tau_0 \sim 1/\kappa$.

\subsection{Pressure evolution}
\label{Sec:p_comp}

Now we turn to the evolution of the (average) pressure. In ideal hydrodynamics 
$(K_0 \to \infty)$
the pressure drops rapidly with time $p_{id} \propto \tau^{-4/3}$.
Therefore it is
more convenient to study dissipative effects 
{\em relative} to ideal hydrodynamics through the ratio 
$p(\tau)/p_{id}(\tau)$.

\begin{figure}
\epsfysize=7cm
\epsfbox{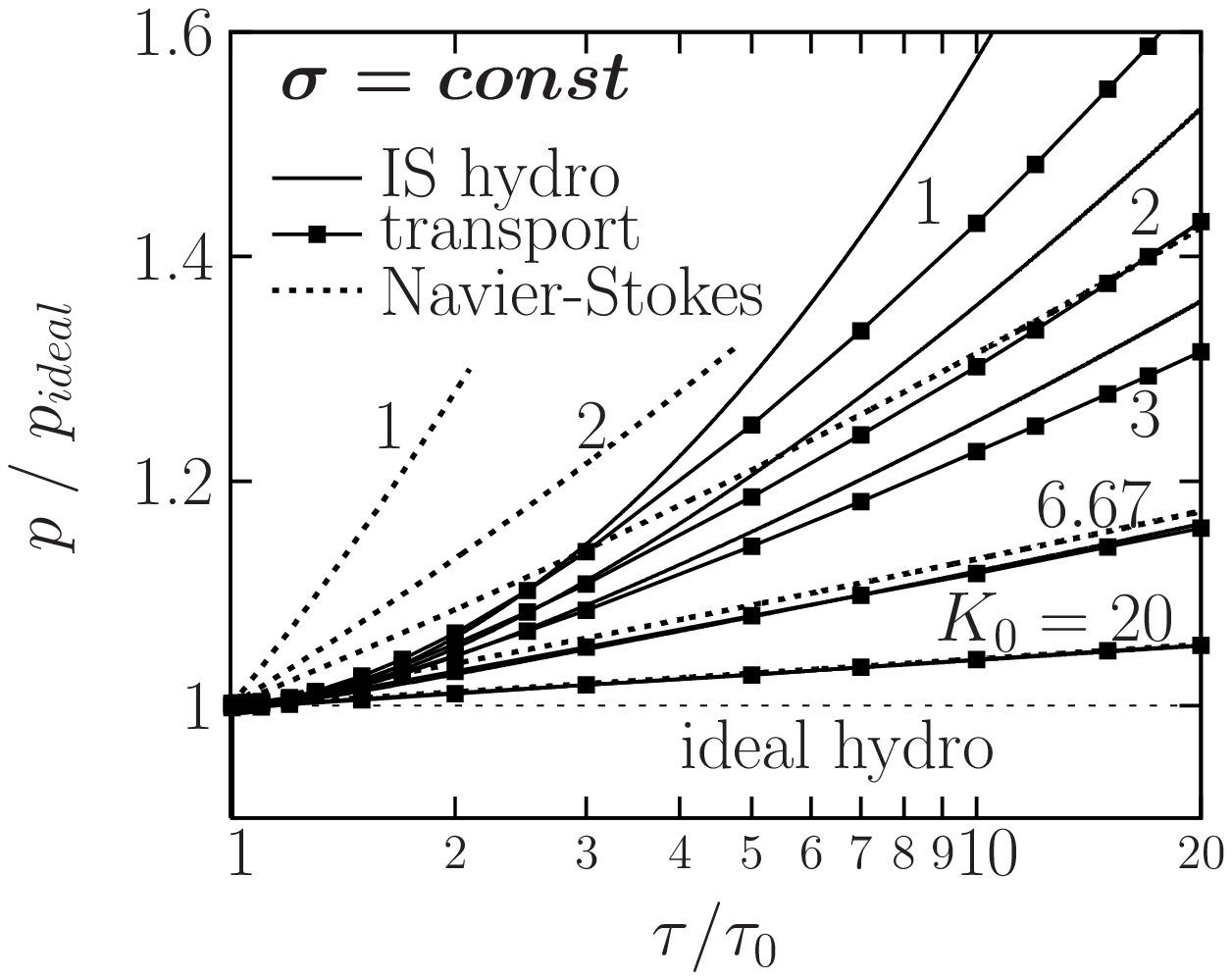}
\epsfysize=7cm
\epsfbox{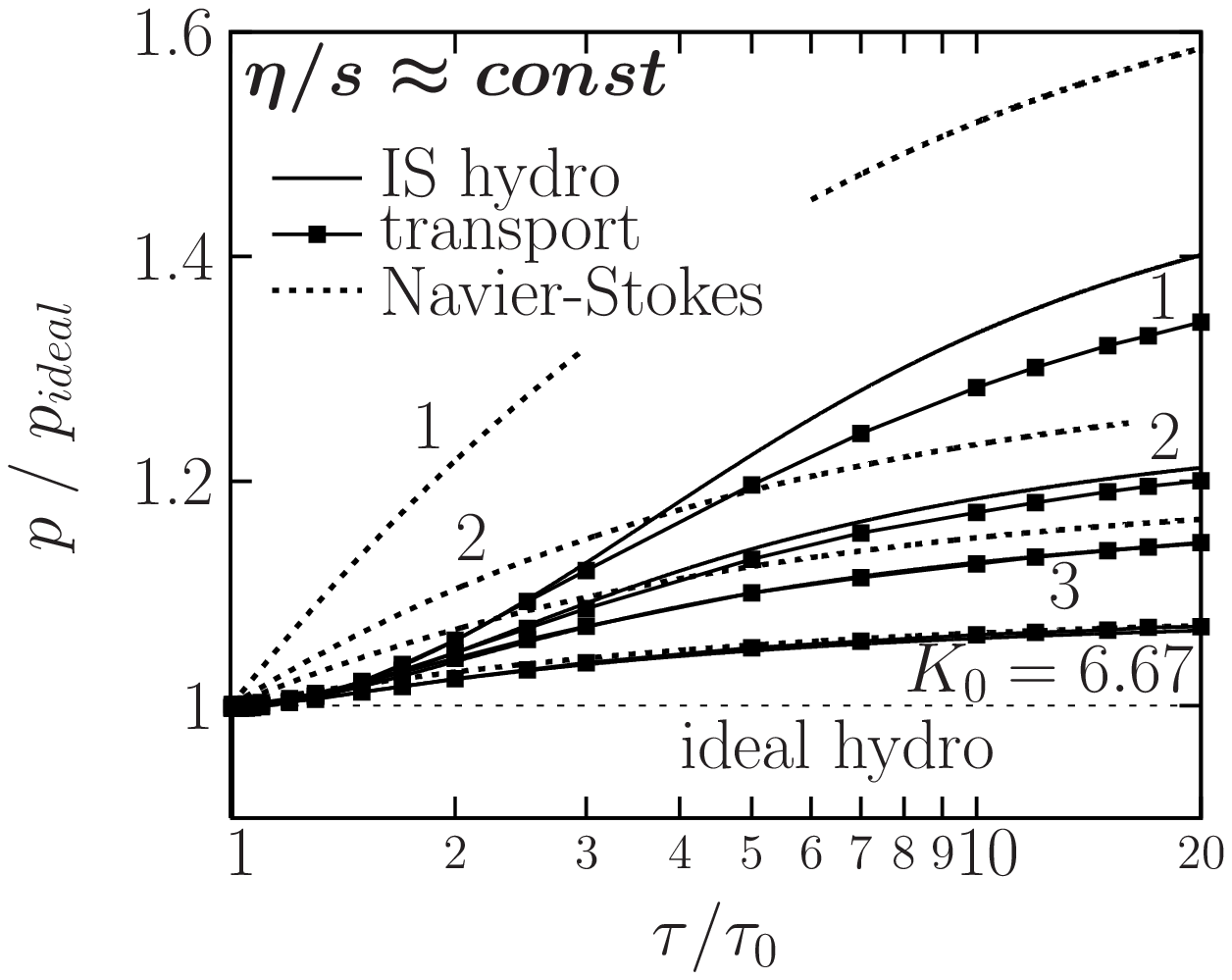}
\caption{Same as Fig.~\ref{Fig:R_comp} except for the time evolution of 
the (average) pressure. The pressure is plotted normalized to the pressure 
$p_{ideal}(\tau) = p_0 (\tau_0/\tau)^{4/3}$ in ideal 
hydrodynamics.}
\label{Fig:p_comp}
\end{figure}

Figure~\ref{Fig:p_comp} shows the pressure relative to that in ideal
hydrodynamics as a function of the rescaled proper time $\ttau = \tau/\tau_0$
from the transport (solid lines with symbols) and Israel-Stewart 
hydrodynamics (solid lines without symbols) with local equilibrium initial
condition. 
The left panel shows calculations for the $\sigma=const$ scenario. 
For all $K_0$ values, the evolution starts out the same between IS hydro and 
transport but then the hydro starts to accumulate deviations 
because it follows the shear stress evolution only approximately.
For $K_0 = 1$, IS hydro maintains $10$\% accuracy in the magnitude of 
dissipative {\em corrections} (i.e., $p/p_{id} - 1$) only 
up to $\tau \approx 4\tau_0$. As 
$K_0$ increases, the situation improves gradually, for $K_0 = 3$, $10$\%
accuracy holds up to $\tau \approx 10\tau_0$, 
and by $K_0 \approx 7$ the hydro stays within a few percent of the transport 
even until
$\tau = 20 \tau_0$.

The right panel shows the same but for the growing cross section 
scenario with $\etaOs \approx const$. This scenario is more favorable
for the hydrodynamic approximation 
because $K \sim \tau^{2/3}$ grows with time. For $K_0 = 1$, 
the error in the dissipative correction $(p/p_{id}-1)$ 
is less than $10$\% up to $\tau \approx 5\tau_0$, and already
for $K_0 = 2$ IS hydro is accurate to within better than $10$\% throughout
the whole range $\tau \le 20 \tau_0$ studied. The pressure evolution
results therefore reinforce the regions of validity found in the
previous Section ($K_0 \gton 3$ for 
$\sigma=const$, and $K_0 \gton 2$ for $\etaOs \approx const$)

Clearly, the region of applicability for
Navier-Stokes is
more limited (Figure~\ref{Fig:p_comp}, dotted lines without symbols). 
For low $K_0$, it overestimates the pressure corrections not only at
late times but also at early $\tau \sim few \times \tau_0$.
$K_0 \approx 7$ is barely sufficient 
for $10$\% accuracy in viscous corrections 
for $\etaOs \approx const$, but it is not enough in case of $\sigma = const$.
Based on the trends with increasing $K_0$, we estimate
that $K_0 \gton 9-10$ is needed for Navier-Stokes with $\sigma=const$ 
to deviate less than $10$\%  
from the viscous effects calculated with the transport.
{\em Therefore, for local equilibrium initial conditions, 
Navier-Stokes theory becomes applicable at about three times shorter
mean free paths, or equivalently three times larger longitudinal 
proper time $\tau$ 
(i.e., three times slower longitudinal expansion), than Israel-Stewart theory}.

\subsection{Entropy}
\label{Sec:dS_comp}

Now we proceed with results on entropy production. In transport theory, 
the entropy current is defined as
\be
S^\mu(x) 
= - \int \frac{d^3p}{p^0} \, p^\mu f(x,\vp) 
\left[\ln \left(\frac{(2\pi)^3}{g} f(x,\vp)\right) - 1\right]
\ee
where $g$ is the number of internal degrees of freedom. This nonlinear 
function of the phasespace density $f$ is cumbersome to evaluate 
with the MPC code, and therefore we here opt for an approximate result
based on the truncated Israel-Stewart expression (\ref{Smu1D}),
evaluated using the pressure and shear stress from the transport.
This includes dissipative corrections to the entropy 
up to quadratic order in $\phi$.

\begin{figure}
\epsfysize=7cm
\epsfbox{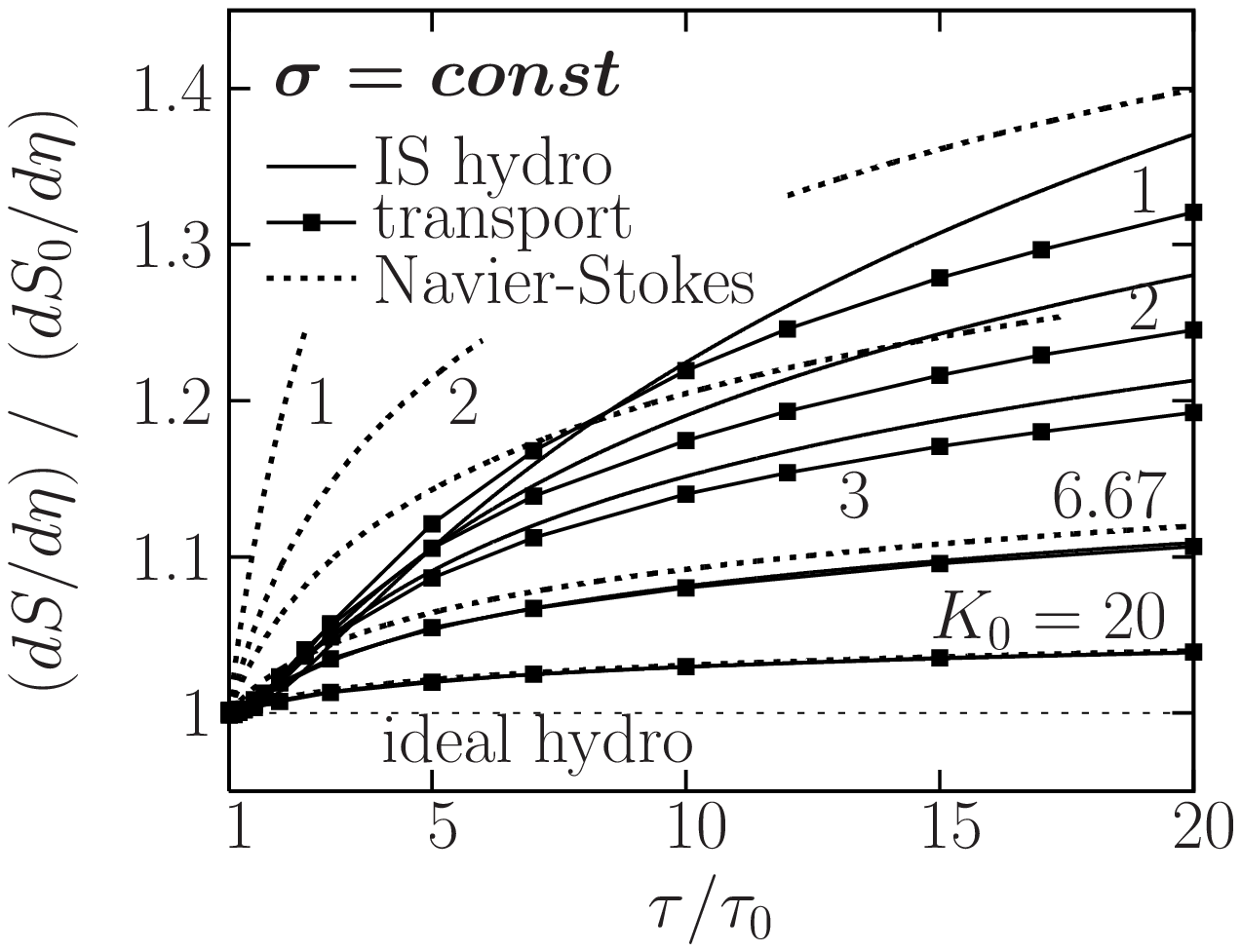}
\epsfysize=7cm
\epsfbox{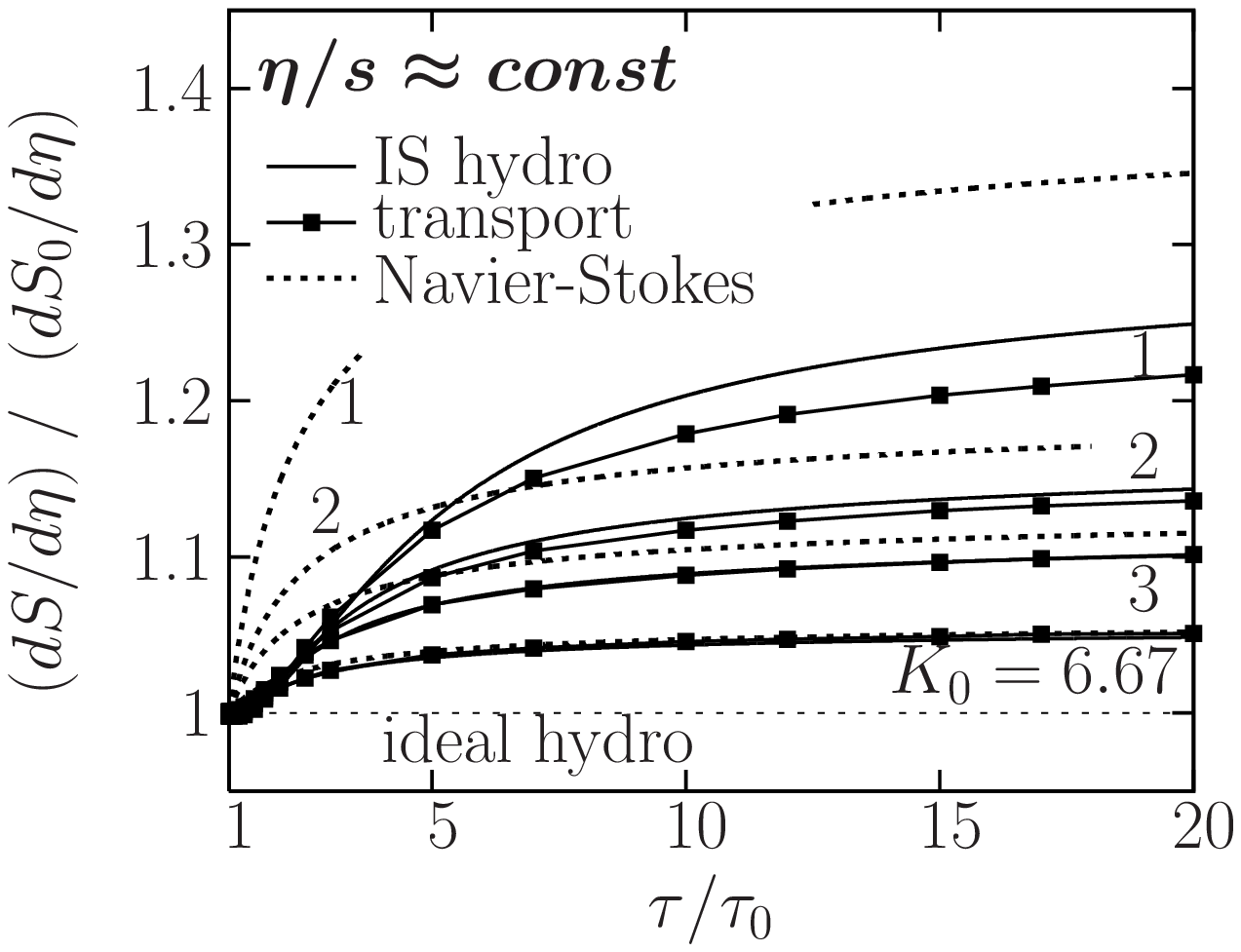}
\caption{Same as Fig.~\ref{Fig:R_comp} except for the time evolution of 
the entropy per unit rapidity, normalized by its initial value (note the
linear time axis used this time). For the transport, entropy was calculated
approximately using the Israel-Stewart entropy expression (\ref{Smu1D}).
Chemically equilibrated initial conditions 
(i.e., $\chi_0 = 0$) were assumed.}
\label{Fig:dS_comp}
\end{figure}

In the most dissipative
$\sigma= const$ scenario with $K_0 = 1$, 
there is about $30$\% additional entropy 
produced by late times $\tau/\tau_0 \sim 10-20$ 
as can be seen in Figure~\ref{Fig:dS_comp} (left plot).
For $\etaOs \approx const$ (right plot), the same $K_0=1$ yields
only about 20\% extra entropy.
With increasing $K_0$ entropy generation gradually weakens and by 
$K_0 \sim 7$ it is only 10 and 5 \%, respectively.

The Israel-Stewart results are within $15$\% of the approximate transport 
results already for $K_0 = 1$, and about $10$\% accuracy in 
the calculated dissipative effect is achieved for
$K_0 \gton 3$ (for $\sigma = const$) and $K_0 \gton 2$ 
(for $\etaOs \approx const$). 
In contrast, the Navier-Stokes strongly overpredicts the entropy,
unless $K_0$ exceeds about $6$ for $\sigma = const$ or 
$\approx 3$ for $\etaOs \approx const$. The bounds for $10$\% accuracy 
are in agreement with
those found previously in Sec.~\ref{Sec:p_comp}.

\subsection{Limitations of the 'naive' Israel-Stewart approximation}
\label{Sec:naive}

\begin{figure}
\epsfysize=6.3cm
\epsfbox{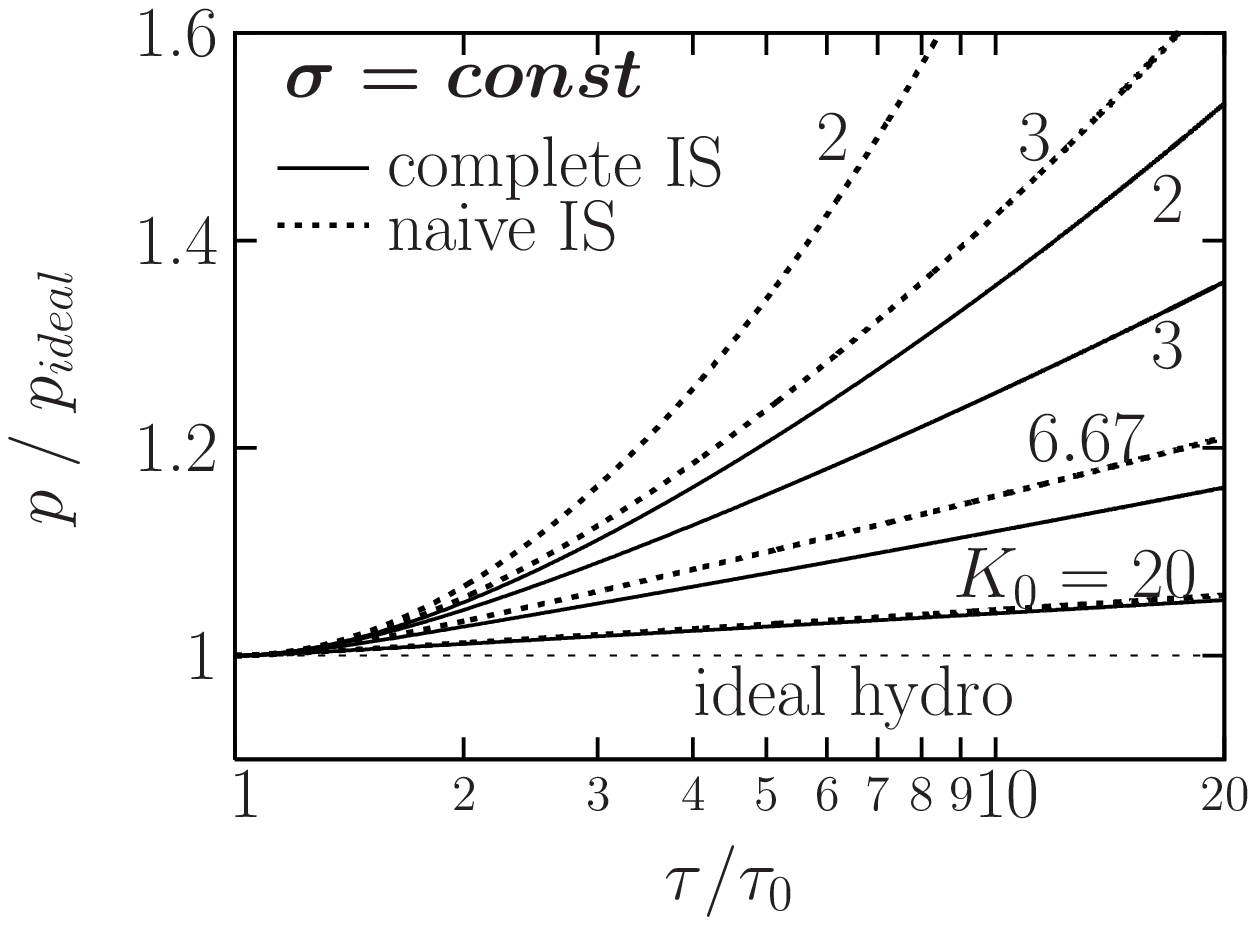}
\epsfysize=6.3cm
\epsfbox{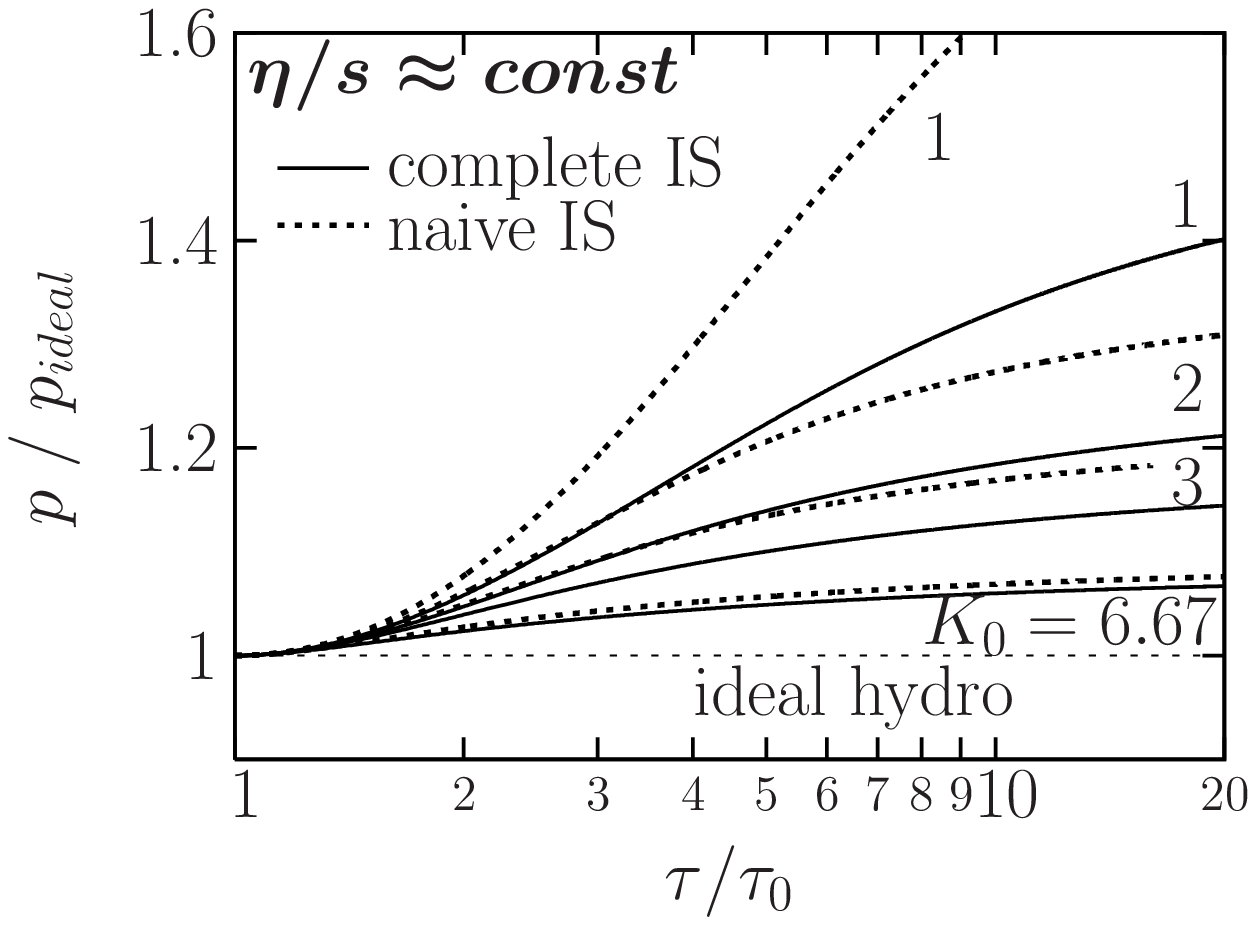}
\caption{Time evolution of the (average) pressure
from complete Israel-Stewart theory (solid lines)
and the 'naive' Israel-Stewart approximation (dotted)
as a function of $K \equiv \tau / \lambda_{tr}(\tau)$,
for local equilibrium initial conditions $\pi_L(\tau_0) = 0$.
The pressure is plotted normalized to the pressure 
$p_{ideal}(\tau) = p_0 (\tau_0/\tau)^{4/3}$ in ideal 
hydrodynamics.
{\em Left: } $\sigma = const$ scenario, in which case 
$K (\tau) = const = K_0 = 2$, 3, 6.67 and 20. 
{\em Right:} $\sigma \propto \tau^{2/3}$ scenario, 
for which $\etaOs \approx const$ and the curves
are labeled by the initial $K_0 = K(\tau_0) = 1$, 2, 3, and 6.67. 
}
\label{Fig:p_naive}
\end{figure}

Now we discuss the applicability of the 'naive' Israel-Stewart equations.
Figure~\ref{Fig:p_naive} compares the pressure evolution in complete
Israel-Stewart theory to that in the naive approximation, for local equilibrium
initial conditions ($\xi_0 = 0$), 
as a function of the rescaled proper time $\tau/\tau_0$.
Clearly, the naive result overshoots the 
pressure both for the constant cross section scenario and for 
$\etaOs \approx const$, unless
$K_0$ is large. This confirms expectations based on the analytic 
solutions in
App.~\ref{App:IS}. Though the 'naive' theory converges to the correct result
at large enough $K_0 \sim 7-20$, 
comparison with Fig.~\ref{Fig:p_comp} tells that it is even less accurate
than Navier-Stokes theory.

\begin{figure}
\epsfysize=6.3cm
\epsfbox{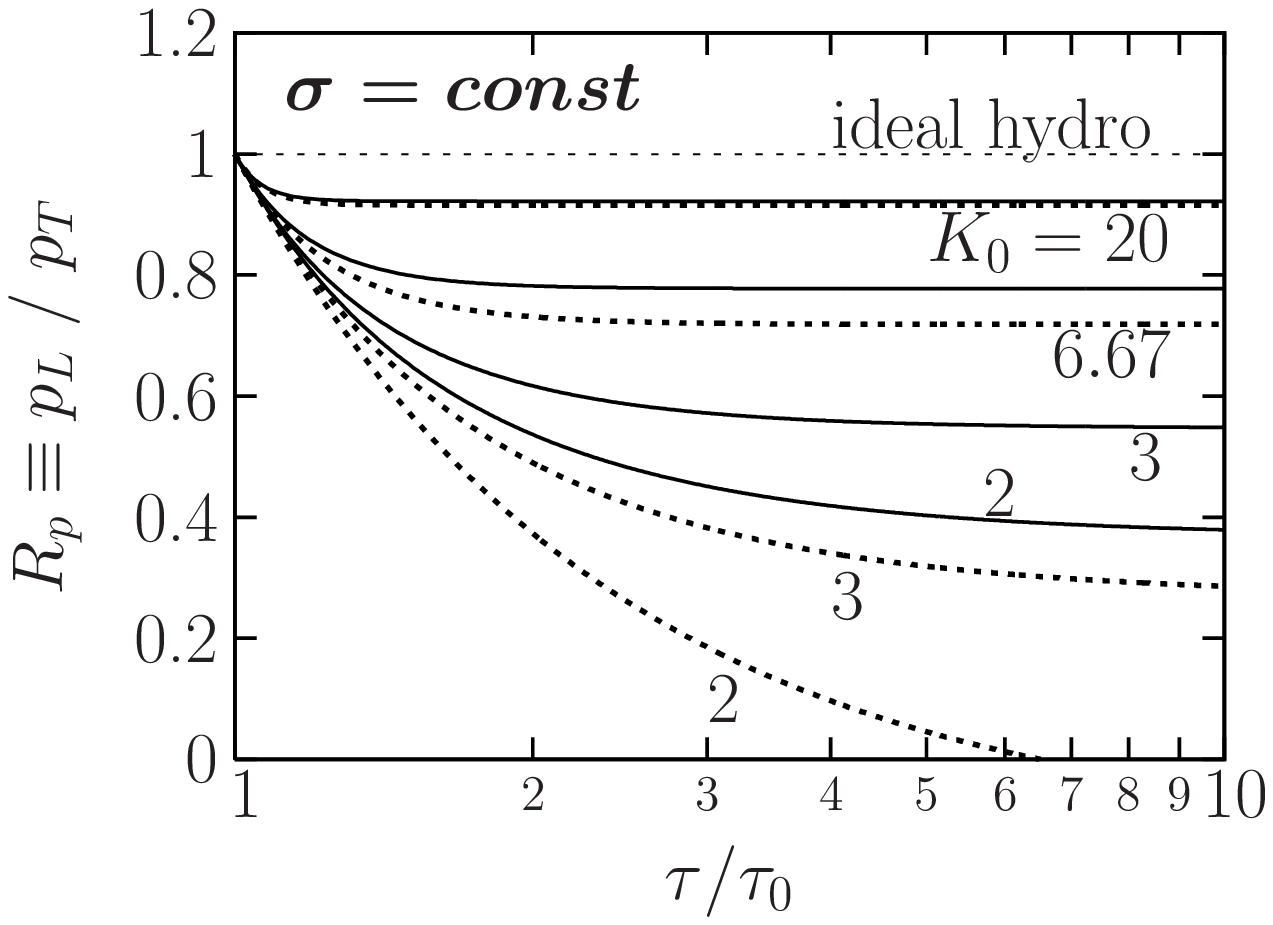}
\epsfysize=6.3cm
\epsfbox{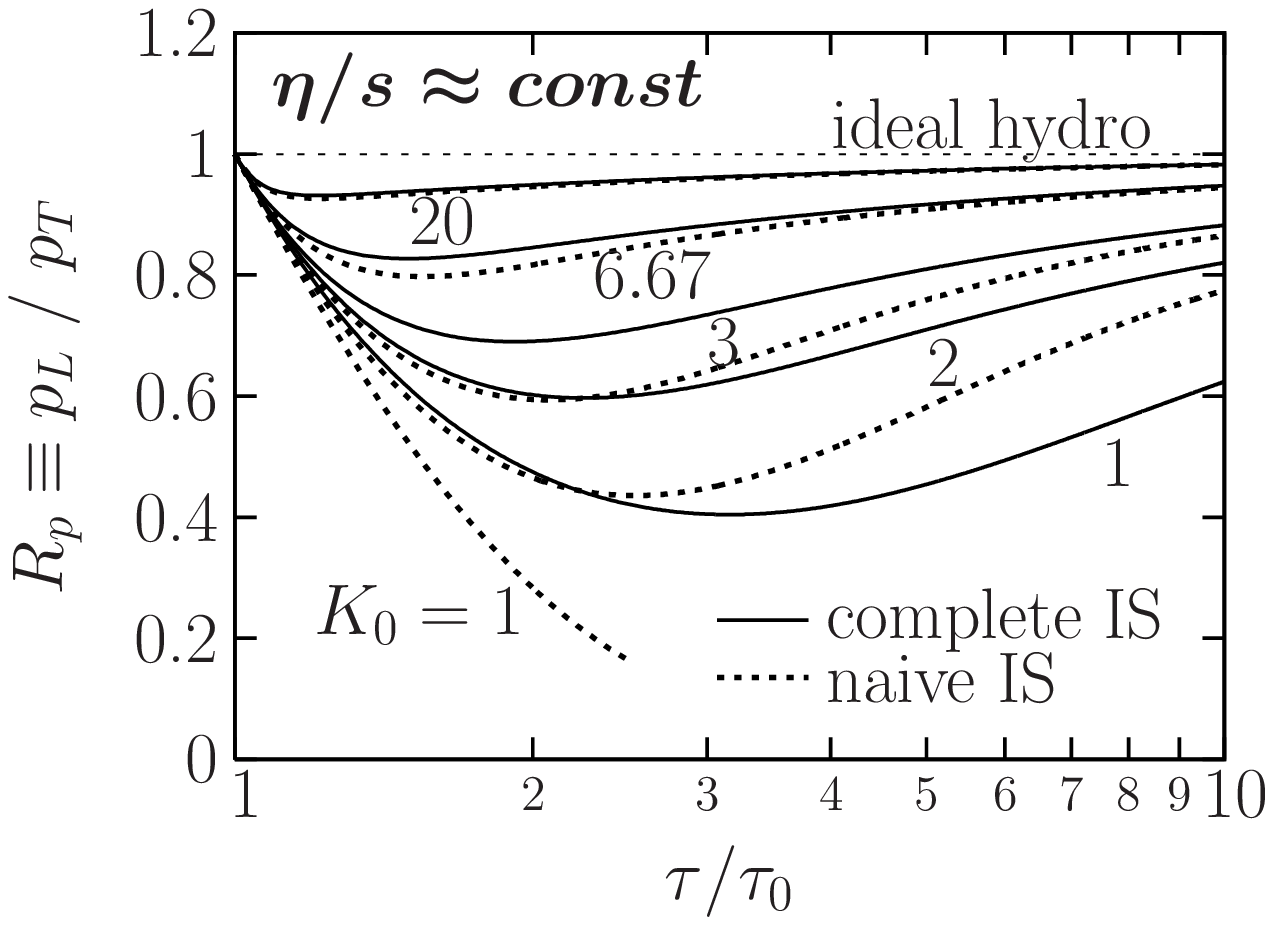}
\caption{Same as Fig.~\ref{Fig:p_naive} but for the
time evolution of the pressure anisotropy.}
\label{Fig:R_naive}
\end{figure}

The reason for the large errors is
that away from local equilibrium 
the 'naive' approach drives the shear stress more negative
(compare (\ref{EOMpiL}) and (\ref{EOMpiL_naive}), and note that typically 
$\pi_L < 0$).
This is demonstrated in Fig.~\ref{Fig:R_naive} where
we plot the pressure anisotropy $R_p$, which is a monotonic function of 
$\xi = \pi_L/p$. 
For $\sigma = const$, we find that 
the naive approach saturates the anisotropy
at a lower value than the complete theory, 
confirming analytic expectations in App.~\ref{Sec:IS_const}.
For $\etaOs \approx const$, the system does approach
ideal hydrodynamic behavior
eventually, however that occurs on a much longer timescale than 
from complete Israel-Stewart theory. This is in agreement with expectation
based on the analytic solutions (\ref{latetime_form})-(\ref{naiveIS_coeffs}).

The pressure anisotropy results further reinforce our conclusion that 
the 'naive' Israel-Stewart approximation is poorer 
than Navier-Stokes (cf. Fig.~\ref{Fig:R_comp}). In heavy-ion collisions,
gradients are large, at least initially, and therefore 
cannot be ignored even if dissipative corrections 
(e.g., $\pi_L / p$) are small.

\subsection{Implications for heavy-ion physics}
\label{Sec:HI}

Having determined the region of validity 
(defined as $10$\% accuracy in dissipative effects)
for Israel-Stewart and 
Navier-Stokes hydrodynamics in terms of the initial ratio of the expansion
and scattering timescales $K_0 = \tau_0 / \lambda_{tr,0}$ 
\bea
&& K_0^{IS} \gton 3 
\qquad\qquad  K_0^{NS} \gton 9
\qquad\qquad (\sigma = const) \\
&& K_0^{IS} \gton 2 
\qquad\qquad K_0^{NS} \gton 6
\qquad\qquad (\etaOs \approx const) \ ,
\eea
 we now turn
to implications for heavy-ion collisions. 
From (\ref{def_K}), (\ref{muT}), and (\ref{neq}),
\be
\kappa_0 = \frac{T_0 \tau_0}{4-\chi_0} \frac{s_0}{\eta_{s,0}} \approx
15.9 \times
\frac{1}{1-\chi_0/4} \left(\frac{T_0}{{\rm 1\ GeV}}\right)
 \left(\frac{\tau_0}{{\rm 1\ fm}}\right) 
\left(\frac{1/(4\pi)}{\eta_{s_0}/s_0}\right) \ , \qquad
K_0 \approx 0.8 \kappa_0 \ .
\label{kappa0_estimate}
\ee
Therefore, we can place an upper limit on the (initial) shear 
viscosity for which IS or NS reproduces with better than $10$\% accuracy
the viscous corrections to 
basic observables such as pressure and entropy:
\bea
&& \left.\frac{4\pi\eta_{s,0}}{s_{eq,0}}\right|_{IS} \lton 0.8T_0\tau_0 
\qquad  \left.\frac{4\pi\eta_{s,0}}{s_{eq,0}}\right|_{NS} \lton 0.25 T_0\tau_0 
\qquad\qquad (\sigma = const) \\
&& \left.\frac{4\pi\eta_{s}}{s_{eq}}\right|_{IS} \lton 1.2T_0\tau_0 
\qquad\quad  \left.\frac{4\pi\eta_{s}}{s_{eq}}\right|_{NS} \lton 0.4T_0\tau_0
\qquad\qquad (\etaOs \approx const)
\eea
where we assumed chemical equilibrium initial conditions $(\chi_0 = 0)$.
If the shear viscosity of dense quark-gluon matter
is bounded from below by $4\pi\etaOs \gton 1$, as has been conjectured
recently, then 
the situation for Israel-Stewart is close to marginal.
For $\etaOs = 1/(4\pi)$, typical parton transport initial conditions 
($T_0 = 0.7$ GeV, $\tau_0 = 0.1$ fm) translate into 
$K_0 \lton 1$, for which IS is not applicable for either of 
Scenario I or II,
while for typical hydrodynamic initial conditions
($T_0 \sim 0.38$ GeV, $\tau_0 = 0.6$ fm)
we have $K_0 \lton 3$, sufficient for both scenarios
(barely for $\sigma = const$).

On the other hand, Navier-Stokes may be marginally applicable only if
$\etaOs \lton 0.5/(4\pi)$ throughout the whole evolution, at least
based on this 0+1D study, 
where acausal artifacts and instabilities do not arise. We emphasize that
the bound quoted here
is for initial conditions close to local equilibrium.
The accuracy of the Navier-Stokes approximation strongly depends on how far 
the initial shear stress is from the Navier-Stokes value. If the evolution
starts out near the Navier-Stokes limit, we expect
Navier-Stokes to be accurate up to higher viscosities.

Within the region of applicability of Israel-Stewart, dissipative
corrections to the average pressure and the entropy are modest and stay
below $\sim 20$\% even up to late times $\tau \le 10 \tau_0$. This may
serve as a useful ``rule of thumb'' applicability condition for
hydrodynamics: if dissipative corrections to average pressure and the    
entropy calculated from hydrodynamics are significantly larger than 20\%, the
validity of hydrodynamics is questionable.

The above findings reinforce a recent calculation\cite{QM2008} in 2+1D
that found good agreement
between IS hydrodynamics and $2\to 2$ transport, for conditions
expected in $Au+Au$ at $\sqrt{s_{NN}} \sim 200$~GeV/nucleon at RHIC, in case
of a small
shear viscosity to entropy density ratio $\etaOs \approx 1/(4\pi)$ 
(on average). 
The same study also found good agreement between
the two theories for a large
constant transport cross section $\sigma_{tr} \approx 13$~mb.
That is also in line with our results here because it 
corresponds to $4\pi \etaOs(\tau_0) \approx 0.25$ in the center of the 
collision zone, i.e., initially $\etaOs \lton 1/(4\pi)$ in most of the system.

Finally we note that the applicability of the hydrodynamic approach on 
very short time and length scales is another important question.
In typical real-life problems
$T_0\tau_0 \gg 1$ because the hydrodynamic expansion timescale $\tau$
is by orders of magnitude larger than 
the quantum (energy) timescale $1/T$. 
This also leaves ample room to make
hydrodynamics applicable ($\kappa_0 \gg 1$) even for appreciable viscosities.
In the heavy-ion case, however, the two timescales are comparable
$T_0 \tau_0 \sim \cO(1)$, and therefore a macroscopic treatment may be 
marginal.

\section{Conclusions}

Based on comparison to covariant transport theory, we explore
the region of validity of Israel-Stewart and Navier-Stokes hydrodynamics
in heavy-ion physics applications. 
We follow the evolution of the average pressure, pressure anisotropy, 
and entropy for a massless ideal gas in 0+1D
longitudinally expanding Bjorken geometry. 
Binary $2\to 2$ interactions are considered for two main scenarios, 
a fixed cross section $\sigma = const$ 
(Scenario I, pessimistic for hydrodynamics) and a scale invariant 
system with $\etaOs \approx const$ (Scenario II, optimistic for hydrodynamics).

We find (Sec.~\ref{Sec:IS_validity})
that dissipative effects calculated from Israel-Stewart
hydrodynamics 
reproduce those from the transport to within 10\%, provided
initially
the expansion timescale is three (for Scenario I) or two (for Scenario II)
times larger than the transport mean free path,
i.e., the initial Knudsen number $K_0 = \tau_0 / \lambda_{tr,0} \gton 3$ or 2.
When this criterion is fulfilled,
Israel-Stewart is accurate even if initial pressure anisotropies 
are large $p_L/p_T \sim 0.4-1.7$ - there is
no need to start near the Navier-Stokes limit.
On the other hand, same accuracy from Navier-Stokes requires three times 
larger $K_0$, if the expansion starts from local thermal equilibrium 
(unlike for Israel-Stewart, 
the applicability of Navier-Stokes depends strongly on how far the 
initial shear stress is from its Navier-Stokes value).
We emphasize that these findings apply only when {\em initial} 
viscous corrections are of the quadratic form suggested by Grad (\ref{phiG}).

These results imply that (Sec.~\ref{Sec:HI}), 
for typical heavy ion initial conditions at RHIC
energies, 
Israel-Stewart hydrodynamics is accurate up to $\etaOs \lton 1.5/(4\pi)$,
while for Navier-Stokes $\etaOs \lton 0.5/(4\pi)$ is needed. This is supported
by a recent 2+1D calculation\cite{QM2008} 
that finds good agreement between Israel-Stewart
and transport for $\etaOs \approx 1/(4\pi)$, and also for a 
large $\sigma_{tr} \approx 13$ mb.

In addition, we test the accuracy of the 
naive Israel-Stewart approximation (Sec.~\ref{Sec:naive}) that
neglects products of gradients and dissipative quantities in the equations of 
motion, and find that it has an even more limited applicability 
than Navier-Stokes.

We also compare in detail (App.~\ref{App:IS}) 
Israel-Stewart and Navier-Stokes solutions in 0+1D
for four scenarios, $\sigma = const$, $\sigma \propto 1/T^2$, 
$\sigma \propto \tau^{2/3}$ and $\etaOs = const$, and find that 
results for the latter two are almost identical, even at low
initial Knudsen numbers $K_0 \sim 1$.
Moreover, we obtain 
analytic Israel-Stewart and Navier-Stokes solutions in 0+1D,
which are useful for quick estimates (Secs.~\ref{Sec:work} and \ref{Sec:S}) 
and 
to test numerical solution techniques. 
We also derive additional
tests (App.~\ref{App:ConsLaws}) 
based on generalized conservation laws for conserved currents, 
energy-momentum, and entropy, which can be utilized to verify the accuracy of
numerical Israel-Stewart solvers in 1+1, 2+1, and 3+1 dimensions.

Finally we emphasize that 
the current study is limited to a massless ideal gas
with particle number conserving interactions in 0+1D Bjorken geometry.
The influence of the transverse expansion will be quantified in
a future paper (requires at minimum a 1+1D approach).
It will be also important to check
how the results depend on the equation of state and the presence of particle
non-conserving processes, such as radiative 
$2\leftrightarrow 3$. For a nonconformal equation of state, bulk
viscosity may become important\cite{Kharzeevbulk,Friesbulk}. 
Ideally, one should also test the 
accuracy of the hydrodynamic
approximation for nonequilibrium theories other than covariant transport.

\acknowledgements
We thank RIKEN, 
Brookhaven National Laboratory and
the US Department of Energy [DE-AC02-98CH10886] for providing facilities
essential for the completion of this work.
We also thank the hospitality of INT/Seattle (D.M., P.H.), KFMI/RMKI (D.M.),
and Iowa State University (P.H.)
where parts of this work have been 
completed. Computational
resources managed by RCAC/Purdue are also gratefully acknowledged.

\appendix

\section{Origin of $a_0$, $a_1$, $a'_0$, $a'_1$ in the 
Israel-Stewart equations of motion}
\label{App:ai}

The equations of motion (\ref{IS_EOMPi})-(\ref{IS_EOMpi}) 
reproduce the entropy production
rate (\ref{SdivIS}) only approximately, up to typically small
{\em quartic and higher-order corrections in dissipative quantities}.
With $a_i \equiv 0 \equiv a'_i$, a contribution
\be
\Pi q^\mu T\nabla_\mu(\alpha_0/T) - q_\nu \pi^{\nu\mu} T \nabla_\mu(\alpha_1/T)
\ee
would be {\em missing} from $T\partial_\mu S^\mu$ in (\ref{SdivIS}). 
These terms are bilinear in the dissipative quantities and,
therefore, can be split arbitrarily between the bulk pressure and heat, and 
heat and shear equations of motion.
I.e., with
\be
T\nabla^\mu(\alpha_i/T) \equiv a_i^\mu + {a'_i}^\mu  
\ee
(\ref{SdivIS}) is identically satisfied but contributions to the 
equations of motion depend on $a_i$
\bea
\beta_0 D\Pi &=& (...) + {a'_0}^\mu q_\mu  \\
\beta_1 Dq^\mu &=& (...) + \Delta^\mu_{\ \nu} {a_0}^\nu \Pi 
  - a_1^\nu \pi^\mu_{\ \nu} \\
\beta_2 D\pi^{\mu\nu} &=& (...) - {a'_1}^{\langle \nu} q^{\mu\rangle} 
\eea

Only components orthogonal to $u^\mu$ contribute but apart from that
constraint
$a_0^\mu$ and $a_1^\mu$ are arbitrary functions of the hydrodynamic fields 
and their derivatives, and potentially new scalar functions
$\{a_i^{(k)} (e,n)\}$ characterizing an isotropic matter. 
However, ignoring the dependence on
the dissipative quantities is consistent with
the truncation of the
entropy current (\ref{SmuIS}) 
at quadratic order in those. Moreover, for small deviations from 
equilibrium one may seek to include only
the leading contributions coming from first derivatives of the 
ideal hydrodynamic fields, i.e.,
\be
a_i^\nu = a_i^{(1)} Du^\nu + a_i^{(2)} T \nabla^\nu \frac{1}{T} + a_i^{(3)} 
\nabla^\nu \frac{\mu}{T}
\ee
where we chose $1/T$ and $\mu/T$ as the two independent variables instead
of $e$ and $n$.
But the three terms are not independent - energy-momentum conservation
(\ref{hydroeq}) and the Gibbs-Duham relation $s\, dT = dP - n\, d\mu$
provide one constraint 
\be
\frac{1}{T} \Delta^{\nu\alpha} Du_\alpha + \nabla^\nu \frac{1}{T} 
= \frac{n}{\varepsilon + p} \nabla^\nu \frac{\mu}{T} \ ,
\ee
and $\nabla^\nu(\mu/T)$ may be ignored, at least parametrically, 
because it is 
proportional to the heat flow (\ref{NS_N}) in 
the first-order (Navier-Stokes) theory. 
Therefore, to leading accuracy only one scalar function enters and we 
can write
\be
a_i^\mu = -a_i (e,n) Du^\mu \ .
\ee

Analogous arguments give
\be
T\nabla^\nu (\alpha_i/T) \approx 
T \frac{\partial (\alpha_i/T)}{\partial (1/T)} \nabla^\nu \frac{1}{T}
\approx 
- \frac{\partial (\alpha_i/T)}{\partial (1/T)} \nabla^{\nu\alpha} Du_\alpha
\ee
from which (\ref{ai_prime}) follows. 

We plan to revisit the above approximations in a future study.
In any case, they do not influence our 0+1D calculations here because 
the $a_i$ terms do not play a role (heat flow vanishes by symmetry).

\section{Generalized conservation laws}
\label{App:ConsLaws}

Here we present general relations of the form
\be
 \frac{\mathrm{d}\cal A(\tau)}{\mathrm{d}\tau} = \cal B(\tau)
\label{test}
\ee
that can be used to test the accuracy of numerical dissipative
hydrodynamics solutions in any dimensions. 
${\cal A}$ and ${\cal B}$ only depend on the
hydrodynamic fields at the given $\tau$. Evaluating them at each
time step, one can either numerically differentiate ${\cal A}(\tau)$ or
integrate ${\cal B}(\tau)$ and check how accurately the solutions
satisfy~(\ref{test}).

Consider a four-divergence $\partial_\mu A^\mu(x)$
(in regular Minkowski coordinates). Integration over a four-volume $V_4$ gives
\be
\int\limits_{V_4} \mathrm{d}^4 x\, \partial_\mu A^\mu(x) 
= \int\limits_{\sigma(V_4)} \mathrm{d}\sigma_\mu(x)\, A^\mu(x)
\ee
where $\sigma(V_4)$ is the 3D boundary (``surface'') of $V_4$.
Now take the special case of a Bjorken `` box''
$V_4 = \Delta\tau \times \Delta \eta \times A_T$ 
with an infinite transverse area $A_T\to \infty$ but infinitesimal
proper time and finite coordinate rapidity extensions
$\Delta\tau \to 0$, $\Delta \eta = \eta_2-\eta_1$. 
Assuming $A^\mu(x)$ drops faster than
$1/x_T^2$ at large $|\vx_T|$, we can neglect surface terms
at $|\vx_T| \to \infty$ and keep only contributions on $\tau = const$ and
$\eta = const$ hypersurfaces:
\be
\int \mathrm{d}\tau\, \tau\, \mathrm{d}\eta\, \mathrm{d}x_T^2\, 
                                            \partial_\mu A^\mu(x) 
= \left[\int\limits_{\sigma(\tau+d\tau)} \mathrm{d}\sigma_\mu^{(\tau)} 
- \int\limits_{\sigma(\tau)} \mathrm{d}\sigma_\mu^{(\tau)} 
+ \int\limits_{\sigma(\eta_2)}  \mathrm{d}\sigma_\mu^{(\eta)} - 
\int\limits_{\sigma(\eta_1)}  \mathrm{d}\sigma_\mu^{(\eta)}
\right] 
\, A^\mu(x) 
\ee
where the surface normals are
\be
\mathrm{d}\sigma^\mu_{(\tau)} 
  = \tau\, \mathrm{d}x_T^2\, \mathrm{d}\eta\, u_B^\mu \ ,
\qquad  
\mathrm{d}\sigma^\mu_{(\eta)} 
  = -\mathrm{d}\tau\, \mathrm{d}x_T^2 \, u_3^\mu \ ,
\quad {\rm with\ }
u_B^\mu \equiv (\ch \eta, \vo, \sh \eta) \ , 
\quad 
u_3^\mu \equiv (\sh \eta, \vo, \ch \eta)
\ee
and we used
$\mathrm{d}^4 x = \mathrm{d}\tau \, \tau\, \mathrm{d}\eta \, \mathrm{d}x_T^2$. 
Here, $u_B$ is the longitudinal Bjorken flow velocity, while $u_3$ is
its orthonormal counterpart in the $t-z$ plane. Note that the actual
flow velocity does \emph{not} need to be $u_B$. Dividing by
$\Delta \tau$ and taking the limit we arrive at
\be
\tau \int \mathrm{d}\eta\,\mathrm{d}x_T^2\, \partial_\mu A^\mu 
= \partial_\tau
\left(\tau \, \int \mathrm{d}\eta\, \mathrm{d}x_T^2\, u_B^\mu A_\mu \right) 
- \int \mathrm{d}x_T^2 \, u_3^\mu \left(A_\mu(\eta_1) - A_\mu(\eta_2)\right) 
\ ,
\label{conslaw}
\ee 
which is a generalized conservation law for the quantity
\be 
{\cal A} \equiv 
         \tau \, \int \mathrm{d}\eta\, \mathrm{d}x_T^2\, u_B^\mu A_\mu \ .  
\ee
If $\partial_\mu A^\mu \equiv 0$, and the surface term
$u_3^\mu(A_\mu(\eta_1)-A_\mu(\eta_2))$ vanishes, we have
${\cal A}(\tau) = const$.

In a boost-invariant calculation the longitudinal extension of the
system is formally infinite and thus a generalized conservation law
for a quantity per unit rapidity is more practical. It can be obtained
in a similar fashion if one divides by $\Delta\eta$ and takes the limit
$\Delta\eta \to 0$. The result is
\be
 \tau \int \mathrm{d}x_T^2\, \partial_\mu A^\mu
  = \partial_\tau \frac{\mathrm{d}{\cal A}}{\mathrm{d}\eta}
    - \partial_\eta \int \mathrm{d}x_T^2\, u_3^\mu A_\mu \ ,
\label{conslaw-rap}
\ee
where
\be
 \frac{\mathrm{d}{\cal A}}{\mathrm{d}\eta} 
   = \tau \int \mathrm{d}x_T^2\, u_B^\mu A_\mu \ .
\ee
Again, if $\partial_\mu A^\mu \equiv 0$ and the $\eta$-derivative term
vanishes, we have $\mathrm{d}{\cal A}/\mathrm{d}\eta = const$.

   \subsection{Charge / Particle number}

We first apply Eq.~(\ref{conslaw}) to a \emph{conserved} current in
Eckart frame: $N^\mu = n_{eq}u^\mu$, where
$u^\mu = \gamma(\ch\,\theta, v\, \ve_R, \sh\, \theta)$
is the flow four-velocity and $\theta$ is the flow rapidity. Now
$u_B^\mu u_\mu = \gamma\ch(\eta-\theta)$ and
$u_3^\mu u _\mu = \gamma\sh(\eta-\theta)$. If the rapidity interval
is so large that $N^\mu(\eta_1)=N^\mu(\eta_2)=0$, or the system
is boost invariant, $\eta \equiv \theta$, the surface terms are zero
and we get a simple conservation law
\be
N = \tau \int \mathrm{d}\eta \, \mathrm{d}x_T^2 \, \gamma\, n\, 
\ch(\eta-\theta)
  = const \ .
\ee
In a boost-invariant case, the coordinate rapidity integral is trivial
and we get
\be
\frac{\mathrm{d}N}{\mathrm{d}\eta} 
  = \tau \int \mathrm{d}x_T^2 \, \gamma n = const \ .
\ee

   \subsection{Entropy}

Second, we apply Eq.~(\ref{conslaw}) to the entropy current (\ref{SmuIS})
and its divergence (\ref{SdivIS}).
If $S^\mu(\eta_1)=S^\mu(\eta_2)=0$, we get
\be
\partial_\tau S 
 = \tau \int \mathrm{d}\eta\, \mathrm{d}x_T^2\, 
             \left(\frac{\Pi^2}{\zeta T} 
                 - \frac{q_\mu q^\mu}{\kappa_q T^2}
                 + \frac{\pi_{\mu\nu}\pi^{\mu\nu}}{2\eta_s T} \right)
 \ge 0 \ ,
\ee
where the entropy of the system is
\be
S = \tau \int \mathrm{d}\eta \, \mathrm{d}x_T^2\,  u_B^\mu S_\mu \ ,
\ee
and the last inequality follows from the general properties
$q^\mu q_\mu \le 0$ and $\pi^{\mu\nu} \pi_{\mu\nu} \ge 0$.

For longitudinally boost invariant dynamics, 
it is more natural to follow entropy per unit rapidity:
\be
\frac{dS}{d\eta} = \tau \int \mathrm{d}x_T^2\,  u_B^\mu S_\mu \ ,
\qquad \partial_\tau \left(\frac{dS}{d\eta}\right) 
 = \tau \int \mathrm{d}x_T^2\, 
             \left(\frac{\Pi^2}{\zeta T} 
                 - \frac{q_\mu q^\mu}{\kappa_q T^2}
                 + \frac{\pi_{\mu\nu}\pi^{\mu\nu}}{2\eta_s T} \right) \ge 0 \ .
\ee

   \subsection{Energy-momentum}

Finally we derive the conservation equation corresponding to
energy-momentum conservation $\partial_\mu T^{\mu\nu} = 0$.
Contraction of the energy-momentum tensor with $u_B^\mu$ gives the
conservation of energy. In a case where the entire system is within
the interval $[\eta_1,\eta_2]$,
\be 
 \partial_\tau E \equiv 
    \partial_\tau \left( \tau \int \mathrm{d}\eta\, \mathrm{d}x_T^2\, 
                               u_B^\mu T_{\mu\nu} u_B^\nu \right) = 0 \ .
\ee
Contraction with $u_R^\mu \equiv (0, \ve_R, 0)$ gives the change in
transverse radial momentum. Substituting
\be
\partial_\mu (T^{\mu\nu} u_{R,\nu}) = 0 + T^{\mu\nu} \partial_\mu u_{R,\nu} 
\ee
into Eq.~(\ref{conslaw}) results in
\be
 \partial_\tau M_r \equiv
   \partial_\tau \left(\tau \int \mathrm{d}\eta\, \mathrm{d}x_T^2\,
                           u_B^\mu T_{\mu\nu} u_R^\nu \right)
 = \tau \int \mathrm{d}\eta\, \mathrm{d}x_T^2\,
                T^\mu{}_\nu\partial_\mu u_R^\nu \ .
\ee

To be more specific, we also show as an example a boost-invariant,
cylindrically symmetric case. In Landau frame
\be
T^{\mu\nu} = (\varepsilon + p + \Pi) u^\mu u^\nu - (p+\Pi) g^{\mu\nu} 
             + (-\tpi_2 - \tpi_3) u_1^\mu u_1^\nu 
             + \tpi_2 u_2^\mu u_2^\nu
             + \tpi_3 u_3^\mu u_3^\nu \ ,
\ee
where $u_1$ is the orthonormal counterpart of the flow velocity in the
time-radial plane, while $u_2$ and $u_3$ are orthonormal 
counterparts of these in the axial and beam (rapidity) direction
\be
u^\mu = \gamma(\ch\,\eta, v\,\ve_R, \sh\,\eta) \ , 
\quad u_1^\mu = \gamma(v\,\ch\,\eta, \ve_R, v\,\sh\,\eta) \ ,
\quad u_2^\mu = (0, \ve_\phi, 0) \ ,
\quad u_3^\mu = (\sh\,\eta, \vo, \ch\,\eta) \ .
\ee
These vectors are normalized to $u^2 = 1$, $u_1^2 = u_2^2 = u_3^2 = -1$.
The viscous pressure tensor components in the fluid rest frame are
$\pi^{\mu\nu}_{LR} = diag(0,-\tpi_2-\tpi_3,\tpi_2,\tpi_3)$.
It is important to notice that the surface terms in
Eq.~(\ref{conslaw}) or the $\eta$-derivative term in
Eq.~(\ref{conslaw-rap}) are now nonzero.  Contraction by $u_B^\mu$ as
above and substitution into Eq.~(\ref{conslaw-rap}) gives the evolution
of the energy per unit rapidity:
\be
\partial_\tau \left(\frac{\mathrm{d}E}{\mathrm{d}\eta}\right) \equiv
\partial_\tau
\left(\tau \, \int \mathrm{d}x_T^2\,  T^{00}(\eta{=}0) \right) 
= - \int \mathrm{d}x_T^2\, (p + \Pi + \tpi_3) \ .
\ee
Contraction by $u_R^\mu$ gives the evolution of transverse radial momentum per
unit rapidity:
\be
\partial_\tau \left(\frac{\mathrm{d}M_r}{\mathrm{d}\eta}\right) \equiv
\partial_\tau \left(\tau \int \mathrm{d}x_T^2\,  T^{01}(\eta{=}0,\phi{=}0)
\right)
= \tau \, \int \mathrm{d}x_T^2\,  \frac{p + \Pi + \tpi_2}{R} \ ,
\ee
where we have used the relations
\be
 u_B^\mu T_{\mu\nu} u_R^\nu = - T^{01}(\eta{=}0,\phi{=}0) \ , \qquad
\partial_\mu u_R^\nu = -\frac{1}{R} u_{2,\mu} u_2^\nu \ .
\ee

The above results reflect general expectations. Particle number,
per unit rapidity $\mathrm{d}N/\mathrm{d}\eta$, is strictly
conserved in both the ideal and the dissipative case. 
Entropy per unit rapidity $\mathrm{d}S/\mathrm{d}\eta$ 
is conserved for an ideal
fluid but increases if there is dissipation. In both cases, the energy
per unit rapidity $\mathrm{d}E/\mathrm{d}\eta$ decreases due to longitudinal
work, while the radial momentum per unit rapidity 
$\mathrm{d}M_r/\mathrm{d}\eta$ increases due to build-up of radial
flow, {\em as long as the system stays near equilibrium} (i.e., the
total pressure is dominated by the ideal part).

\section{Viscous solutions for various cross-section scenarios}
\label{App:IS}

In this Section we analyze viscous Israel-Stewart and Navier-Stokes
solutions for four different types of cross section:
constant, $\sigma \propto 1/T^2$, $\sigma \propto \tau^{2/3}$, and 
$\etaOs = const$.
For convenience, we will often
use normalized quantities
\be
   \tilde A(\tau/\tau_0) \equiv \frac{A(\tau)}{A(\tau_0)} \ .
\ee
We will show that for typical observables of interest (average pressure,
 pressure anisotropy, entropy, shear viscosity to entropy ratio),
$\etaOs = const$ dynamics is well approximated by 
$\sigma\propto \tau^{2/3}$ already for $K_0 = 1$. 

In analytic considerations,
it will be often convenient to drop the
$\pi_L^2$ term in the equations of motion (\ref{EOMp})-(\ref{EOMpiL}), 
which is a good approximation
for $|\pi_L| \ll p$, i.e., the general
region of validity of viscous hydrodynamics. This should not
be confused with the ``naive'' Israel-Stewart
approximation, which also ignores the $4/3$ factor
in (\ref{EOMpiL}).
For the $\sigma \propto \tau^{2/3}$ and $\sigma = const$ scenarios
we obtain this way accurate approximate analytic 
Israel-Stewart solutions. We also derive analytic Navier-Stokes solutions 
for $\sigma = const$, $\sigma \propto \tau^{2/3}$ and $\sigma \propto 1/T^2$.

\subsection{Solutions for ultra-relativistic gas with constant \boldmath 
$2\to 2$
cross section}
\label{Sec:IS_const}

For a {\em constant} cross section, 
\be
\lambda_{tr}(\tau) \propto \tau \qquad \Rightarrow \qquad
K(\tau) = \frac{\tau_0}{\lambda_{tr}(\tau_0)} 
\equiv K_0 = const \ .
\ee
If we ignore $\pi_L^2$ term, the {\em linear}
equations of motion (\ref{EOMp})-(\ref{EOMpiL})
can be solved in a straightforward manner:
\bea
\pi_L(\ttau) &=& 
  \ttau^{-\frac{4}{3}-\frac{\kappa_0}{3}} 
\left[
  \frac{\pi_{L,0}}{2} \,T_+\!\!\left(\ttau\right)
  - \frac{1}{2D}\left( \kappa_0 \pi_{L,0} + \frac{8p_0}{3} \right) 
\,T_-\!\!\left(\ttau\right)
\right]
\label{piL_constsi}
\\
p(\ttau) &=& 
  \ttau^{-\frac{4}{3}-\frac{\kappa_0}{3}} 
\left[
  \frac{p_0}{2} \,T_+\!\!\left(\ttau\right)
  + \frac{1}{2D} \left( \kappa_0 p_0 - \pi_{L,0} \right) 
\,T_-\!\!\left(\ttau\right)
\right]
\label{p_constsi}
\eea
where
\be
\kappa_0 \equiv \frac{K_0}{C} \ , \qquad 
D \equiv \sqrt{\frac{8}{3} + \kappa_0^2} \ , \qquad
T_{\pm}(x) \equiv x^{D/3} 
                  \pm x^{-D/3}  \ , \quad
p(\tau_0) \equiv p_0 \ , \qquad \pi_L(\tau_0) \equiv \pi_{L,0} \ .
\ee
For a practical approximate formula for the pressure evolution,
see (\ref{p_approx_constsi}).

In the ideal hydrodynamic ($\eta_s \to 0$, or equivalently 
$\kappa_0 \to \infty$) 
limit we recover
\be
\pi_L(\tau > \tau_0) = 0 \ , 
\qquad p(\tau) = p_0 \left(\frac{\tau_0}{\tau}\right)^{4/3} \ .
\ee
At late times the pressure anisotropy,
irrespectively of its initial value $R_{p,0}$,
approaches a {\em constant} determined
solely by the parameter $\kappa_0$
\be
R_\infty \equiv R_p(\tau \to \infty) 
= \frac{12\kappa_0 - 10}{9D+3\kappa_0 + 14} < 1 \ .
\label{Rinfty}
\ee
For a finite $\kappa_0$,
the final anisotropy is below unity. 

Therefore, with a constant
cross section, the late-time behavior of the system 
does {\em not} become ideal hydrodynamic but instead
the Navier-Stokes limit applies
(cf. (\ref{Rdot_IS}) and (\ref{R_NS})).
Indeed, for large $\kappa_0$, 
(\ref{piL_constsi})-(\ref{p_constsi}) reproduce the NS solution
\be
p^{NS}(\tau) = p_0 
\left(\frac{\tau_0}{\tau}\right)^{4/3-4/(9\kappa_0)}  \qquad , \qquad
\pi_L^{NS}(\tau) = - \frac{4p^{NS}(\tau)}{3\kappa_0} \ ,
\label{p_constsi_NS}
\ee
and the final IS and NS anisotropies (\ref{Rinfty}) 
and (\ref{R_NS}) agree, 
$R_\infty = 1 - 2/\kappa_0 + 4/(3\kappa_0^2) + \cO(1/\kappa_0^3)$.

Because $R_\infty$ 
is a monotonically increasing function of $\kappa_0$, the final 
pressure anisotropy is a measure of the viscosity.
Inverting (\ref{Rinfty}),
\be
\kappa_0 = \frac{5 + 14 R_\infty - R_\infty^2}{6 - 3 R_\infty - 3 R_\infty^2}
\ee
i.e., near equilibrium ($\kappa_0 \gg 1$)
\be
\frac{\eta_s(\tau)}{n(\tau)} = \frac{T \tau}{\kappa} 
\approx \frac{1-R_\infty}{2} T_0 \tau_0 
\left(\frac{\tau}{\tau_0}\right)^{\gamma}
\ , \qquad \gamma = \frac{2}{3}+\frac{4}{9}\frac{\eta_s(\tau_0)}{n_0} 
\frac{1}{T_0 \tau_0} \ ,
\ee
where in the last step we approximated the temperature evolution
using the leading NS term (\ref{p_constsi_NS}). It is natural
to measure viscosity relative to the density, which up to a factor
$(4-\chi)$ is the same as $\etaOs$.

The exact analytic solutions to
the 'naive' Israel-Stewart equations are analogous to
(\ref{p_constsi})-(\ref{piL_constsi}) but involve different powers of $\ttau$
\be
 \ttau^{\delta_\pm} \ , \qquad \delta_\pm^{naive} 
                          = -\frac{2}{3} - \frac{\kappa_0}{3}  
              \pm \frac{\sqrt{\kappa_0^2 - 4\kappa_0 + \frac{20}{3}}}{3} \ .
\ee
The late time behavior is governed by the exponent
\be
\delta_+^{naive} = -\frac{4}{3} + \frac{4}{9\kappa_0} + \frac{8}{9\kappa_0^2} 
               + \cO\!\left(\frac{1}{\kappa_0^3}\right) \ ,
\ee
which does incorporate correctly the 
ideal hydrodynamic limit $(-4/3)$ and the Navier-Stokes correction
$4/(9\kappa_0)$ but is in general higher, the smaller the $\kappa_0$,
than the complete IS result 
$\delta_+ = -4/3+4/(9\kappa_0) - 8/(27\kappa_0^3) 
+ \cO(1/\kappa_0^5)$. Therefore, the 'naive' approach overestimates 
the pressure.
In addition, it underestimates the
asymptotic pressure anisotropy 
$R_\infty^{naive} = 1 - 2/\kappa_0 - 8/(3\kappa_0^2) + \cO(1/\kappa_0^3)$,
and therefore,
overpredicts the magnitude of the shear stress to pressure ratio $|\xi|$.

\subsection{Solutions for ultra-relativistic gas with \boldmath
$\sigma_{2\to 2}\propto 1/T^2$}
\label{Sec:IS_T2}

A constant cross section implies the existence of some external scale in the
problem.
For a scale-invariant system, however, the only scale available (in thermal
and chemical equilibrium) is the temperature, and therefore the cross section
behaves as $\sigma \propto 1/T^2$. (\ref{def_K}), 
(\ref{EOS}) and (\ref{n_bj1D}) 
then give
\be
K(\tau) = K_0 \frac{T_0^2}{T^2} 
= \frac{K_0}{\tilde p^{\;\!2} \ttau^2} \ ,
\ee
i.e., even without the $\pi_L^2$ term, 
the equations of motion become nonlinear 
(but are easy to solve numerically). 

For ideal hydrodynamic evolution,
 $p\propto \tau^{-4/3}$ and thus, 
unlike for the case of a constant cross section,
\be
K(\tau) = K_0 \ttau^{2/3}
\label{Ktau_T2_ideal}
\ee 
increases with increasing $\tau$.
$K(\tau)$ must grow in general in the viscous hydrodynamic case as well
because dissipative corrections, namely the $\pi_L/\tau$ term in 
(\ref{EOMp}), are assumed to be small
(or else hydrodynamics is not applicable any longer).
Consequently, the system gets {\em closer and closer} 
to ideal hydrodynamic behavior as
time evolves (as long as the expansion is only one-dimensional).
For example, the pressure anisotropy approaches
unity at late times, for {\em any} $\kappa_0 > 0$ and initial $\pi_{L,0}/p_0$,
\be
  R_p(\tau \to \infty) \to 1 \ .
\ee

The exact Navier-Stokes solution
\be
p^{NS}(\tau) = \left(\frac{\tau_0}{\tau}\right)^{4/3} 
\frac{p_0}{\sqrt{1+\frac{4}{3\kappa_0}
         \left[\left(\frac{\tau_0}{\tau}\right)^{2/3} - 1\right]}}
\ee
behaves similarly. At late times $p \propto \tau^{-4/3}$ 
as in the ideal case, therefore,
$\kappa(\tau\to \infty) 
= \kappa_0 / (\tilde p^{\;\! 2} \ttau^2) \to \infty$, i.e., 
$R_\infty = 1$. The rate of approach to unity is controlled by 
the viscosity 
\be
R_p^{NS}(\tau) = 1 - \frac{2}{\kappa_0} 
\left(\frac{\tau_0}{\tau}\right)^{2/3} \, 
\left[ 1 + \cO(1/\kappa_0^2) +  \cO((\tau_0/\tau)^{2/3})\right]
\approx 1 - \frac{2}{T_0\tau_0}\frac{\eta_s}{n}
\left(\frac{\tau_0}{\tau}\right)^{2/3} \ .
\label{R_NS_sigmaT}
\ee
Viscosity also increases the pressure relative to the ideal case
\be
\frac{p^{NS}}{p_{ideal}}(\tau \gg  \tau_0) \to   
\frac{1}{\sqrt{1 - \frac{4}{3\kappa_0}}} 
\approx 1 + \frac{2}{3T_0\tau_0}\frac{\eta_s}{n} \ .
\label{p_ratio_NS_sigmaT}
\ee

\subsection{Solutions for ultra-relativistic gas with \boldmath 
$\sigma_{2\to 2} \propto \tau^{2/3}$}
\label{Sec:IS_tau23}

Near the ideal hydro limit (i.e., for small viscosities and 
$\pi_{L,0}/p_0$), 
one may substitute the approximate result
(\ref{Ktau_T2_ideal}) in the equations of motion (\ref{EOMp})-(\ref{EOMpiL})
directly. Provided we drop the $\pi_L^2$ term, these can be converted 
to a second-order linear differential equation, e.g., for $p(\tau)$:
\be
\tau \ddot p + \frac{11}{3} \dot p + \frac{40}{27} \frac{p}{\tau} 
+ \frac{2K(\tau)}{3C} \left(\dot p + \frac{4}{3}\frac{p}{\tau}\right) = 0 \ ,
\ee
with initial conditions 
\be
p(\tau_0) = p_0 \ , 
\qquad {\dot p}(\tau_0) = - \frac{4 p_0 + \pi_{L,0}}{3\tau_0}  \ .
\label{initcond_Ktau}
\ee
The general solution with $K(\tau)$ from (\ref{Ktau_T2_ideal}) is%
\footnote{
First substitute $p(\ttau) \equiv \bar p(\ttau) \ttau^{-4/3}$,
then switch to a new variable $x \equiv -\kappa_0 \ttau^{2/3}$, finally
look for the solution in the form $\bar p(x) \equiv x^a q(x)$, and
choose a suitable $a$. 
}
\bea
p(\ttau) &=& \ttau^{-4/3} 
          \left[C_-\, \ttau^{-\frac{2\sqrt{6}}{9}} 
                   \ F_{-}(\kappa_0\, \ttau^{2/3})
              + C_+\, \ttau^{\frac{2\sqrt{6}}{9}} 
                   \ F_{+}(\kappa_0\, \ttau^{2/3})
          \right] 
\label{p_tau23}
\\
\pi_L(\ttau) 
&=& - 3\,\ttau^{-1/3}\, \frac{d [\ttau^{4/3}\, p(\ttau)]}{d\ttau} \ ,
\label{piL_tau23}
\eea
where 
\be
F_{\pm}(x) \equiv \Foo(\pm a, 1 \pm 2 a; -x)
\ , \qquad a = \sqrt{\frac{2}{3}}
\ee
are shorthands for confluent hypergeometric functions of the first 
kind,
while $C_{\pm}$ are matched%
\footnote{
Note that 
$$
\frac{d}{dx} \Foo(a, b, x) \equiv \frac{a}{b}\  \Foo(a+1, b+1, x) \ ,
$$
and from the Wronskian
$$
G_-(x) F_+(x) - G_+(x) F_-(x) = 4 a \ e^{-x}
$$
(cf. $W\{1,2\}$ in Eq. (13.1.20) in \cite{Abramowitz}).
}
to the initial conditions 
(\ref{initcond_Ktau})
\bea
C_{\pm} &=& \pm \frac{e^{\kappa_0}}{4 a} \left[
p_0 \, G_{\mp}(\kappa_0) - \pi_{L,0}\, F_{\mp}(\kappa_0)
\right]
\\
G_{\pm}(x) &\equiv& \pm 2 a \left[
\frac{x}{1 \pm 2 a} \, 
\Foo( 1 \pm a, 2 \pm 2 a, -x)
- \Foo( \pm a, 1 \pm 2 a, -x)
\right] \ .
\eea

A very practical approximate formula for the pressure evolution 
is given by (\ref{p_approx_etas}),
which comes from
the asymptotic forms (cf. (13.5.1) in \cite{Abramowitz})
\be
\Foo(a, b; -x) 
= \frac{\Gamma(b)}{\Gamma(b-a)}\, x^{-a}\, S(a,1+a-b,x) 
     + \frac{\Gamma(b)}{\Gamma(a)}\, e^{-x}\, (-x)^{a-b}\, S(b-a, 1-a, -x)
\label{F11_asympt} \ ,
\ee
where
\be
S(c,d,x) \equiv 1 + \frac{c\,d}{1!\, x} + \frac{c(c+1)d(d+1)}{2!\,x^2} 
+ \frac{c(c+1)(c+2)d(d+1)(d+2)}{3!\,x^3} + \cdots 
\ee
Note that the $e^{-x}$ term in (\ref{F11_asympt}) is crucial. 
For large $\kappa_0$, $C_\pm$ are exponentially large, 
however, the $e^{\kappa_0}$ factors drop out%
\footnote{
For example,
$$
\frac{\Gamma(1+2a)}{\Gamma(1+a)}\, F_-(\kappa_0)\, \kappa_0^{-a} 
- \frac{\Gamma(1-2a)}{\Gamma(1-a)}\, F_+(\kappa_0)\, \kappa_0^{a}
= 2 a \ \frac{e^{-\kappa_0}}{\kappa_0} 
\left[1 - \frac{1}{3\kappa_0} + \cO\!\left(\frac{1}{\kappa_0^2}\right)\right]
$$
and
$$
\frac{\Gamma(1+2a)}{\Gamma(1+a)} \,G_-(\kappa_0)\, \kappa_0^{-a} 
- \frac{\Gamma(1-2a)}{\Gamma(1-a)}\, G_+(\kappa_0)\, \kappa_0^{a} 
= 4 a e^{-\kappa_0}
\left[1+\frac{2}{3\kappa_0} - \frac{1}{9\kappa_0^2}
+ \cO\!\left(\frac{1}{\kappa_0^3}\right)\right]
$$
($a = \sqrt{2/3}$).
}
in linear combinations relevant for the pressure
and shear stress.

At late times $\tau \gg \tau_0 / \kappa_0^{3/2}$ 
the IS solutions recover ideal hydrodynamics for any initial condition,
\be
p(\tau)
\propto \left(\frac{\tau_0}{\tau}\right)^{4/3} \ , 
\quad
\pi_L(\tau)
\propto \left(\frac{\tau_0}{\tau}\right)^2
\quad \Rightarrow
\quad 
\frac{\pi_L}{p}(\tau) 
\propto \left(\frac{\tau_0}{\tau}\right)^{2/3} \to 0 \qquad
{\rm for} \ \  \tau \gg \frac{\tau_0}{\kappa_0^{3/2}} \ ,
\ee
as can be inferred from (\ref{F11_asympt}).
The Navier-Stokes solution
\be
p^{NS}(\tau) = p_0 \left(\frac{\tau_0}{\tau}\right)^{4/3} \, 
\exp\!\left\{\frac{2}{3\kappa_0} 
  \left[1 - \left(\frac{\tau_0}{\tau}\right)^{2/3}\right]\right\}
\label{p_tau23_NS}
\ee
exhibits the same features (as the Reader can easily verify).
For the late-time evolution,
this scenario gives smaller viscous corrections to the pressure and 
the pressure anisotropy than $\sigma \propto 1/T^2$. However,
in the large $\kappa_0$ limit 
we recover the same results (\ref{R_NS_sigmaT}) and (\ref{p_ratio_NS_sigmaT}).

Analogous derivation gives the exact solutions in the 'naive' 
Israel-Stewart case:
\bea
p(\ttau) &=& C'_- \, \ttau^{-2(1+a')/3} \, F'_-(\kappa_0 \ttau^{2/3})
         + C'_+  \, \ttau^{-2(1-a')/3} \, F'_+(\kappa_0 \ttau^{2/3})
\ , \qquad\qquad a' = \sqrt{\frac{5}{3}} 
\\ 
\pi_L(\ttau) &=& -3 \ttau^{-1/3}\, \frac{d[\ttau^{4/3}\, p(\ttau)]}{d\ttau}
\eea
where
\bea
C'_\pm &=& \pm \frac{e^{-\kappa_0}}{4a'} \, 
[p_0\, G'_\mp(\kappa_0) - \xi_0 \, F'_\mp(\kappa_0)] \\
F'_\pm(x) &\equiv& \Foo(1\pm a', 1\pm 2a',-x)
\ , \qquad G'_\pm(x) \equiv 2 x \frac{1\pm a'}{1\pm 2a'} F'_\pm(x) 
                   - 2(1\pm a') F'_\pm(x) \ .
\eea
With the help of (\ref{F11_asympt}) it 
is straightforward (but somewhat lengthy) to determine the 
late-time behavior
\be
\frac{p}{p_{ideal}} = T(\ttau) \left[P(\kappa_0) + \xi_0\, X(\kappa_0)\right]
\label{latetime_form}
\ee
where in the 'naive' case
\bea
T^{naive}(\ttau) &=& 1-\frac{2}{3\kappa_0\, \ttau^{2/3}}
                   -\frac{7}{9\kappa_0^2\,\ttau^{4/3}} 
                    + \cO\!\left(\frac{1}{\kappa_0^3\, \ttau^2}\right) 
\\
P^{naive}(\kappa_0) &=& 1 + \frac{2}{3\kappa_0} + \frac{5}{9\kappa_0^2} 
                      + \cO\!\left(\frac{1}{\kappa_0^3}\right) \ ,
\qquad
X^{naive}(\kappa_0) = - \frac{1}{2\kappa_0} - \frac{5}{6\kappa_0^2}
                      + \cO\!\left(\frac{1}{\kappa_0^3}\right) \ .
\label{naiveIS_coeffs}
\eea
Comparing to the 'complete' Israel-Stewart result (\ref{p_tau23}) 
(obtained in the small $\xi$ limit)
\bea
T^{IS}(\ttau) &\approx& 1-\frac{2}{3\kappa_0\, \ttau^{2/3}}
               -\frac{1}{9\kappa_0^2\,\ttau^{4/3}} 
               + \cO\!\left(\frac{1}{\kappa_0^3\, \ttau^2}\right) 
\\
P^{IS}(\kappa_0) &\approx& 
                       1 + \frac{2}{3\kappa_0} - \frac{1}{9\kappa_0^2} 
                      + \cO\!\left(\frac{1}{\kappa_0^3}\right) \ ,
\qquad
X^{IS}(\kappa_0) \approx - \frac{1}{2\kappa_0} + \frac{1}{6\kappa_0^2}
                      + \cO\!\left(\frac{1}{\kappa_0^3}\right) 
\label{IS_coeffs}
\eea
we see that for the 'naive' approximation the evolution
approaches ideal hydrodynamic 
$p / p_{ideal} \sim const$ behavior {\em later} 
(deviation of $T$ from unity is larger), 
and for near-equilibrium initial conditions ($\xi_0 \approx 0$)
the pressure saturates at a {\em higher} value ($P$ is larger).

\subsection{Solutions for ultra-relativistic gas with \boldmath 
$2\to 2$ cross section and $\etaOs = const$}
\label{Sec:IS_etas}

The last scenario we consider is when the cross section is dynamically 
adjusted to maintain a {\em constant} shear viscosity to
equilibrium entropy density ratio $\etaOs$, 
such as the conjectured lower bound
of $1/(4\pi)$.
From (\ref{n_bj1D}), (\ref{EOS}), (\ref{muT}), and (\ref{neq}),
\bea
\tilde s_{eq} =
\frac{1}{\ttau} 
         \left(1 + \frac{\ln \left[\ttau^4 
                                   \, \tilde p^3(\ttau)\right]}
         {4 - \chi_0} \right) \ ,
\label{seq}
\eea
and thus
\bea
\frac{\eta_s}{s_{eq}} 
= \frac{\eta_{s,0}}{s_{eq,0}}
  \, \frac{\tilde p(\ttau)\, \ttau^2}{\tilde K(\ttau)}
  \, \frac{4 - \chi_0} 
       {4 - \chi_0 
               + \ln \left[\ttau^4 \, \tilde p^3(\ttau)\right]} \ ,
\label{etas}
\eea
where
\be
\frac{\eta_{s,0}}{s_{eq,0}} =
\frac{T_0 \tau_0}{\kappa_0 (4-\chi_0)} \ .
\ee 
Therefore, $\etaOs = const$ requires
\be
K(\ttau) = K_0 \,  \tilde p(\ttau)\, \ttau^2
  \, \frac{4 - \chi_0} 
       {4 - \chi_0 
               + \ln \left[\ttau^4 \, \tilde p^3(\ttau)\right]}  \ .
\label{K_etas}
\ee
Within the generic region of validity for viscous hydrodynamics,
$|\pi_L| \ll p$,
this scenario also implies a growing $K(\tau) \sim \tau^{\approx 2/3}$
and therefore convergence to the ideal limit at late times.
Note that the double ratio $(\etaOs)/(\eta_{s,0}/s_{eq,0})$ as a 
function of $\tau/\tau_0$ depends only on $\pi_{L,0}/p_0$, $\kappa_0$,
the type of cross section
(encoded in $\tilde K$), and $\chi_0$.

We now analyze the time evolution of $\etaOs$ in the three earlier
scenarios.
Compared to the entropy density, $\etaOs$ 
contains an additional multiplicative term 
that comes from the time evolution of the shear 
viscosity. Assume first, for simplicity, 
that we are very close to the ideal hydro limit,
in which case 
$\etaOs \propto \tau^{2/3} / K(\tau)$. For a constant cross 
section, this results in a {\em growing} $\etaOs \propto \tau^{2/3}$;
while
for the other two cases, $\sigma \propto \tau^{2/3}$ or $\sigma \propto 1/T^2$
we obtain $\etaOs \approx const$.

In reality, there are of course viscous effects. 
Because 
\be
\frac{\tilde p(\ttau) \, \ttau^2}{\tilde K(\ttau)} =
\left\{\matrix{ 
\tilde p(\ttau) \ttau^{4/3} \times \ttau^{2/3}  
\quad {\rm for \ } \sigma = const \cr 
\left[\tilde p(\ttau) \ttau^{4/3}\right]^3 \ \qquad {\rm for \ }
\sigma \propto 1/T^2 \cr 
\tilde p(\ttau) \ttau^{4/3} \quad \, \ \qquad {\rm for \ } 
\sigma \propto \tau^{2/3} \cr} \right. \ ,
\label{firstterm}
\ee
the relevant quantity that determines the evolution of $\etaOs$ 
is $\tilde p \, \ttau^{4/3}$.
The last term in (\ref{etas}) is only a logarithm.
Therefore, the first term, (\ref{firstterm}), dominates the 
behavior. Typically, $\pi_L < 0$ and thus 
dissipation generates an increasing $\tilde p \, \ttau^{4/3}$.
The increase in $\etaOs$ is then fastest for the
constant cross section case.
The other two cases,
$\sigma \propto 1/T^2$ and 
$\sigma \propto \tau^{2/3}$, are not equivalent when there is dissipation
because for the latter the prefactor (\ref{firstterm})
is only linear in $\tilde p(\ttau) \ttau^{4/3}$
 and, therefore, $\etaOs$ grows much slower.

\subsection{Comparison of the various cross section scenarios}

After exploring the general behavior, we compare numerical 
solutions for the four scenarios. 
Unless stated otherwise, for the $\etaOs = const$ case we start 
the evolution from
chemical equilibrium, i.e., take $\chi_0 = 0$. For the other three scenarios,
the pressure and shear stress evolution does not depend on $\chi_0$.
For simplicity, we start the evolution
from $\pi_L(\tau_0) = 0$, and consider two extremes $K_0 = 1$, i.e., equal
expansion and scattering timescales,
and $K_0 = 6.67$, i.e., 6.67 times slower expansion than 
the timescale for scattering. On all figures, the dotted curves correspond
to the approximation when the $\pi_L^2$ term in (\ref{EOMpiL}) is ignored.

\begin{figure}
\epsfysize=6.5cm
\epsfbox{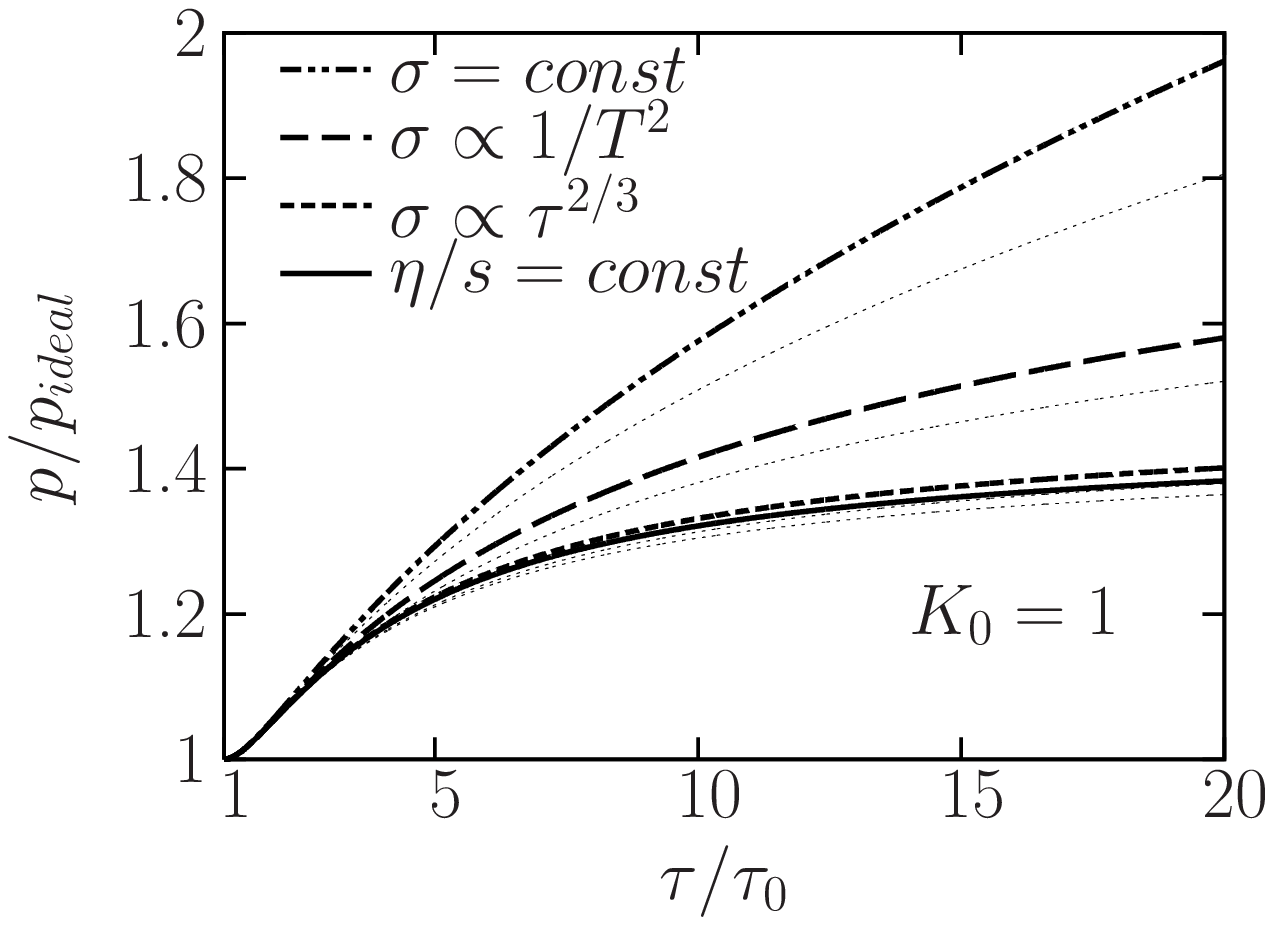}
\epsfysize=6.5cm
\epsfbox{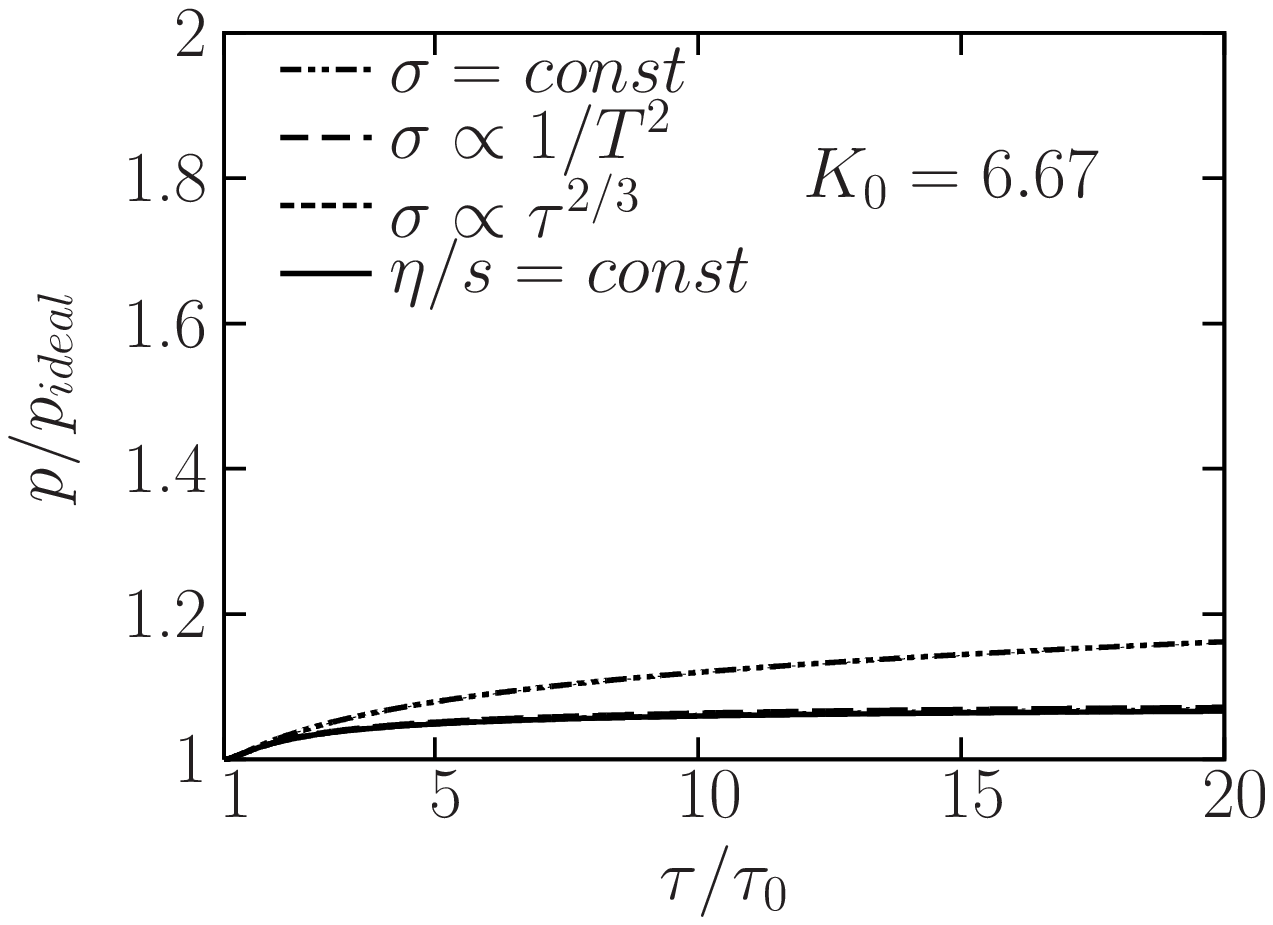}
\caption{Pressure evolution from viscous hydrodynamics relative
to the ideal hydrodynamic $p = p_0 (\tau_0/\tau)^{-4/3}$ result
 in 0+1D
Bjorken geometry for an ultrarelativistic gas with $2\to 2$ interactions.
Four scenarios are compared for $K_0 = 1$ (left) and 5 (right): 
$\sigma = const$ (dash-dot-dot), 
$\sigma \propto 1/T^2$ (long dash), $\sigma \propto \tau^{2/3}$ (short dash), 
and $\etaOs = const$ (solid).
Approximate results with dropping $\pi_L^2$ terms
in the equation of motion are also shown (thin dotted lines). }
\label{fig:p}
\end{figure}

Figure~\ref{fig:p} shows the evolution of the pressure relative to the 
ideal hydrodynamic $p\sim \tau^{-4/3}$ result (for a comparison
of the same observable between hydrodynamics and transport 
see Figure~\ref{Fig:p_comp}).
Dissipation increases
the pressure because it reduces the $pdV$ work. The effect is largest
for the $\sigma = const$ scenario, while smallest for $\etaOs = const$
and $\sigma \propto \tau^{2/3}$, which two give basically the same result.
For $K_0 = 1$ the fourth scenario $\sigma\propto T^2$ is in between these 
limits but by $K_0 = 6.67$ it becomes equivalent 
to $\sigma \propto \tau^{2/3}$. Dropping $\pi_L^2$ terms in (\ref{EOMpiL}) 
(thin dotted lines) is a fair $10-15$\% 
approximation for $\sigma = const$ and $\sigma \propto 1/T^2$ at $K_0 = 1$, 
which improves to an essentially 
exact one by $K_0 = 6.67$. For the other two scenarios,
$\etaOs = const$ and $\sigma\propto \tau^{2/3}$, the nonlinear term
can be safely ignored already for $K_0 = 1$. Note that for $K_0 = 6.67$,
dissipative corrections to the pressure are still 
very modest $10-15$\% at late $\tau/\tau_0 \sim 10-20$ in all four cases
studied.

\begin{figure}
\epsfysize=6.5cm
\epsfbox{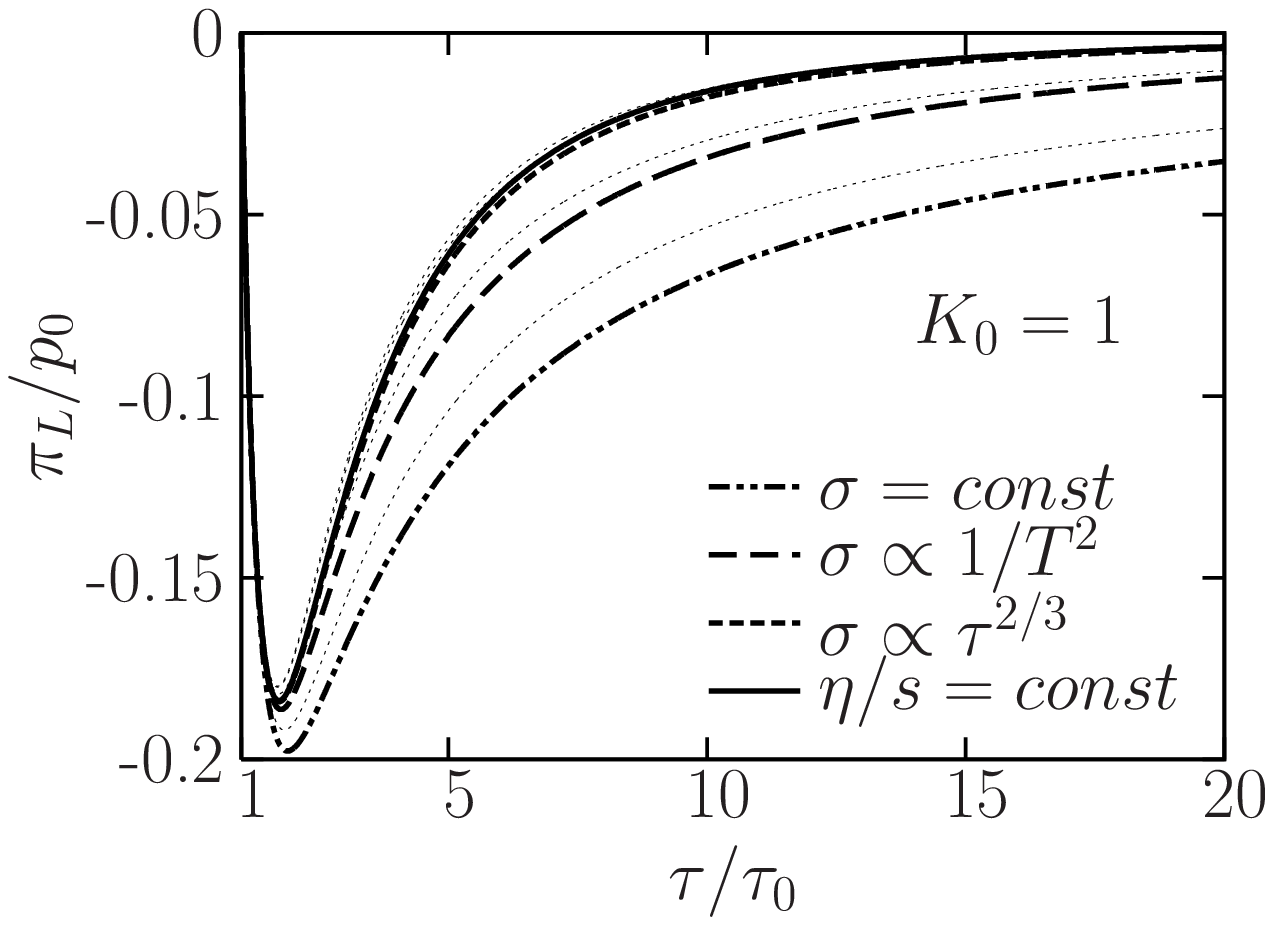}
\epsfysize=6.5cm
\epsfbox{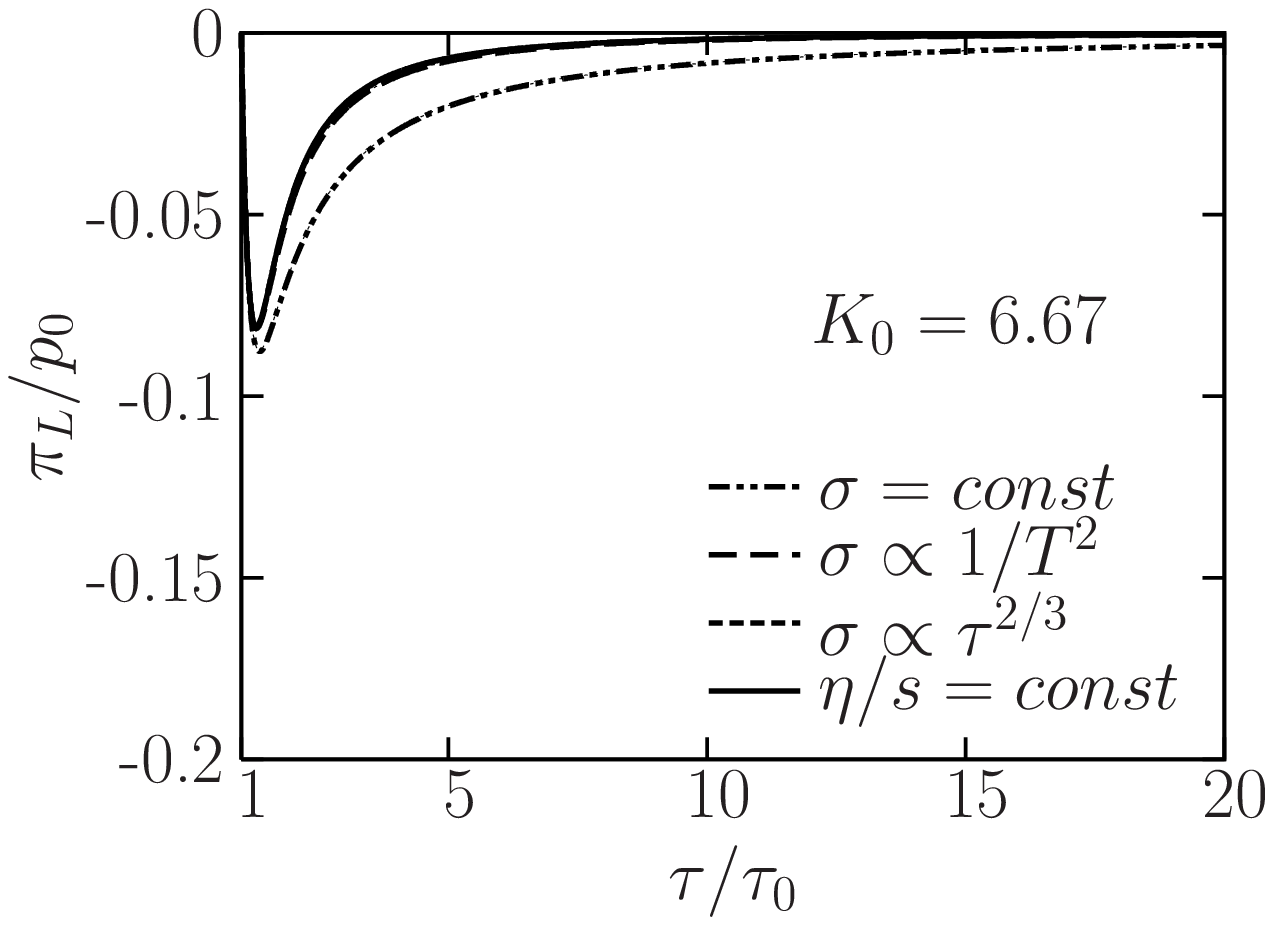}
\caption{Same as Fig.~\ref{fig:p} but 
for the longitudinal viscous shear $\pi_L$ normalized by the 
{\em initial} pressure.}
\label{fig:piL}
\end{figure}

Now we turn to the evolution of the viscous stress $\pi_L$ shown 
in Fig.~\ref{fig:piL}. All four scenarios give very similar results for the
early $\tau/\tau_0 \lton 1.5-2$ growth in magnitude but they differ in 
late-time relaxation. As inferred from the pressure evolution already,
$\etaOs = const$ and $\sigma \propto \tau^{2/3}$ are largely identical
and relax quickly toward the ideal limit. $\sigma = const$ is the one
that stays furthest away from equilibrium. For low $K_0 = 1$, the
$\sigma \propto 1/T^2$ case lies in between but by $K_0 = 6.67$ it 
becomes identical
to $\etaOs = const$ and $\sigma \propto \tau^{2/3}$. 
The $\pi_L^2$ term in the equation of motion affects the pressure and the 
viscous stress similarly, and can be ignored for $K_0 = 6.67$ in all cases -
for $\sigma \propto \tau^{2/3}$ and $\etaOs = const$ even at 
$K_0 = 1$.

\begin{figure}
\epsfysize=6.5cm
\epsfbox{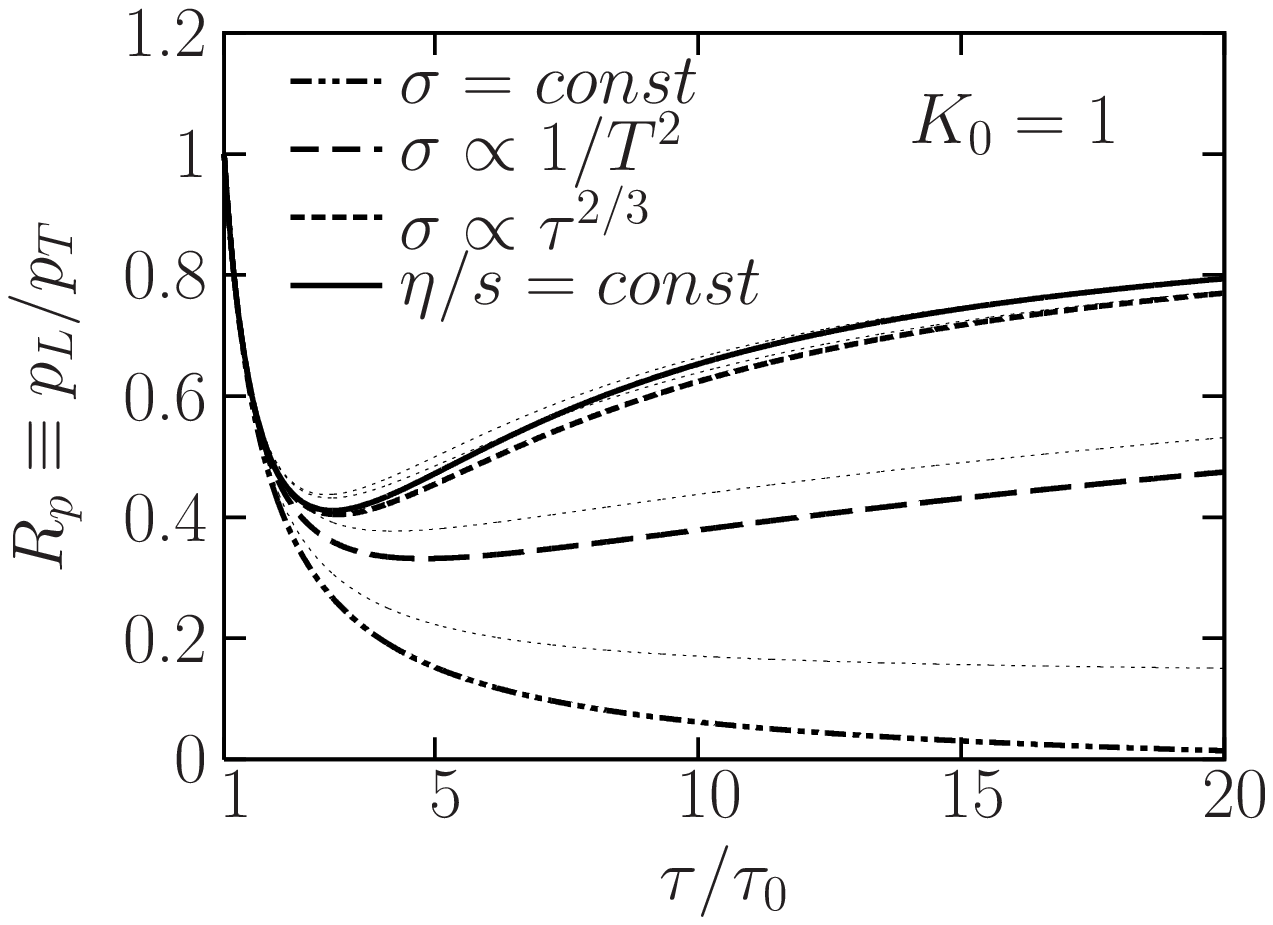}
\epsfysize=6.5cm
\epsfbox{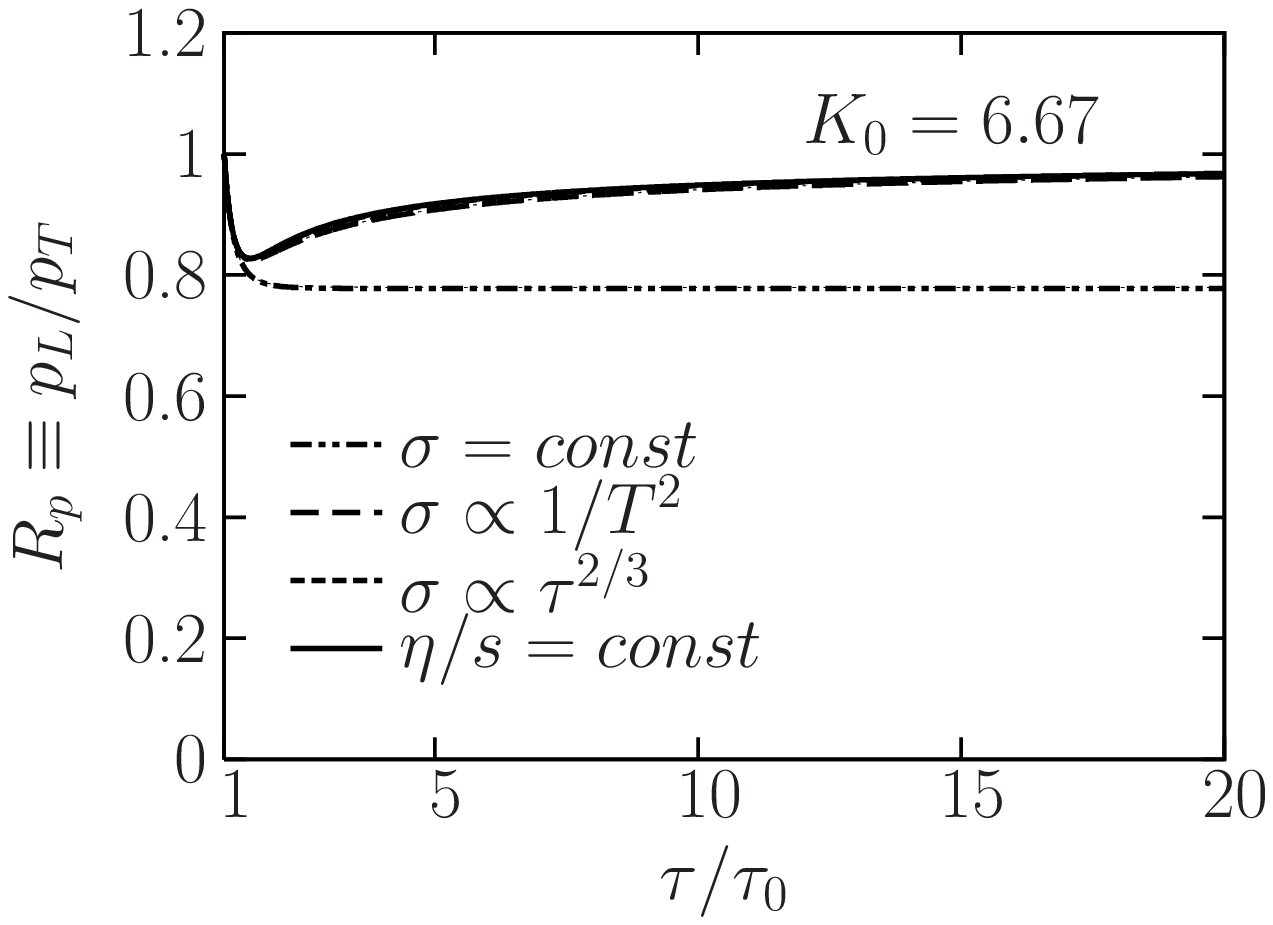}
\caption{Same as Fig.~\ref{fig:p} but 
for the pressure anisotropy evolution.}
\label{fig:R}
\end{figure}

The same observations carry over to the pressure anisotropy
$R_p = p_L/p_T$ shown in Figure~\ref{fig:R}. We plot this quantity
because it is the same one shown in Figure~\ref{Fig:R_comp}
for the hydro-transport comparison in Sec.~\ref{Sec:IS_validity}
(but note the logarithmic time axis there). These results further
confirm that $\sigma \propto \tau^{2/3}$
is a very good approximation to $\etaOs = const$ already for $K_0 = 1$.

\begin{figure}
\epsfysize=6.5cm
\epsfbox{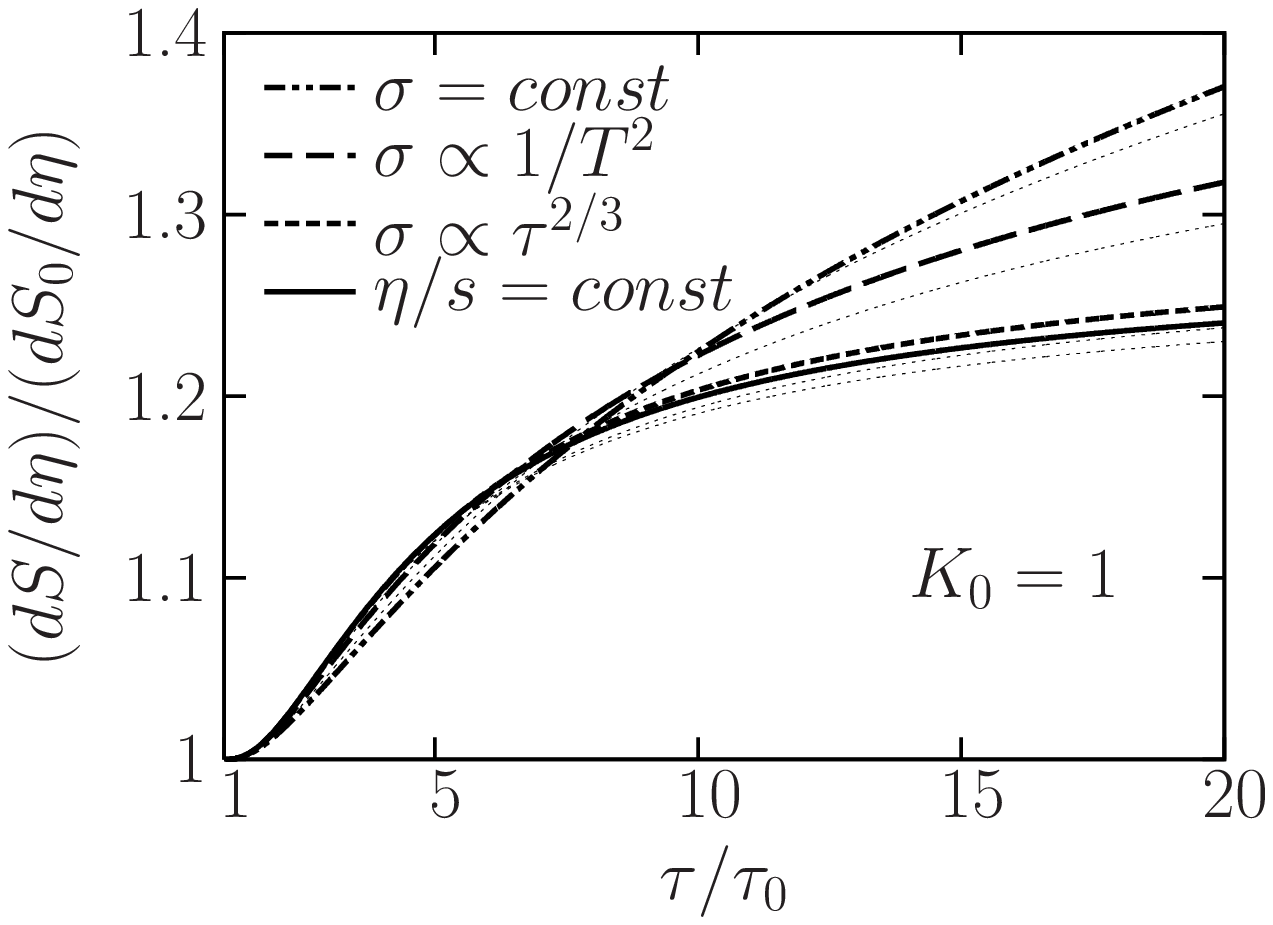}
\epsfysize=6.5cm
\epsfbox{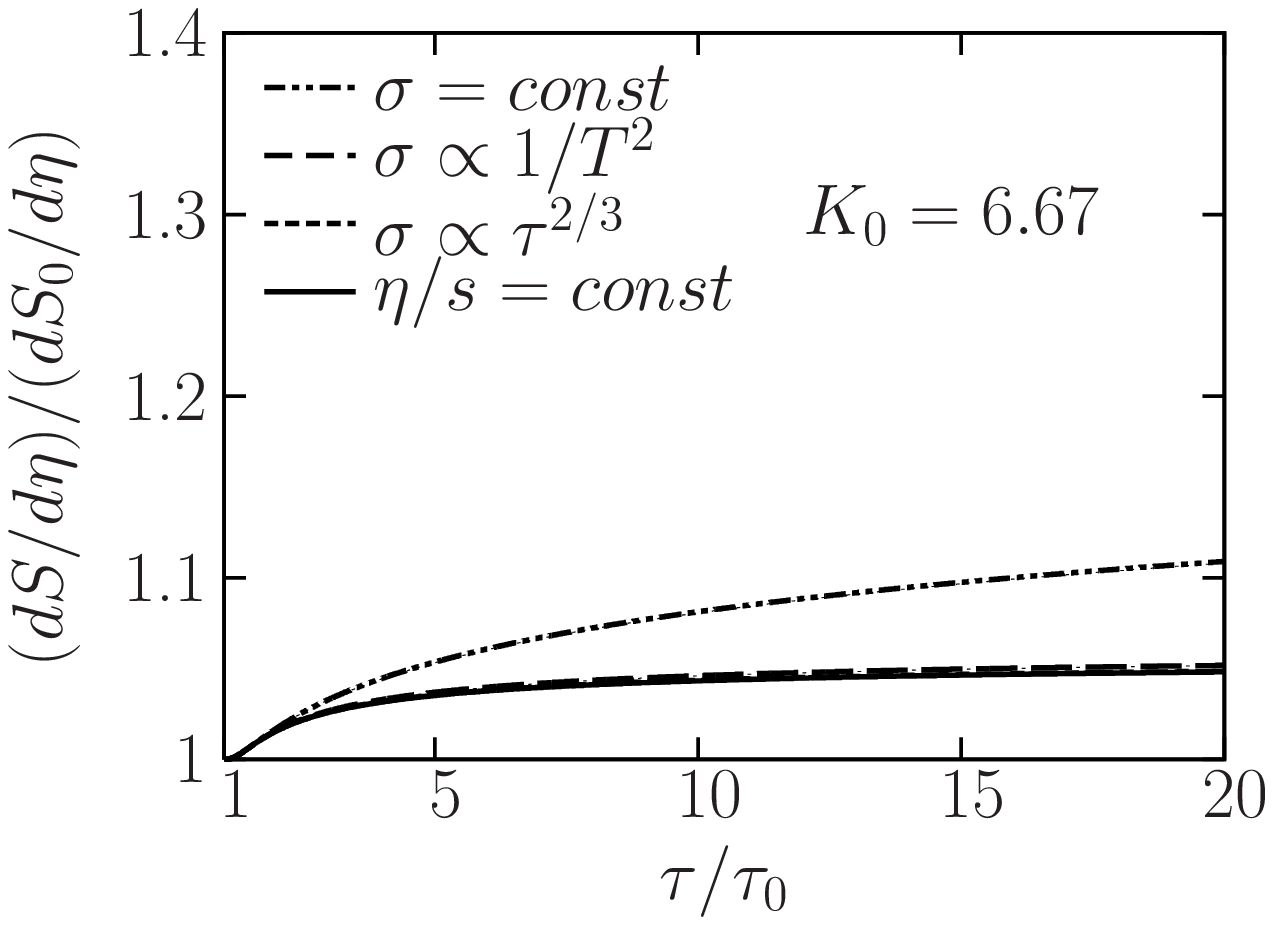}
\caption{Same as Fig.~\ref{fig:p} but 
for the normalized entropy per unit rapidity $(dS/d\eta) / (dS_0/d\eta)$.}
\label{fig:dSdeta}
\end{figure}

\begin{figure}
\epsfysize=6.5cm
\epsfbox{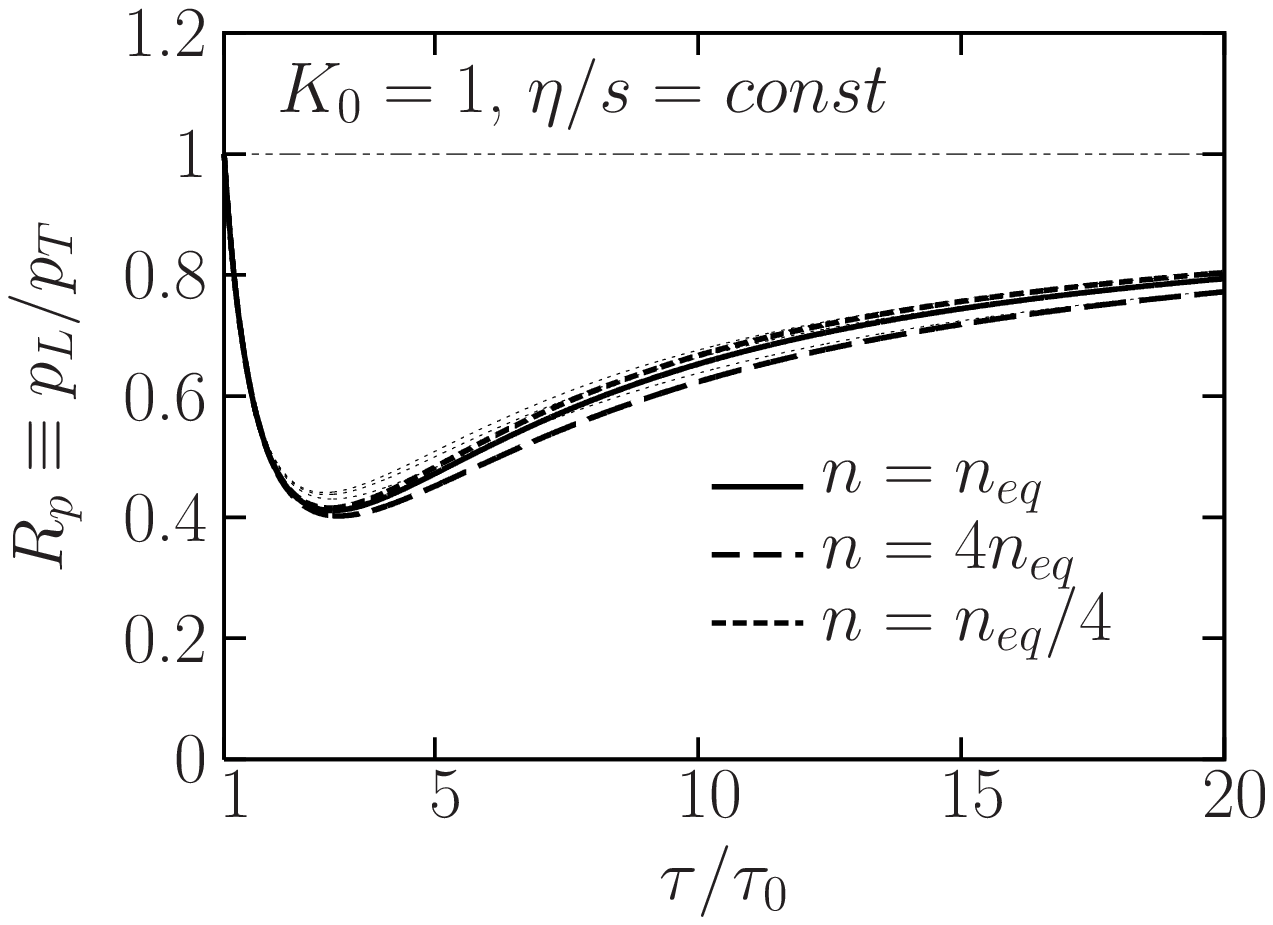}
\epsfysize=6.5cm
\epsfbox{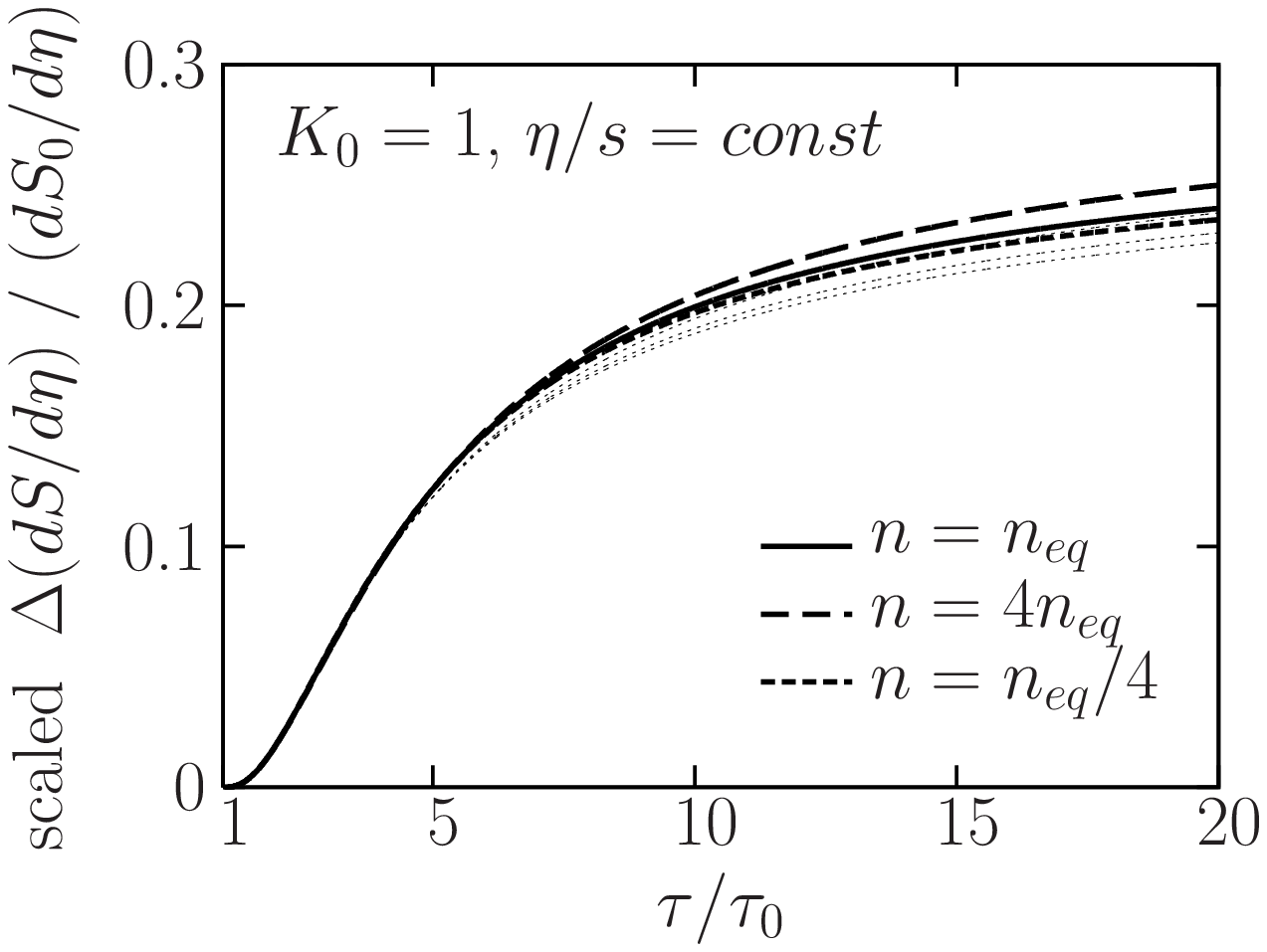}
\caption{Initial density dependence of Israel-Stewart 
viscous hydrodynamic solutions
for an ultrarelativistic gas expanding longitudinally
in 0+1D Bjorken geometry
with $2\to 2$ interactions
that maintain $\etaOs = const$. 
To amplify density effects,
the initial expansion timescale to mean free path ratio is chosen to 
be low, $K_0 = 1$. Three different {\em initial} densities are considered: 
chemical equilibrium $n = n_{eq}$ (solid),
oversaturation at
$n = 4n_{eq}$ (long dash), and undersaturation at $n= n_{eq}/4$ (short dash).
{\em Left:} time evolution of the pressure anisotropy $R_p = p_L/p_T$.
{\em Right:} time evolution of the {\em produced} entropy per unit rapidity,
normalized to the initial entropy per unit rapidity. The produced entropy
is scaled by $(4-\chi_0)/4$ to eliminate trivial 
density effects that do not come
from the shear stress and pressure evolution (see text). 
Approximate results with dropping $\pi_L^2$ terms
in the equation of motion are also shown (thin dotted lines).}
\label{fig:alpha0}
\end{figure}

Figure~\ref{fig:dSdeta} shows entropy production $dS/d\eta$ as a function
of proper time for the four scenarios, with local thermal $(\xi_0=0)$ and 
chemical $(\chi_0 = 0)$ equilibrium initial
conditions. 
Due to scalings, only entropy relative to the initial one
plays a role
\be
\frac{(dS/d\eta)}{(dS_0/d\eta)} = \tilde\tau  \tilde{\bar s}
= 1 + \frac{1}{4-\chi_0}
       \left[\ln(\tilde\tau^4 \tilde p^3) - \frac{9\xi^2(\tau)}{16}\right] \ .
\label{deltaS}
\ee
For $K_0 = 1$, a constant cross section generates about $35$\% extra 
entropy by late $\tau \sim 15-20 \tau_0$. With $\sigma \propto 1/T^2$, 
the increase is only $\sim 30$\%, while $\sigma \propto \tau^{2/3}$ and 
$\etaOs = const$ give the smallest increase of about $25$\%.
For a larger $K_0 \sim 7$, the system is much closer to ideal 
hydrodynamics and therefore entropy generation is slower - about $10$\% for
$\sigma = const$, while only $5$\% for the other three cases.
Note that these results also depend on $\chi_0$, but almost entirely 
through the explicit
$1/(4-\chi_0)$ factor in (\ref{deltaS}). Therefore, results
for arbitrary $\chi_0 \ne 0$ can be obtained 
via straightforward rescaling.
In the $\etaOs = const$ case the shear stress and pressure 
evolution also depend on $\chi_0$ but only very weakly as we show later below
(cf. Figure~\ref{fig:alpha0}).

\begin{figure}
\epsfysize=6.5cm
\epsfbox{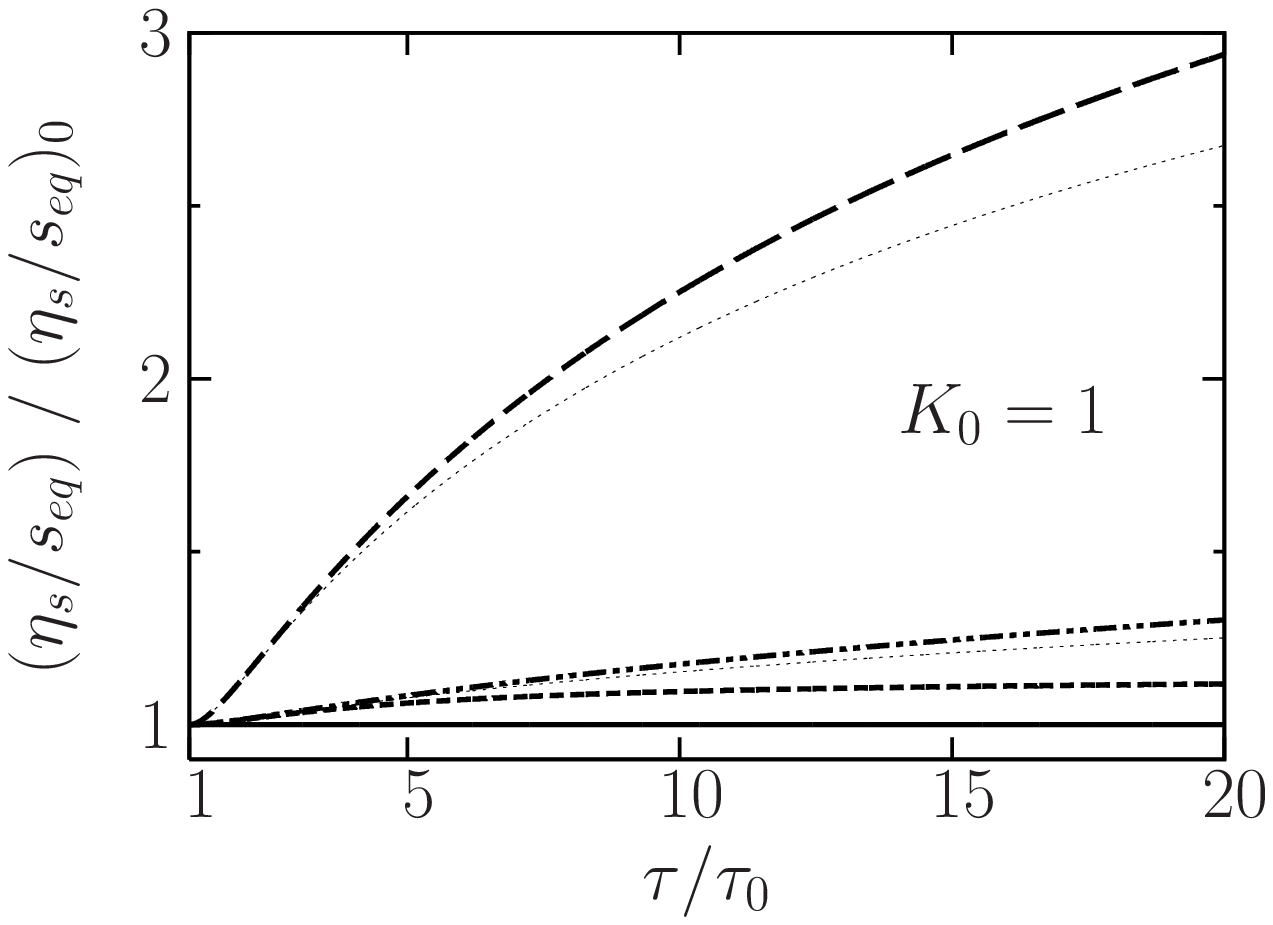}
\epsfysize=6.5cm
\epsfbox{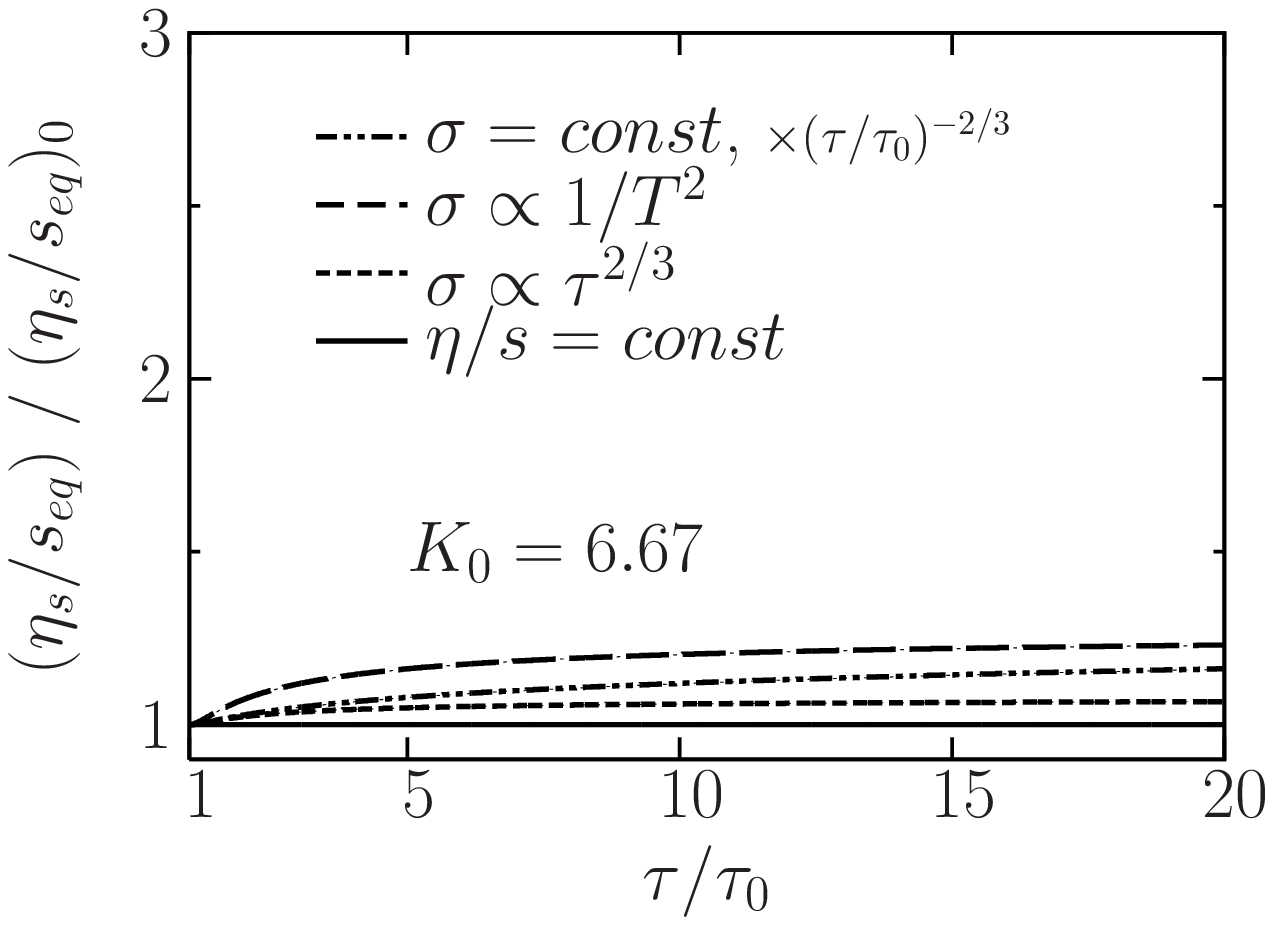}
\caption{Same as Fig.~\ref{fig:p} but 
for the shear viscosity to equilibrium entropy density ratio $\etaOs$.
The results for $\sigma = const$ are divided by $(\tau/\tau_0)^{2/3}$,
otherwise they would quickly grow off the plot.}
\label{fig:etas}
\end{figure}

Figure~\ref{fig:etas} shows the evolution of the shear viscosity to equilibrium
entropy density ratio $\etaOs$, normalized by the initial value of
the ratio. The entropy is calculated for a system starting from chemical
equilibrium $(\chi_0 = 0)$. 
The rough expectations that 
$\etaOs \sim \tau^{2/3}$ for a constant cross section, while 
$\etaOs\sim const$
for both $\sigma \propto 1/T^2$ and $\sigma \propto \tau^{2/3}$,
hold within a factor of three already for $K_0 = 1$ and 
up to $\tau = 20 \tau_0$
(note that the $\tau^{2/3}$ growth in the $\sigma = const$ case has been scaled
out in the plots). Relative to this ``zeroth order'' behavior,
for all three scenarios, $\etaOs$ grows with time, reinforcing
the general results in Sec.~\ref{Sec:IS_etas}. The relative 
growth decreases with increasing $K_0$. The $K_0$ dependence is strongest
for the constant cross section scenario: the factor of three gain
by $\tau = 20 \tau_0$ for $K_0 = 1$ is tamed to an about $25$\% increase
for $K_0 \sim 7$. For the other two scenarios, $\sigma \propto 1/T^2$ and 
$\sigma \propto \tau^{2/3}$, the ratio stays nearly constant much more 
robustly. As expected (cf. end of Sec.~\ref{Sec:IS_etas}), 
{\em of all cases studied 
$\sigma \propto \tau^{2/3}$ approximates $\etaOs = const$ the best,
with only $\sim 10$\% deviation accumulated by late $\tau = 20 \tau_0$ 
even for a small $K_0 = 1$.}

Finally, in Figure~\ref{fig:alpha0} 
we show that the results for $\etaOs = const$ depend only weakly
on the initial density, i.e., $\chi_0$. 
Density dependence in shear stress and pressure evolution
arises in this case because the cross section
is a function of the initial density (see (\ref{K_etas})).
The dependence is weaker, the closer the system is to
ideal hydrodynamics because in that case $p\propto \tau^{-4/3}$ and 
$\chi_0$ drops out from $K(\tau)$. But even for a pessimistic $K_0 = 1$,
the pressure anisotropy (left plot),
varies less than $10$\%
as we change the density by a factor of 4
around chemical equilibrium density ($\chi_0 = \pm \ln 4$).
In fact a decrease in the density has a much weaker effect than an increase.
The right plot shows the effect of the same initial density 
variation on entropy $dS/d\eta$ 
production
normalized to the initial entropy.
Most of the density dependence in the entropy change comes from the trivial
$1/(4-\chi_0)$ prefactor in (\ref{deltaS})
which is there in any cross section scenario 
even if the shear stress and pressure
evolution are independent of the density.
To highlight dynamical density effects, we therefore plot, again for a
pessimistic $K_0 = 1$, the normalized change in entropy
\be
\frac{4-\chi_0}{4} \frac{\Delta (dS/d\eta)}{(dS_0/d\eta)} 
\equiv \frac{4-\chi_0}{4} \left(\frac{(dS/d\eta)}{(dS_0/d\eta)} - 1 \right) 
\ee
(the scaling factor is chosen such that it has 
no effect for chemical equilibrium initial conditions $\chi_0 = 0$).
The results show practically no density dependence,
apart from few-percent changes, even for such a low $K_0$.

\section{Useful relations from covariant transport}
\label{App:transp}

\subsection{Particle number and transverse energy}
\label{App:Et}

The particle number and transverse energy distributions for particles crossing
a 3D hypersurface $\sigma(x) = const$ are given by
\bea
dN &=& dy dp_T^2 \int p^\mu d\sigma_\mu(x) f(x, \vp) \\
dE_T &=& dy dp_T^2 \int p^\mu d\sigma_\mu(x) m_T f(x, \vp)
\eea
where $m_T \equiv \sqrt{p_T^2 + m^2}$, 
$p_T \equiv \sqrt{p_x^2 + p_y^2}$
is the transverse momentum, and $d\sigma_\mu(x)$ is the normal to the 
hypersurface at space time coordinate $x$. 
For our boost-invariant scenario it is natural to follow quantities
per {\em unit coordinate rapidity} as a function of the proper
time $\tau$. For $\tau = const$ hypersurfaces
$p^\mu d\sigma_\mu = m_T\, \tau \ch\omega\, d^2 x_T d\eta$,
and in our 0+1D case, 
$f$ only depends on $\sh \omega$, $p_\perp$, and $\tau$, where 
$\omega \equiv y - \eta$. Thus,
\bea
\frac{dN(\tau)}{d\eta}
&=& \tau A_T \int d^2\, p_T\, d\omega \ m_T\, \ch \omega  \,
f(\tau, \sh \omega, p_T) \\
\frac{dE_T(\tau)}{d\eta}
&=& \tau A_T \int d^2\, p_T\, d\omega \  m_T^2\, \ch \omega  \,
f(\tau, \sh \omega, p_T) \ .
\eea
$A_T$ is the transverse area of the system.
With the local thermal equilibrium distribution for ultrarelativistic 
particles
\be
f(\sh\omega, p_\perp) = \cN\, e^{-p_\perp \ch \omega /T} 
\ , \qquad \cN = \frac{n}{8\pi T^3} 
\label{f0_eq}
\ee
and the quadratic form (\ref{phiGpi}), straightforward integration gives
\bea
\frac{dN}{d\eta} &=& n\,\tau\,  A_T  = const\\
\frac{dE_T(\tau)}{d\eta} 
&=& \frac{3\pi T}{4} \frac{dN}{d\eta} \left(1-\frac{5\xi}{16}\right) \ .  
\eea
Clearly, dissipation slows the decrease of transverse energy 
(for typical $\pi_L < 0$), and $2\to 2$ 
interactions of course course conserve particle number.

Note that
$dE_T/d\eta / (\tau A_T)$ is 
almost identical to the transverse pressure (\ref{Tmunu_tr}), but has an
extra $\ch\omega$ factor in the integrand.

\subsection{Early pressure evolution}
\label{App:early}

Here we evaluate the early transverse and longitudinal pressure evolution
from the transport for a local equilibrium initial condition. The results
hold for {\em any} interaction, not only $2\to 2$.

In local equilibrium the collision term vanishes, thus in the vicinity of 
$\tau = \tau_0$ the evolution is governed by {\em free streaming}. 
In our 0+1D case, free streaming
\be
\left[\ch \omega \,\partial_\tau - \frac{\sh\omega}{\tau} 
\partial_\omega\right] 
f(\sh \omega, p_\perp, \tau) = 0  \qquad\qquad (\omega \equiv y - \eta)
\ee
implies
\be
f(\sh\omega, p_\perp, \tau) = f(\tau \sh\omega / \tau_0, p_\perp, \tau_0) \ .
\label{streaming_solution}
\ee
Substituting a local thermal initial distribution for ultrarelativistic 
particles (\ref{f0_eq}),
the definition of the energy-momentum tensor
\be
T^{\mu\nu}(\eta=0,\tau) = \int \frac{d^3p}{p_0} p^\mu p^\nu\, f = 
\int d^2 p_\perp\, dy\, p^\mu p^\nu\, f(\sh y, p_\perp, \tau)
\label{Tmunu_tr}
\ee
gives the transverse pressure
\bea
p_T(\tau) \equiv T^{xx}(\eta = 0, \tau) 
&=& \cN \int dp_\perp\, p_\perp d\phi \, dy\, 
(p_\perp \cos\phi)^2 \, 
\exp\!\left[-\frac{p_\perp}{T_0} \sqrt{1+a^2 \sh^2 y}\right]
=  \frac{3T_0 n}{2} 
\int\limits_0^\infty \frac{dy}{(1+a^2 \sh^2 y)^2}
\eea
Here $a \equiv \tau/\tau_0$. Change of variables to $q = a\, \sh y$ leads to
\be
p_T(\tau) = \frac{3T_0 n}{2} 
\int\limits_0^\infty \frac{dq}{(1+q^2)^2 \sqrt{q^2 + a^2}} 
= T_0 n \, 
\frac{3 \left[\sqrt{a^2-1} + (a^2 - 2)\, \acos \frac{1}{a} \right]}
     {4 (a^2 - 1)^{3/2}} \ .
\ee
Analogous calculation gives for the longitudinal pressure
\bea
p_L(\tau) \equiv T^{zz}(\eta = 0,\tau) 
= 3T_0 n \int \frac{dy\,\sh^2 y}{(1+a^2 \sh^2 y)^2}
= T_0 n \, \frac{3}{2(a^2-1)} 
 \left[\frac{\acos \frac{1}{a}}{\sqrt{a^2-1}} - \frac{1}{a^2}\right] \ .
\eea
Expanding near $a=1$, 
\bea
p_T(\tau) &=& T_0 n \left[1 - \frac{4 (\tau-\tau_0)}{5\tau_0} 
                                + \cO((\tau-\tau_0)^2)\right] 
\\
p_L(\tau) &=& T_0 n \left[1 - \frac{12 (\tau-\tau_0)}{5\tau_0} 
                                + \cO((\tau-\tau_0)^2)\right]
\eea
and thus (\ref{R_early_tr}) follows.


\end{document}